\documentclass[superscriptaddress,amssymb,twocolumn,aps,longbibliography,prb]{revtex4-1}

\usepackage{graphicx}
\usepackage{dcolumn}
\usepackage{bm}
\usepackage{physics}
\usepackage{xcolor}
\usepackage{mathtools}
\usepackage[justification=RaggedRight, singlelinecheck=false, caption=false]{subfig}
\usepackage[colorlinks=true,linkcolor=blue]{hyperref}

\begin{document}

\title{Periodic, quasi-periodic, fractal, Kolakoski and random binary polymers: \\ Energy structure and carrier transport}

\author{K. Lambropoulos}
 \email{klambro@phys.uoa.gr}
\affiliation{National and Kapodistrian University of Athens, Department of Physics, Panepistimiopolis, 15784 Zografos, Athens, Greece}
\author{C. Simserides}
\email{csimseri@phys.uoa.gr}
\affiliation{National and Kapodistrian University of Athens, Department of Physics, Panepistimiopolis, 15784 Zografos, Athens, Greece}

\date{\today}

\begin{abstract}
We study periodic, quasi-periodic (Thue-Morse, Fibonacci, Period Doubling, Rudin-Shapiro), fractal (Cantor, generalized Cantor), Kolakoski and random binary sequences using a tight-binding wire model, where a site is a monomer (e.g., in DNA, a base pair). We use B-DNA as our prototype system. All sequences have purines, guanine (G) or adenine (A) on the same strand, i.e., our prototype binary alphabet is (G,A). Our aim is to examine the influence of sequence intricacy and magnitude of parameters on energy structure, localization and charge transport. We study quantities such as autocorrelation function, eigenspectra, density of states, Lyapunov exponents, transmission coefficients and current-voltage curves. We show that the degree of sequence intricacy and the presence of correlations decisively affect the aforementioned physical properties. Periodic segments have enhanced transport properties. Specifically, in homogeneous sequences transport efficiency is maximum. There are several deterministic aperiodic sequences that can support significant currents, depending on the Fermi level of the leads. Random sequences is the less efficient category.
\end{abstract}

\maketitle

\section{Introduction} \label{sec:Introduction}
We focus on periodic, aperiodic and random binary sequences, i.e., sequences based on a binary alphabet, like $\{0,1\}$. We use B-DNA as a prototype system and investigate sequences based on the couple $\{$G,A$\}$. This means that in one strand of double helix B-DNA we have either Guanine (G) or Adenine (A), and of course, in the complementary strand we have Cytosine (C) and Thymine (T), respectively. The persistence length $\ell_p$ of a polymer somehow quantifies its stiffness, in the sense that pieces shorter than $\ell_p$ behave rather like a flexible elastic beam, while much longer pieces are more likely to bend. DNA is among the stiffest of known polymers with $\ell_p \approx$ 50 nm or 150 base pairs~\cite{Manning:2006}. This is one of the reasons we chose B-DNA as our prototype system, along with its biological and nanoscientific importance. On the other hand, if we stretch and join the DNA of all chromosomes of a single cell, that would give us a length of the order of a meter and would consist of billions of base pairs.

DNA is fundamental to living organisms because the sequence of its bases (adenine, guanine, thymine, cytosine) carries their genetic code. Its remarkable properties have drawn the interest of a broad interdisciplinary scientific community, beyond molecular biology and genetics. From a physics point of view, its electronic structure and its charge transfer and transport properties properties are studied with the aim to understand its biological functions and their potential applications in nanotechnology (e.g., nanocircuits, molecular wires)~\cite{Wohlgamuth:2013, LewisWasielewski:2013}. The base-pair stack of the double-helix DNA structure creates a nearly one-dimensional $\pi$-pathway that favors charge transfer and transport. The term \textit{transfer} means that a carrier, created (e.g., by oxidation or reduction) or injected at a specific place, moves to a more favorable location, while the term \textit{transport} implies the use of electrodes between which electric voltage is applied.

Recent research has shown that carrier movement through DNA can be manipulated. For example, the carrier transfer rate through DNA can be tuned by chemical modification, e.g. using various natural and artificial nucleobases with different highest occupied molecular orbital (HOMO) levels~\cite{KawaiMajimaBook:2015}. Transfer rates can be increased by many orders of magnitude with appropriate sequence choice~\cite{LChMKTS:2015, LChMKLTTS:2016, LVBMS:2018}. Furthermore, dynamical fluctuations, arising from either solvent fluctuations or base-pair vibrations can gate charge transport, counteracting the intrinsic disordered potential profile of the sequence~\cite{Gutierrez:2010}.

Many external factors (such as aqueousness, counterions, extraction process, electrodes, purity, substrate), influence carrier motion along DNA~\cite{MaciaBook:2008}. Hence, the need for a better understanding of the intrinsic factors that affect charge transfer and transport, such as geometry and base-pair sequence, arises. \textit{Ab initio} calculations \cite{YeShen:2000, YeJiang:2000, Barnett:2003, Artacho:2003, Voityuk:2008, Kubar:2008, TMLS:2017}
and model Hamiltonians \cite{Simserides:2014, LChMKTS:2015,  LChMKLTTS:2016, LVBMS:2018, Cuniberti:2002, Roche-et-al:2003, Roche:2003, Palmero:2004, Yamada:2004, Klotsa:2005, Shih:2008, Joe:2010, Caetano:2005, WangChakrabort:2006} have been used to explore the variety of experimental results and the underlying mechanisms. The former are currently limited to short segments for computational reasons, while the latter allow to address systems of realistic length. Here we study rather long sequences, hence we adopt the latter approach. The aim of this work is a comparative examination of the influence of base-pair sequence on charge transport.

Several works have been devoted to the study of transfer and transport in specific DNA structures (periodic~\cite{LChMKTS:2015, LChMKLTTS:2016, LVBMS:2018, Macia:2005, Sarmento:2009}, quasiperiodic~\cite{Macia:2006, Sarmento:2012, Albuquerque:2014}, random and natural~\cite{Roche-et-al:2003, Roche:2003, Paez:2012, Kundu:2015, Fathizadeh:2018}) using variants of the Tight-Binding (TB) method. Here, we employ the TB wire model, with the sites of the chain being the base pairs, to study the spectral, localization and charge transport properties of periodic, deterministic aperiodic [Thue-Morse ($\textit{TM}$), Fibonacci ($\textit{F}$), Period Doubling ($\textit{PD}$), Rudin-Shapiro ($\textit{RS}$), Cantor set ($\textit{CS}$), generalized Cantor set ($\textit{GCS}$), Kolakoski ($\textit{KOL}$)] and random DNA binary segments. 

We use a TB parametrization that allows for different hopping (or transfer or coupling) parameters (or integrals). Such TB parametrizations for DNA have been derived by many scientists in many works and used with various TB models (wire, ladder, extended ladder, fishbone, etc). For example, 
for coupling parameters and on-site energies cf. Refs.~\cite{Voityuk:2008,Kubar:2008,HKS:2010-2011,Mehrez:2005}, 
for coupling parameters cf. Refs.~\cite{Voityuk:2001,Ivanova:2008,Migliore:2009}, and 
for on-site energies cf. Refs.~\cite{SugiyamaSaito:1996,HutterClark:1996,Zhang:2002,Li:2001,Li:2002,ShuklaLeszczynski:2002}. 
Roughly, the coupling integrals found in the literature are usually \cite{Simserides:2014} in the range 0.001 to 0.200 eV, although sometimes even smaller or larger values have been reported.
In Ref.~\cite{Senthilkumar:2005} there are some nice tables showing the variance of on-site energies and coupling parameters for different triplets (or triads) of base pairs. TB parameters may change at different levels of theory, and their values can tune the results, having both quantitative and qualitative effects. The analysis of what happens changing TB parameters may indicate future research directions.

When dealing with charge transport properties, it is usual in the literature to use only one hopping parameter and/or on-site energy, to simplify the problem. We go beyond these simplifying hypotheses in the present manuscript. This leads to quantitative and qualitative consequences. Our treatment gives a clearer picture, as it will be discussed below. In this spirit, we calculate -among other quantities- autocorrelation functions, integrated density of states, Lyapunov exponents, transmission coefficients and current-voltage (I-V) curves taking into account the different on-site energies as well as the different hopping parameters.

The rest of the paper is organized as follows: In Sec.~\ref{sec:TB-TM} we outline our TB and transfer matrix method (TMM) framework. In Sec.~\ref{sec:Sequences} we present the studied sequences. In Sec.~\ref{sec:SequenceProperties} we focus on the occurrence percentages of on-site energies, hopping parameters and triplets (a site and its previous and next neighbors) in the sequences. In Sec.~\ref{sec:ES-DOS-IDOS}, we discuss eigenspectra, density of states (DOS) and integrated density of states (IDOS). In Sec.~\ref{sec:LE} we present Lyapunov exponents, which characterize the localization length of eigenstates. In Sec.~\ref{sec:TC} we discuss zero-bias transmission coefficients. In Sec.~\ref{sec:IV}, we study I-V characteristics using the Landauer-B\"{u}ttiker formalism. In Sec.~\ref{sec:parametrization} we state some remarks on the effect the parameters have on the results. Finally, in Sec.~\ref{sec:Conclusion}, we state our conclusions.

\section{Tight-Binding and Transfer Matrix Method} \label{sec:TB-TM}
In the present work, we focus on  periodic, deterministic aperiodic and random DNA segments  consisting of different base pairs with their purines (A and G) on the $5'$-$3'$ strand. We will use this strand to denote the segments. For example, the notation GGAG means that we have the GGAG bases in the $5'$-$3'$ strand and the complementary ones, CCTC, in the $3'$-$5'$ strand. All studied sequences start with G.

The TB system of equations for a DNA segment in the Wire Model~\cite{Cuniberti:2007,LChMKLTTS:2016} reads
\begin{equation} \label{Eq:TBsystem}
E \psi_n = E_n \psi_n + t_{n-1} \psi_{n-1} +  t_{n} \psi_{n+1},
\end{equation}
$ \forall n = 1, 2, \dots, N$, where $E$ is the eigenenergy, $E_n$ is the on-site energy of base pair $n$, $|\psi_n|^2$ is the relevant occupation probability, and $t_{\ell}$ is the hopping integral between base pairs $l$ and $l+1$. The on-site energies are taken $E_{A-T}=-8.3$ eV for the A-T base pair and $E_{G-C}=-8.0$ eV for the G-C base pair.~\cite{HKS:2010-2011, Simserides:2014, LChMKTS:2015, LChMKLTTS:2016, LKMTLGTChS:2016} 
The hopping integrals between successive base pairs that are involved in the segments studied here are shown in Table \ref{table:parameters}.\cite{HKS:2010-2011, Simserides:2014, LChMKTS:2015, LChMKLTTS:2016, LKMTLGTChS:2016} The values of the parameters correspond to the HOMO of the base pairs and are discussed in Ref.~\cite{Simserides:2014}.

\begin{table} [h]
	\centering
	\caption{HOMO Hopping integrals, $t_{rc}^{53}$, between successive base pairs involved in the segments studied in this work, in the $5'$-$3'$ direction. $r(c)$ stands for the base pair in the row (column) of the table.}
	\begin{tabular}{c|c|c}
		$t_{rc}^{53}$ (eV) & G & A \\ \hline
		G & $-$0.100 & $-$0.110\\  \hline
		A & $-$0.030 & $-$0.020\\
	\end{tabular}
	\label{table:parameters}
\end{table}

Eq. \eqref{Eq:TBsystem} can equivalently be solved using the TMM, by rewriting it in the matrix form
\begin{equation} \label{Eq:TBsystemTM}
\begin{pmatrix}
\psi_{n+1}\\
\psi_n
\end{pmatrix}
= P_n(E)
\begin{pmatrix}
\psi_n\\
\psi_{n-1}
\end{pmatrix},
\end{equation}
where
\begin{equation} \label{Eq:TM}
P_n(E) =
\begin{pmatrix}
\frac{E-E_n}{t_{n}} & -\frac{t_{n-1}}{t_{n}}\\
1 & 0
\end{pmatrix}
\end{equation}
is the Transfer Matrix of base pair $n$. The product
\begin{equation} \label{Eq:GTM}
M_N(E) = \prod_{n=N}^{1}P_n(E)
\end{equation}
defines the Global Transfer Matrix (GTM) of the segment, containing all the information about its energetics.
The elements of the GTM are recurrently given by
\begin{subequations} \label{Eq:UCTMrec}
	\begin{equation} \label{Eq:UCTMrec1}
	M_N^{11(12)} = \frac{E-E_N}{t_N} M_{N-1}^{11(12)} -   \frac{t_{N-1}}{t_N} M_{N-2}^{11(12)}
	\end{equation}
	\begin{equation} \label{Eq:UCTMrec2}
	M_N^{21(22)} = M_{N-1}^{11(12)}
	\end{equation}
\end{subequations}
with initial conditions $M_1^{11} = (E-E_1)/t_1$, $M_1^{12} = -t_N/t_1$, $M_0^{21} = 1$, $M_0^{22} = 0$. $M^{ij}$ is the element $ij$ of matrix $M$. If we cyclically bound the segment, the GTM is a symplectic matrix, hence it is always unimodular.

\section{Sequences} \label{sec:Sequences}
We denote periodic segments by (XY\dots Z)$_m$, where $m$ is the total number of repetition units. The categories, substitution rules and substitution matrices of the studied deterministic aperiodic sequences can be found below. Using, e.g., the binary alphabet $\{$i, j$\}$, the substitution matrix $S$ of a given sequence has elements $S_{i, j} = n_i[s(j)]$, where $n_i[s(j)]$ is the number of times $i$ is present in the substitution rule $s(j)$. Apart from the fractal (Cantor and Generalized Cantor) and Kolakoski$\{1,2\}$ sequences, the rest of the cases studied here have primitive substitution matrices $S$ (a matrix $S$ is primitive if it is non-negative and there is a $n\in \textit{N}$ such that $S^n$ is positive).
\\
\subsection{Fibonacci} \label{subsec:Fib}

The Fibonacci sequence, named after the Italian mathematician Leonardo Pisano (Fibonacci) who introduced it in his 1212 book \textit{Liber Abaci}, in a study of the population growth of rabbits~\cite{Fibonacci:1212}, is a number sequence the terms of which are generated by the addition of the two previous terms, with given initial conditions. 
However, this sequence appears many centuries before in Indian mathematics, in connection with Sanskrit prosody. For example, the possible ways to arrange short (S) and double, long (L) syllables with given total duration measured as $g$ S syllables is the Fibonacci number of the $g+1$ generation. If $\mathcal{N}_g$ is the Fibonacci number of generation $g$, and we set $\mathcal{N}_0 = \mathcal{N}_1 = 1$, the recurrence relation $\mathcal{N}_g = \mathcal{N}_{g-1} + \textit{N}_{g-2}$ produces the number sequence $1,1,2,3,5,8,13,21,34\dots$. 
Using the two-letter alphabet $\{$G, A$\}$, we can define the Fibonacci word $\textit{F}_g$ by the substitution rule $s$(A) $=$ G, $s$(G) $=$ GA, starting with $\textit{F}_0 =$ A. $\textit{F}_1 =$ G,  $\textit{F}_2 =$ GA,  $\textit{F}_3 =$ GAG,  $\textit{F}_4 =$  GAGGA, etc. Obviously, the length of the word $\textit{F}_g$ is  $\mathcal{N}_g$. The substitution matrix of the Fibonacci sequence is
	\begin{equation}
	S = \begin{pmatrix}
	1&1\\1&0
	\end{pmatrix}.
	\end{equation}

\subsection{Thue-Morse} \label{subsec:ThM}

The Thue-Morse (TM) sequence (aka Prouhet-Thue-Morse sequence) was studied by Eugene Prouhet in the field of number theory~\cite{Prouhet:1851}, defined by Alex Thue in the field of combinatorics~\cite{Thue:1906}, and rediscovered by Marston Morse in the context of differential geometry~\cite{Morse:1921}. It is a binary sequence of $0$s and $1$s, starting with $0$, with its $g^{\text{th}}$ generation constructed by appending the Boolean complement of the previous generation to the sequence. With the two-letter alphabet $\{$G, A$\}$, we can define the TM word $\textit{TM}_g$ by the substitution rule $s$(G) $=$ GA, $s$(A) $=$ AG, starting with $\textit{TM}_0 =$ G. $\textit{TM}_1 =$ GA,  $\textit{TM}_2 =$ GAAG,  $\textit{TM}_3 =$ GAAGAGGA, etc. The length of the word $\textit{TM}_g$ is  $2^g$. The substitution matrix of the TM sequence is
	\begin{equation}
	S = \begin{pmatrix}
	1&1\\1&1
	\end{pmatrix}.
	\end{equation}

\subsection{Period-Doubling} \label{subsec:PD}

The Period-Doubling (PD) sequence is closely connected with the TM sequence. Specifically, its elements are given by the first differences of the elements of the TM binary sequence modulo $2$. Using the two-letter alphabet $\{$G, A$\}$, we can define the PD word $\textit{PD}_g$ by the substitution rule $s$(G) $=$ GA, $s$(A) $=$ GG, starting with $\textit{PD}_0 =$ G. $\textit{PD}_1 =$ GA,  $\textit{PD}_2 =$ GAGG,  $\textit{PD}_3 =$ GAGGGAGA, etc. The length of the word $\textit{PD}_g$ is  $2^g$. The substitution matrix of the PD sequence is
	\begin{equation}
	S = \begin{pmatrix}
	1&2\\1&0
	\end{pmatrix}.
	\end{equation}

\subsection{Rudin-Shapiro} \label{subsec:RS}

The Rudin-Shapiro (RS, aka Golay-Rudin-Shapiro) sequence is the sequence of the appended coefficients of the RS polynomials~\cite{Shapiro:1951,Rudin:1959}. It contains only $\pm1$ and is generated by starting with $+1,+1$ and employing the rules
\begin{align*}
+1,+1 &\rightarrow +1,+1,+1,-1\\
+1,-1 &\rightarrow +1,+1,-1,+1\\
-1,+1 &\rightarrow -1,-1,+1,-1\\
-1,-1 &\rightarrow -1,-1,-1,+1.
\end{align*} 
Using the four-letter alphabet $\{$$i$ = GG, $j$ = GA, $k$ = AG, $\ell$ = AA$\}$, we can define the RS word $\textit{RS}_g$ by the substitution rule $s$(GG) $=$ GGGA, $s$(GA) $=$ GGAG, $s$(AG) $=$ AAGA, $s$(AA) $=$ AAAG, starting with $\textit{RS}_0 =$ GG. 
$\textit{RS}_1 =$ GGGA,  $\textit{RS}_2 =$ GGGAGGAG, etc. The length of the word $\textit{RS}_g$ is  $2^{g+1}$. The substitution matrix of the RS sequence is
	\begin{equation}
	S = \begin{pmatrix}
	1&1 &0&0\\1&0 &1&0\\0&1 &0&1\\0&0 &1&1
	\end{pmatrix}.
	\end{equation}

\subsection{Cantor Set} \label{subsec:CS}

The Cantor Set (CS), named after mathematician Georg Cantor who introduced it~\cite{Cantor:1883}, is one of the most well-known deterministic fractals. It is obtained as follows: given the continuous interval $[0,1]$, the middle third, $(\frac{1}{3},\frac{2}{3})$ is deleted, resulting in the union $[0,\frac{1}{3}]\cup[\frac{2}{3},1]$. Then, the open middle third of each remaining interval is deleted, and the process is repeated \textit{ad infinitum}. 
Using the two-letter alphabet $\{$G, A$\}$, we can define the CS word $\textit{CS}_g$ by the substitution rule $s$(G) $=$ GAG, $s$(A) $=$ AAA, starting with $\textit{CS}_0 =$ G. $\textit{CS}_1 =$ GAG, $\textit{CS}_2 =$ GAGAAAGAG, etc. All generations are palindromic words. The length of the word $\textit{CS}_g$ is $3^{g}$. The (non-primitive) substitution matrix of the CS sequence is
	\begin{equation}
	S = \begin{pmatrix}
	2&0\\1&3
	\end{pmatrix}.
	\end{equation}

\subsection{Generalized Cantor Set} \label{subsec:GCS}

In accordance with the rationale described above, one can imagine the construction of a generalized CS word, $\textit{GCS}_g(t,d)$, produced by the two-letter alphabet $\{$G, A$\}$, where $t$ is the total number of letters substituting each letter of the sequence in the next generation and $d$ is the number of letters that correspond to the ``deleted" middle segment ($t>d$). $t$ and $d$ are mutually odd or even, to preserve the palindromicity of the words. For example, the generalized word $\textit{GCS}_g(4,2)$ is given by the rule $s$(G) $=$ GAAG, $s$(A) $=$ AAAA, starting with $\textit{GCS}_0(4,2) =$ G. The length of the word $\textit{GCS}_g(t,d)$ is $t^{g}$. The (non-primitive) substitution matrix of the generalized CS sequence is
	\begin{equation}
	S = \begin{pmatrix}
	t-d&0\\d&t
	\end{pmatrix}.
	\end{equation}

\subsection{Kolakoski} \label{subsec:Kol}

The Kolakoski $\{p,q\}$ sequences are a family of sequences of the integers $p \neq q$ that are their own run length encodings (a \textit{run} is defined here as the maximal subsequence of identical numbers). The classic and most well known sequence of this class, Kolakoski$(1,2)$~\cite{OEIS:A000002}, also referred to as Oldenburger-Kolakoski sequence, was popularized by recreational mathematician William Kolakoski~\cite{Kolakoski:1965}, but it was independently introduced by Rufus Oldenburger~\cite{Oldenburger:1936}. This family of sequences possesses different properties in different cases. For example, for specific values of $p$ and $q$, they may show pure-point or continuous diffraction spectra~\cite{Sing:2004}.  Each generation, Kol$_g(p,q)$, of the sequences can be seen as the run length encoding of the next generation, starting with Kol$_0(p,q) = q^p$ and following the substitution rule
\begin{align*}
s(q) &= p^q \qquad \text{if $q$ was at odd $n$,}\\
s(q) &= q^q \qquad \text{if $q$ was at even $n$,}\\
s(p) &= p^p \qquad \text{if $p$ was at odd $n$,}\\
s(p) &= q^p \qquad \text{if $p$ was at even $n$.}\\
\end{align*} 
For example 
$\textit{KOL}_0(1,2) = 2$, 
$\textit{KOL}_1(1,2) = 11$, 
$\textit{KOL}_2(1,2) = 12$,  
$\textit{KOL}_3(1,2) = 122$, 
$\textit{KOL}_4(1,2) = 12211$, 
$\textit{KOL}_5(1,2) = 1221121$, etc. 
Accordingly, using the two-letter alphabet $\{$G, A$\}$, we can define the $\textit{KOL}(p,q)$ word $\textit{KOL}_g(p,q)$ by assigning G to $p$ and A to $q$. Thus, e.g., $\textit{KOL}_5(1,2)$ = GAAGGAG. The length of $\textit{KOL}(1,2)$ as the generation increases is given by the OEIS sequence A001083~\cite{OEIS:A001083}. Generally, the length of the word $\textit{KOL}_g(p,q)$ is equal to the sum of the terms of $\textit{KOL}_{g-1}(p,q$). 

Here, we focus on the Kolakoski$(1,2)$ and $(1,3)$ sequences. The former (and generally cases where $p$ and $q$ have different parity) cannot be associated with a primitive substitution matrix. The latter (and generally cases where $p = 2\mu+1$ and $q = 2\nu+1$) can alternatively be constructed by the three-letter alphabet $\{$$i$ = $pp$, $j$ = $pq$, $k$ = $qq$$\}$ and the substitution rule $s(i) = i^\mu j k^\mu$, $s(j) = i^\mu j k^\nu$, $s(k) = i^\nu j k^\nu$. Hence, we arrive at the substitution matrix
	\begin{equation}
	S = \begin{pmatrix}
	\mu & 1 & \mu\\ \mu & 1 & \nu\\ \nu & 1 & \nu
	\end{pmatrix}.
	\end{equation}

\subsection{Substitution matrices and letter frequencies} \label{subsec:SM}

For deterministic aperiodic segments with primitive substitution rules [i.e., all cases studied here apart from the fractals and $\textit{KOL}(1,2)$], the frequencies of the letters G and A in each sequence can explicitly be determined. From the Perron-Frobenius theorem it follows that the substitution matrix $S$ has a unique, real, positive eigenvalue, $\lambda_{PF}$, and the corresponding eigenvector can be chosen to have strictly positive entries. The normalized components (such as their sum is one) of the right eigenvector associated with $\lambda_{PF}$ give the asymptotic relative frequencies of the letters G and A~\cite{Baake:2013}. In the fractal sequences, these frequencies asymptotically reach 100\% and 0\%, respectively; in $\textit{KOL}(1,2)$ sequence, they are conjectured to be 50\% for each letter~\cite{Keane:1991} [cf. Fig. \ref{fig:percentage}(h)].

For segments with primitive substitution rules, we can also determine the frequencies of the legal $k$-letter words in the sequence. We present a way to analytically obtain these frequencies, based on the following proposition~\cite{Baake:2013}: If $s$ is the primitive substitution rule of a sequence based on the alphabet $A$, $W = \{w = w_1w_2\dots w_k, \forall w_i \in A\}$ is the set of the legal $k$-letter words in the sequence, and $s(w) = w_1^\prime w_2^\prime\dots w_n^\prime$, then the induced substitution $s_k(w) = (w_1^\prime w_2^\prime\dots w_k^\prime)(w_2^\prime w_3^\prime\dots w_{k+1}^\prime)\dots(w_l^\prime w_{l+1}^\prime\dots w_{l+k-1}^\prime)$, where $l$ is the length of the word $s(w_1)$, is also primitive.

Hence, a primitive induced substitution matrix $S_k$ can be defined, from which we can find the asymptotic relative frequencies of the legal $k$-letter words from the Perron-Frobenius theorem. For sequences in which $S$ is defined with the help of another alphabet [i.e., $\textit{RS}$, and $\textit{KOL}(1,3)$, cf. Subsec.~\ref{subsec:RS}, Subsec.~\ref{subsec:Kol}], these frequencies can be deduced from the possible $2k$-letter words of the helping alphabet.

Below, we demonstrate the procedure to determine the induced substitution matrix $S_3$ of the possible 3-letter words of the Period-Doubling sequence, for illustration. In this case, $k=3$, $W = \{\text{GGG, GGA, GAG, AGG, AGA}\}$ (cf. Sec.~\ref{sec:SequenceProperties}) and $l$ is always $2$. Hence,
\begin{align*}
s(\text{GGG}) = \text{GAGAGA} &\rightarrow s_3(\text{GGG}) = \text{(GAG)(AGA)},\\
s(\text{GGA}) = \text{GAGAGG} &\rightarrow s_3(\text{GGA}) = \text{(GAG)(AGA)},\\
s(\text{GAG}) = \text{GAGGGA} &\rightarrow s_3(\text{GAG}) = \text{(GAG)(AGG)},\\
s(\text{AGG}) = \text{GGGAGA} &\rightarrow s_3(\text{AGG}) = \text{(GGG)(GGA)},\\
s(\text{AGA}) = \text{GGGAGG} &\rightarrow s_3(\text{AGA}) = \text{(GGG)(GGA)}.
\end{align*}
So, the induced primitive substitution matrix $S_3$ reads
\begin{equation} \label{Eq:S3}
S_3 = \begin{pmatrix}
0 &0 &0 &1 &1\\
0 &0 &0 &1 &1\\
1 &1 &1 &0 &0\\
0 &0 &1 &0 &0\\
1 &1 &0 &0 &0\\
\end{pmatrix}.
\end{equation}

\section{Sequence properties} \label{sec:SequenceProperties}

\begin{figure}[!h]
	\centering
	\includegraphics[width=\columnwidth]{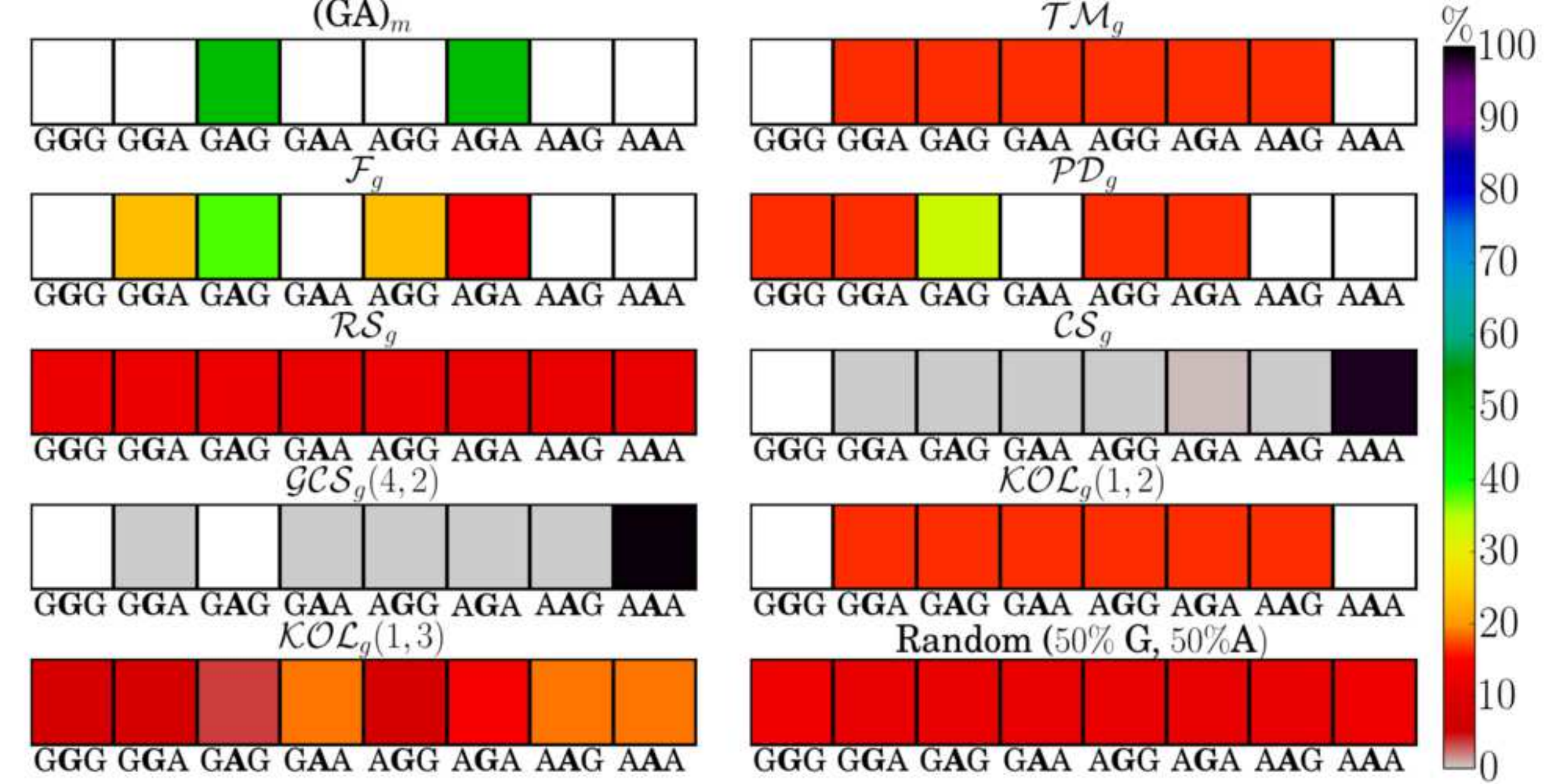}
	\caption{Classification of the DNA segments studied in this work based on the number and occurrence percentage of base-pair triplets. The boxes correspond to each of the 8 possible triplets. For each segment, white boxes correspond to forbidden triplets, and the color of the rest corresponds to their occurrence percentage (calculated for large $N$).}
	\label{fig:triplets}
\end{figure}

To obtain a clear picture of the interplay between sequence intricacy and energy profile of the segments, as well as its effect on localization and transport properties, we present some details on the sequence characteristics of the studied segments. 
We deal with binary sequences, that is sequences based on a binary alphabet, like $\{0,1\}$ or \{G,A\} in our case. Therefore, a useful classification of their properties can be done through the study of the different base-pair triplets that are found in each category~\cite{Macia:2017}. A triplet is made of a base pair and its next and previous neighbors. Since in a realistic treatment we need to simultaneously consider the difference in the on-site energies and the hopping integrals (as done here), the total number of possible triplets ($2^3$ for a binary sequence) corresponds to the total number of different transfer matrices that can be found in the GTM; cf. Eq.~\eqref{Eq:TM}. The number of triplets in each category of DNA segments as well as the occurrence percentage of each triplet (for large $N$) are depicted in Fig.~\ref{fig:triplets}. Finally, we notice, it has been claimed that the on-site energy of a base depends on its flanking bases, an idea beyond the scope of our present calculations \cite{Senthilkumar:2005}.

From Fig.~\ref{fig:triplets} it is obvious that the periodic (GA)$_m$ segment represents the most ordered case (2 triplets with equal occurrence percentages). 
$\textit{F}$ and $\textit{PD}$ segments possess 4 and 5 different triplets, respectively, and have one dominant GAG triplet. 
$\textit{TM}$ and $\textit{KOL}(1,2)$ segments posses 6 equidistributed triplets. 
$\textit{RS}$, random and $\textit{KOL}(1,3)$ segments posses all possible triplets; 
in the first two cases they are equidistributed; in the latter there are some predominant triplets. 
Finally, the Cantor Set family segments posses many of the possible triplets (7 for $\textit{CS}$, 6 for $\textit{GCS}(4,2)$). However, the AAA triplets are predominant, asymptotically reaching $100\%$ occurrence percentage as $N$ increases. For segments with primitive substitution rules, the values at which the occurrence percentage of each possible triplet converge can be found from the procedure described in Subsec.~\ref{subsec:SM}. For example, the occurrence percentages of the possible triplets in $PD$ segments converge to the components of the normalized right eigenvector corresponding to $\lambda_{PF} = 2$ of matrix $S_3$ [Eq.~\eqref{Eq:S3}], i.e. $[\frac{1}{6}, \frac{1}{6}, \frac{1}{3}, \frac{1}{6},\frac{1}{6}]^T$.

\begin{figure*}
	\centering
	\includegraphics[width=0.39\textwidth]{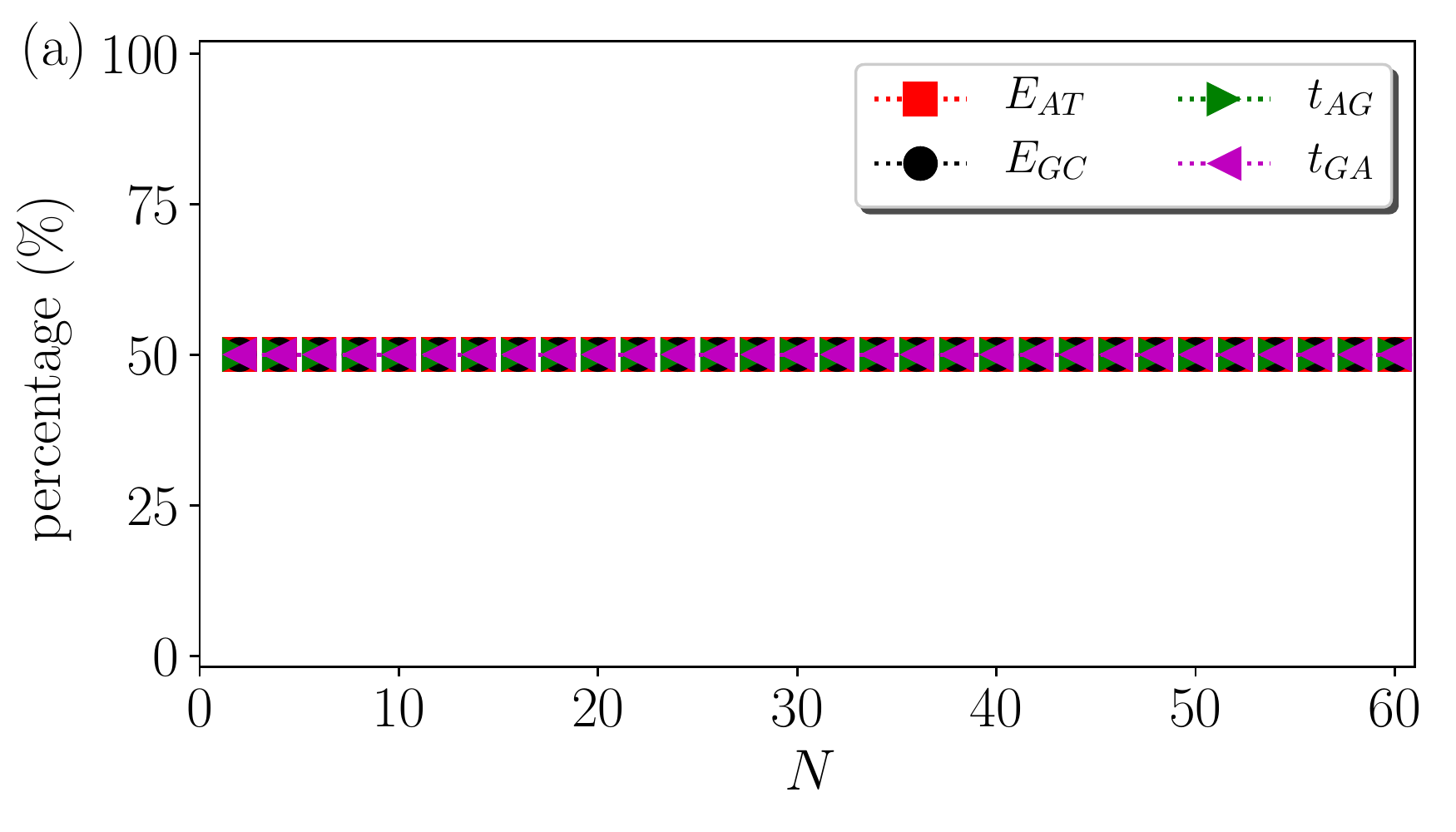}
	\includegraphics[width=0.39\textwidth]{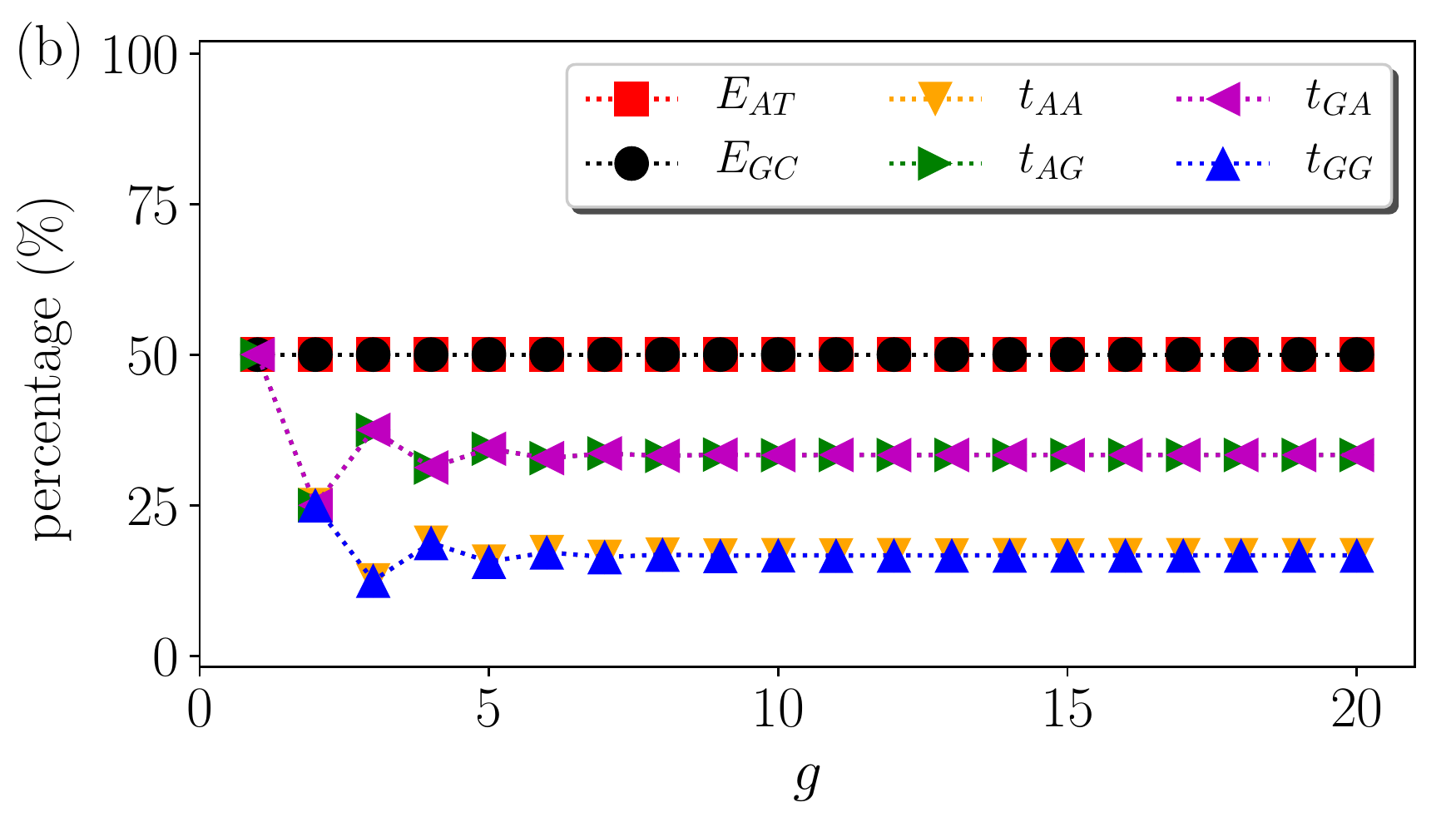}\\
	\includegraphics[width=0.39\textwidth]{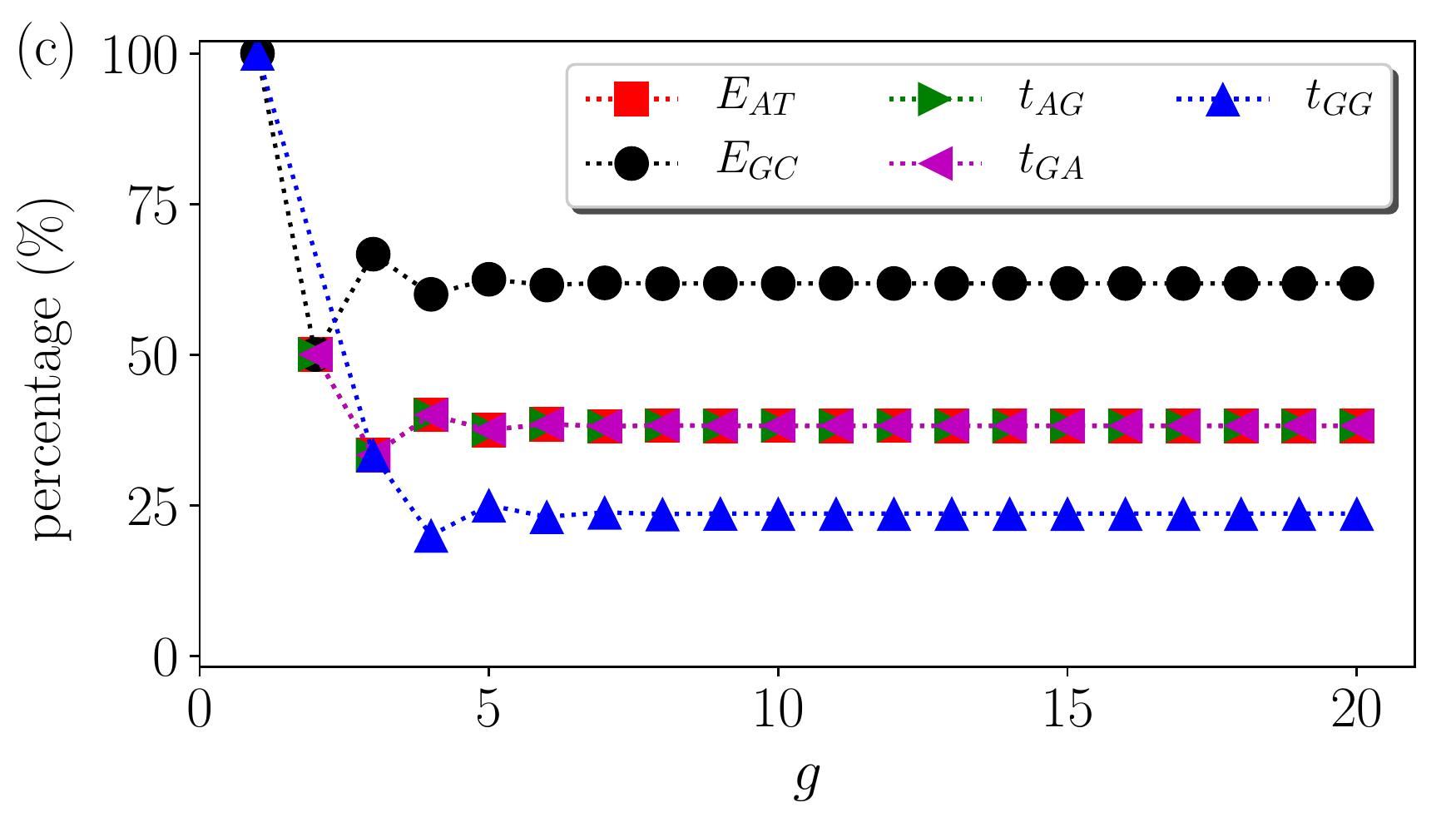}
	\includegraphics[width=0.39\textwidth]{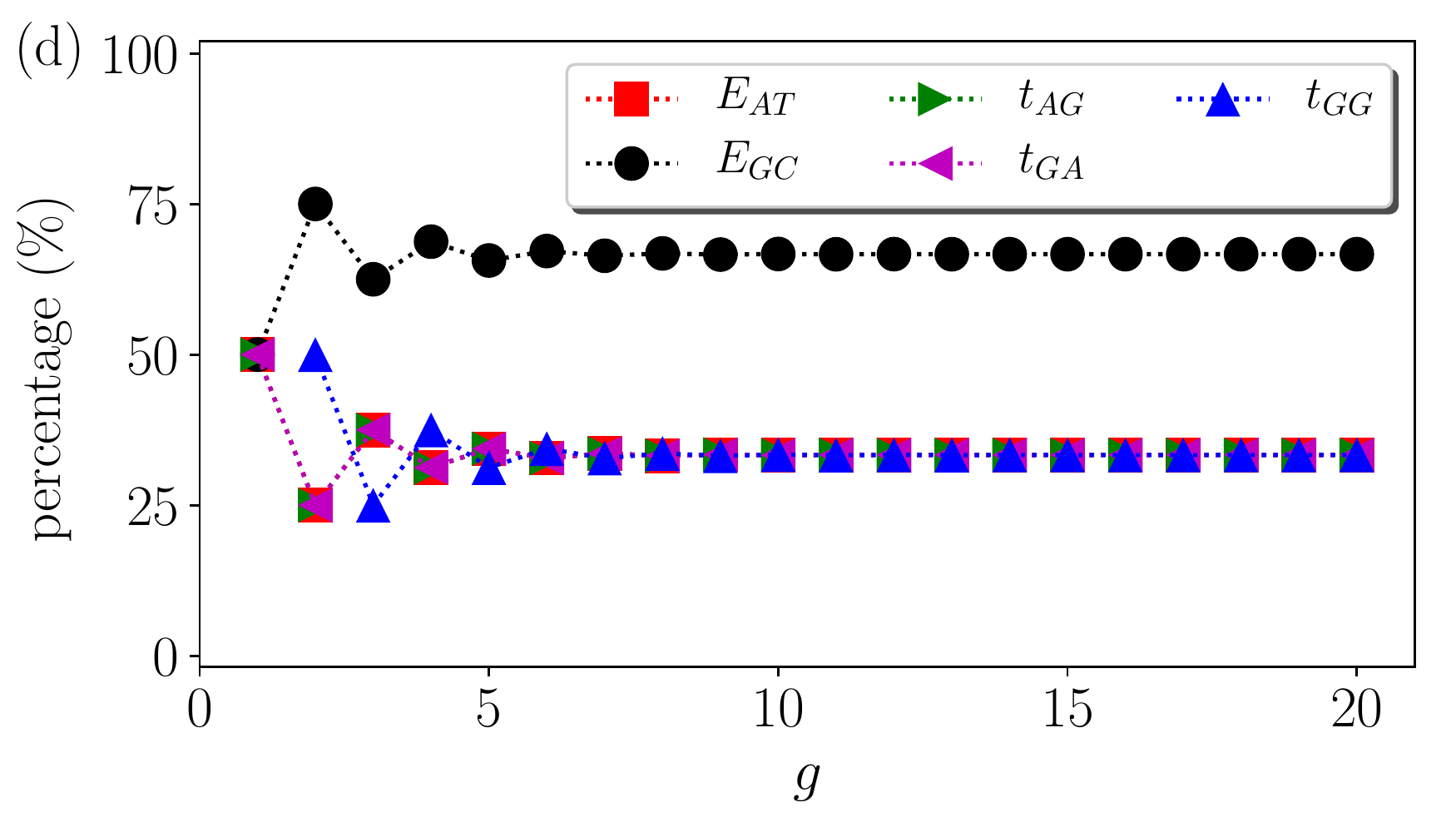}\\ 
	\includegraphics[width=0.39\textwidth]{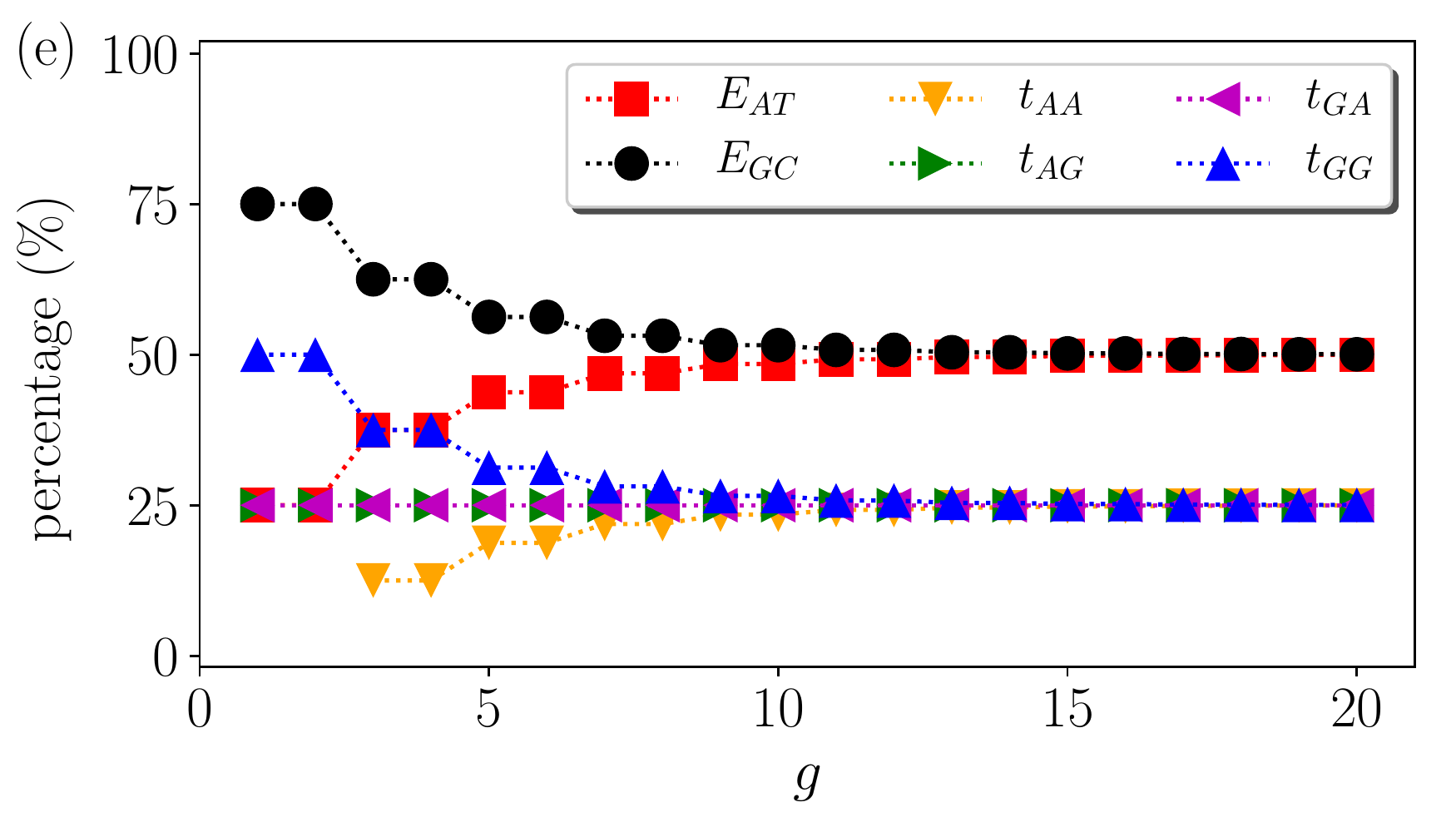}
	\includegraphics[width=0.39\textwidth]{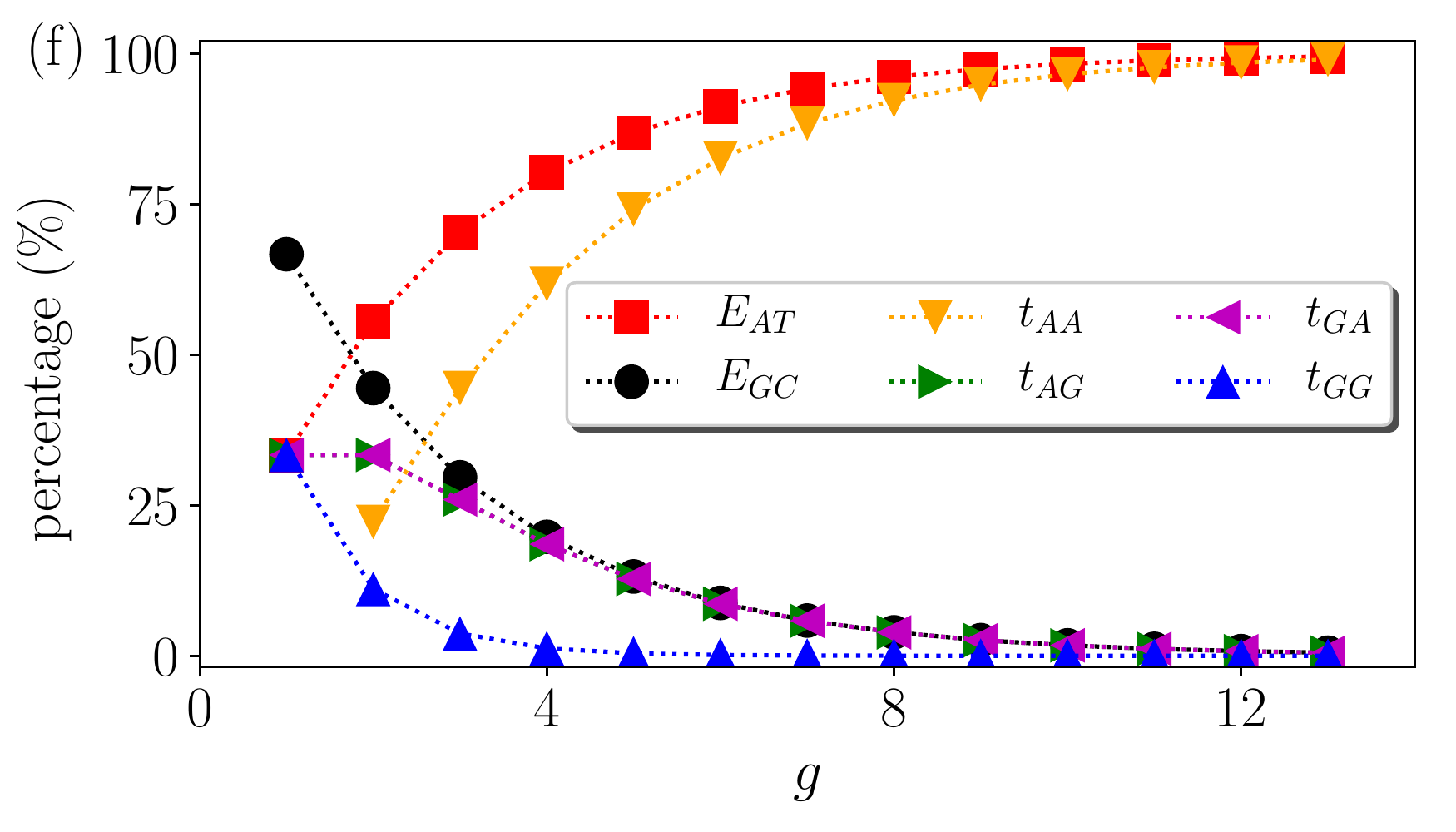}\\ 
	\includegraphics[width=0.39\textwidth]{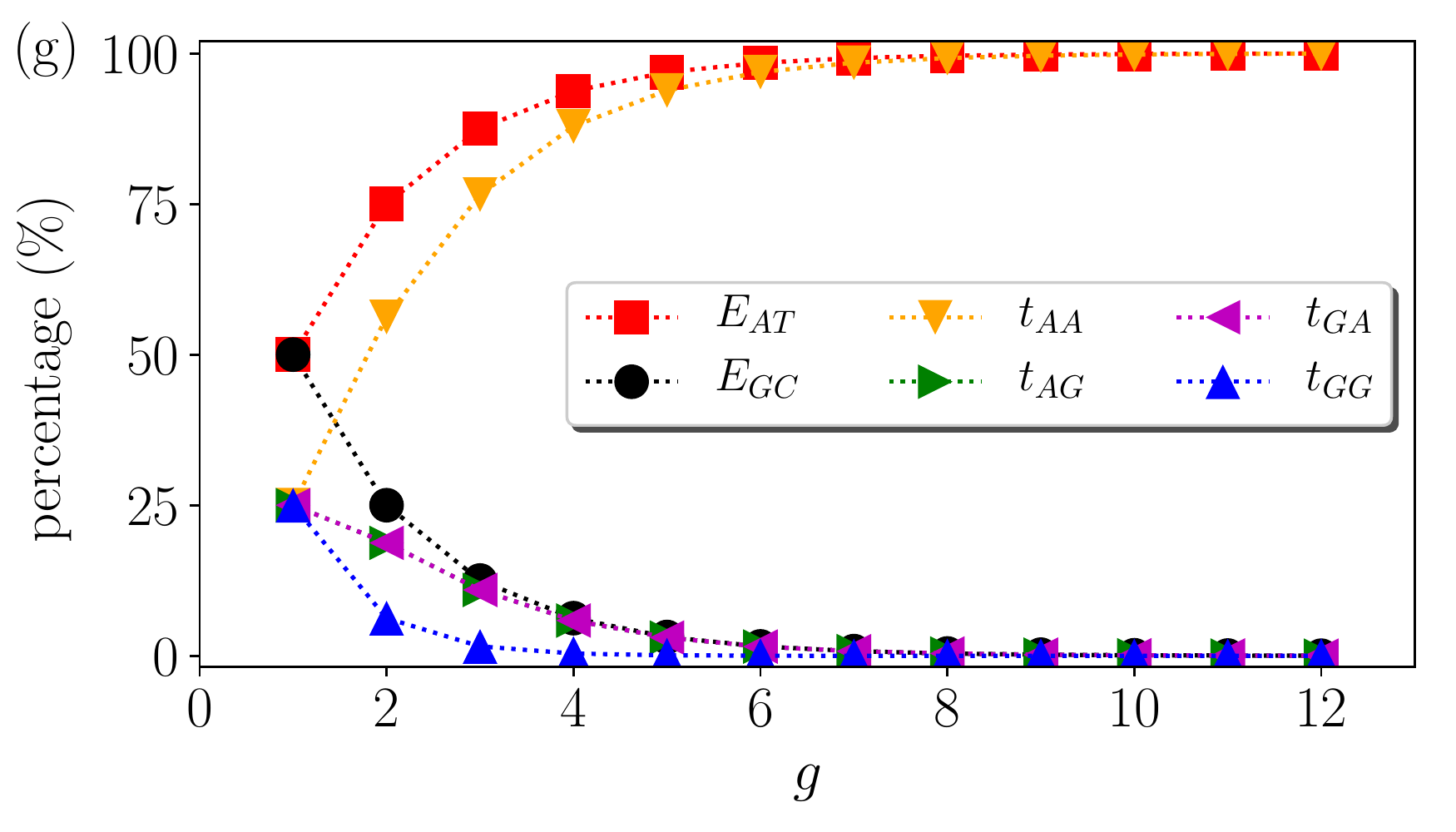}
	\includegraphics[width=0.39\textwidth]{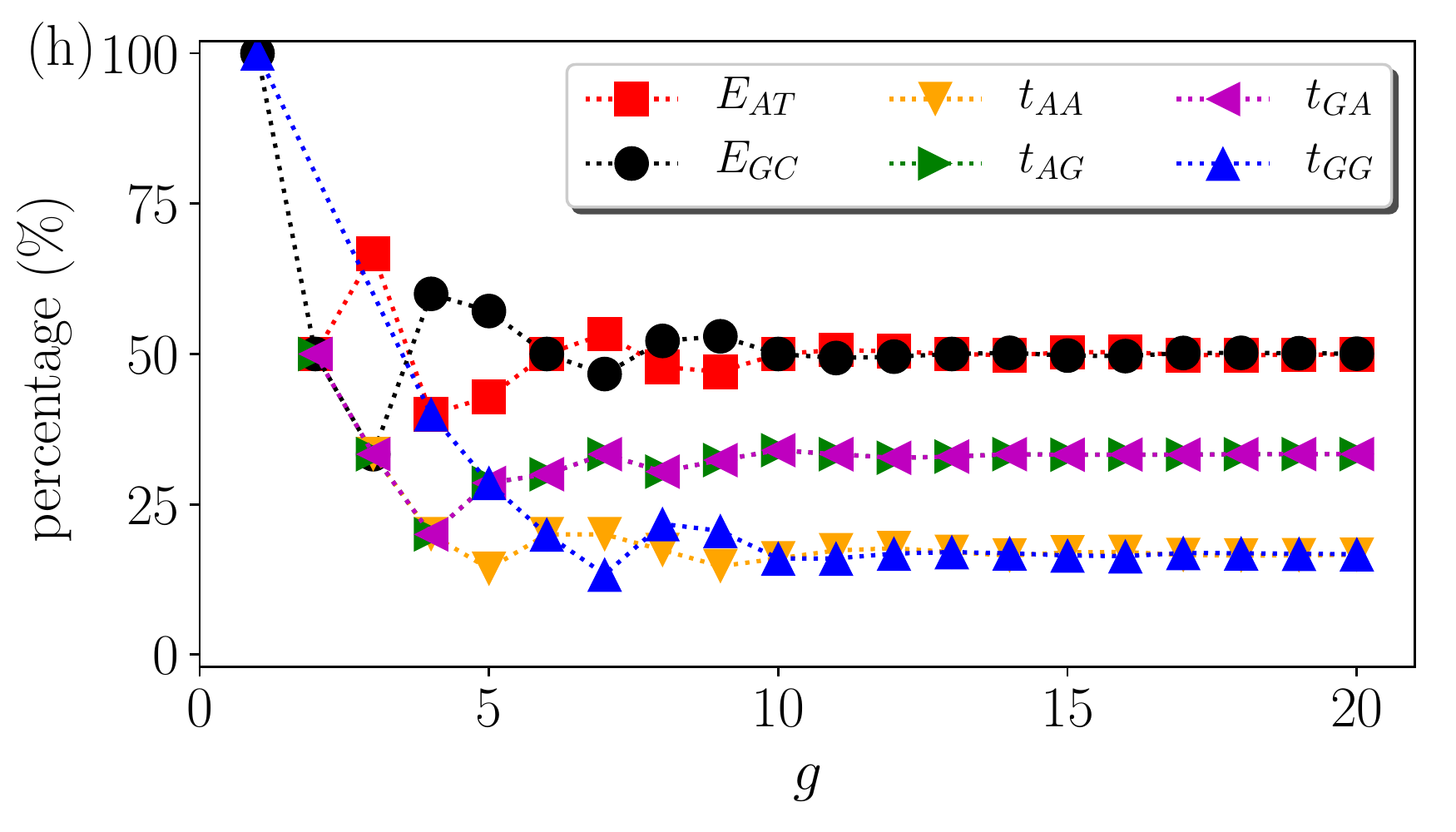}\\
	\includegraphics[width=0.39\textwidth]{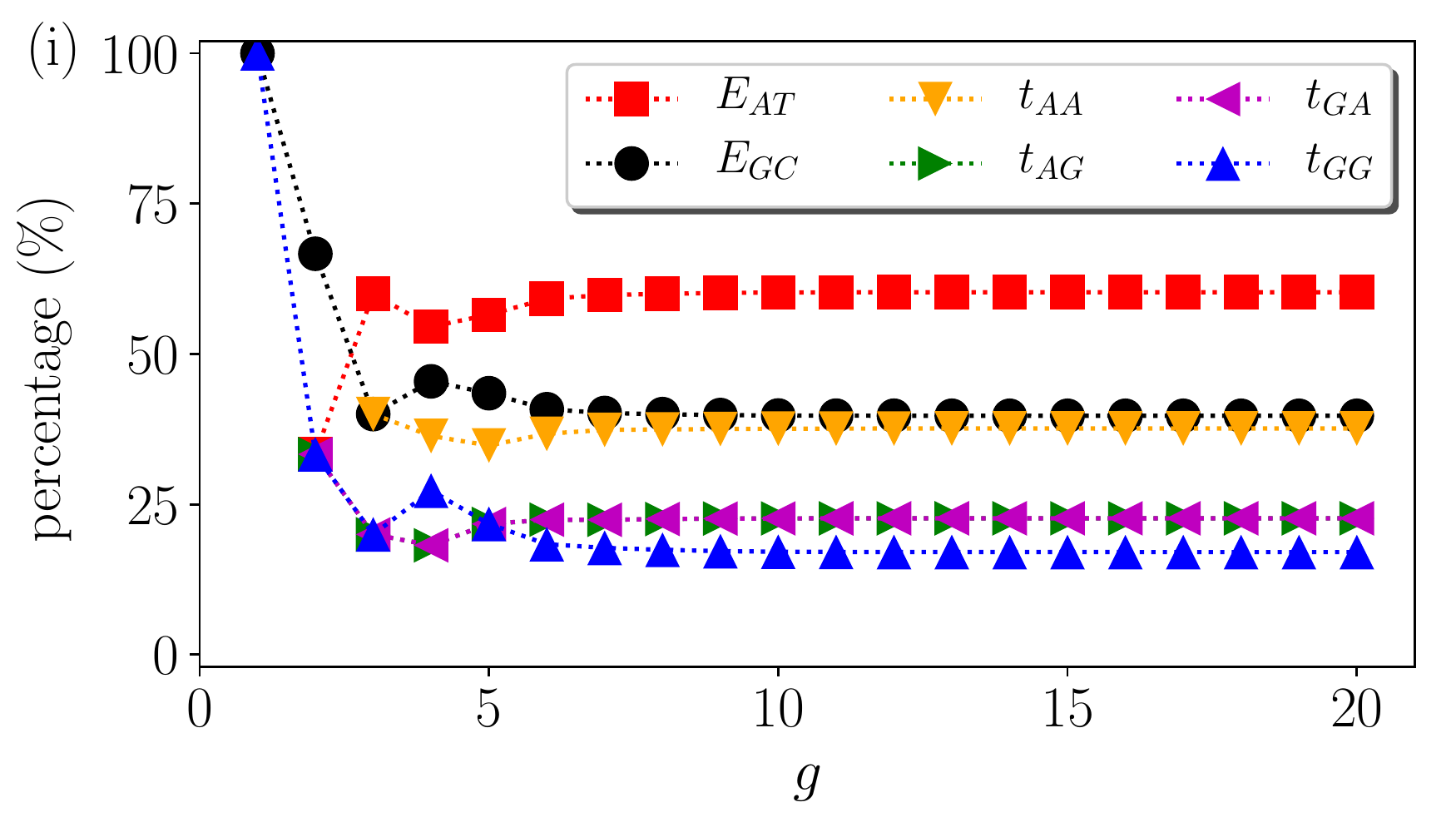}
	\includegraphics[width=0.39\textwidth]{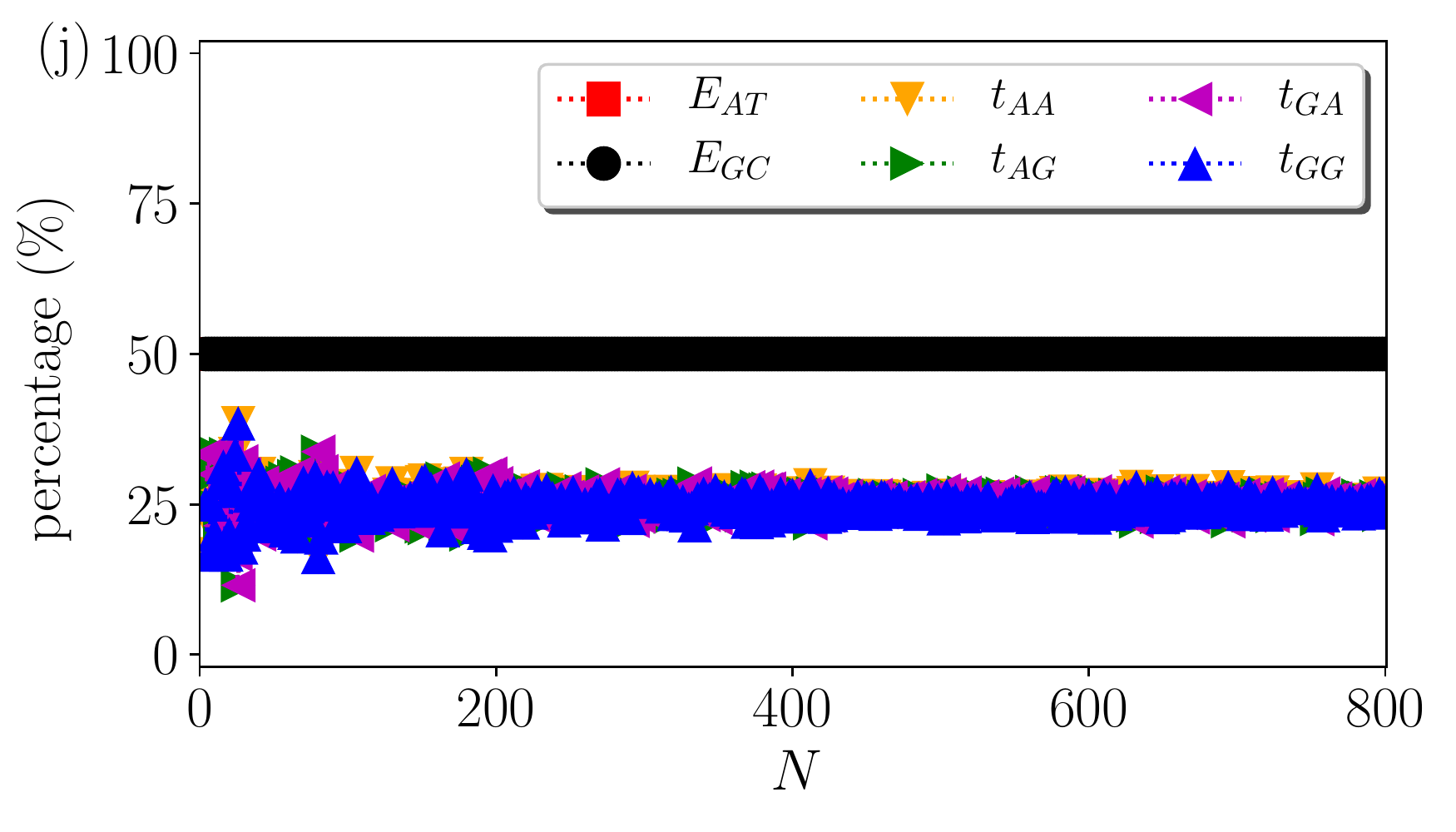}
	\caption{Scaling of the occurrence percentage of each TB parameter in various categories of DNA segments. (a) (GA)$_m$. (b) $\textit{TM}_g$. (c) $\textit{F}_g$. (d) $\textit{PD}_g$. (e) $\textit{RS}_g$. (f) $\textit{CS}_g$. (g) $\textit{GCS}_g$. (h) $\textit{KOL}_g(1,2)$. (i) $\textit{KOL}_g(1,3)$. (j) Random ($50\%$ G, $50\%$ A).}
	\label{fig:percentage}
\end{figure*}

The intricacy of the sequence determines the total number of TB parameters (on-site energies and hopping integrals) and the occurrence percentage of each inside a given segment. In 
Fig.~\ref{fig:percentage}, we present the scaling of each TB parameter occurrence percentage for all the categories of studied segments. Among other things, we observe: 
The occurrence percentage of $t_{GA}$ is always equal to that of $t_{AG}$. 
In all deterministic aperiodic cases, the occurrence percentages reach specific values as the generation, $g$, increases. 
Comparing $\textit{F}$ and $\textit{PD}$ sequences, although the former sequence is simpler (cf.~Fig.~\ref{fig:triplets}), it has the same total number of TB parameters with the latter, since it has the additional triplet GGG. Again, we notice that, for sequences with primitive substitution rules, the values at which the occurrence percentage of each on-site energy and hopping integral converge coincide with the letter frequencies of the possible one- and two-letter words in the sequence, which can be found from the procedure described in Subsec.~\ref{subsec:SM}.

Having obtained an estimate of the intricacy of the sequences, we move to the estimation of the correlations of their energy landscape. We will do this by calculating the autocorrelation function (ACF)~\cite{Albuquerque:2005} for the quantities $\frac{E_n}{t_n}, n=1,\dots,N$. 
This ratio is used to fully capture the energy intricacy of the sequences. 
The lag-$j$ normalized ACF, $ACF(j)$, of $\frac{E_j}{t_j}$, $j=1,2,\dots,N-1$, expresses the degree the base pairs are correlated with their j-$th$ neighbors. Using the notation $y_k = \frac{E_k}{t_k}$, it is given by the expression

\begin{figure*}[t!] 
	\centering
	\includegraphics[width=0.95\columnwidth]{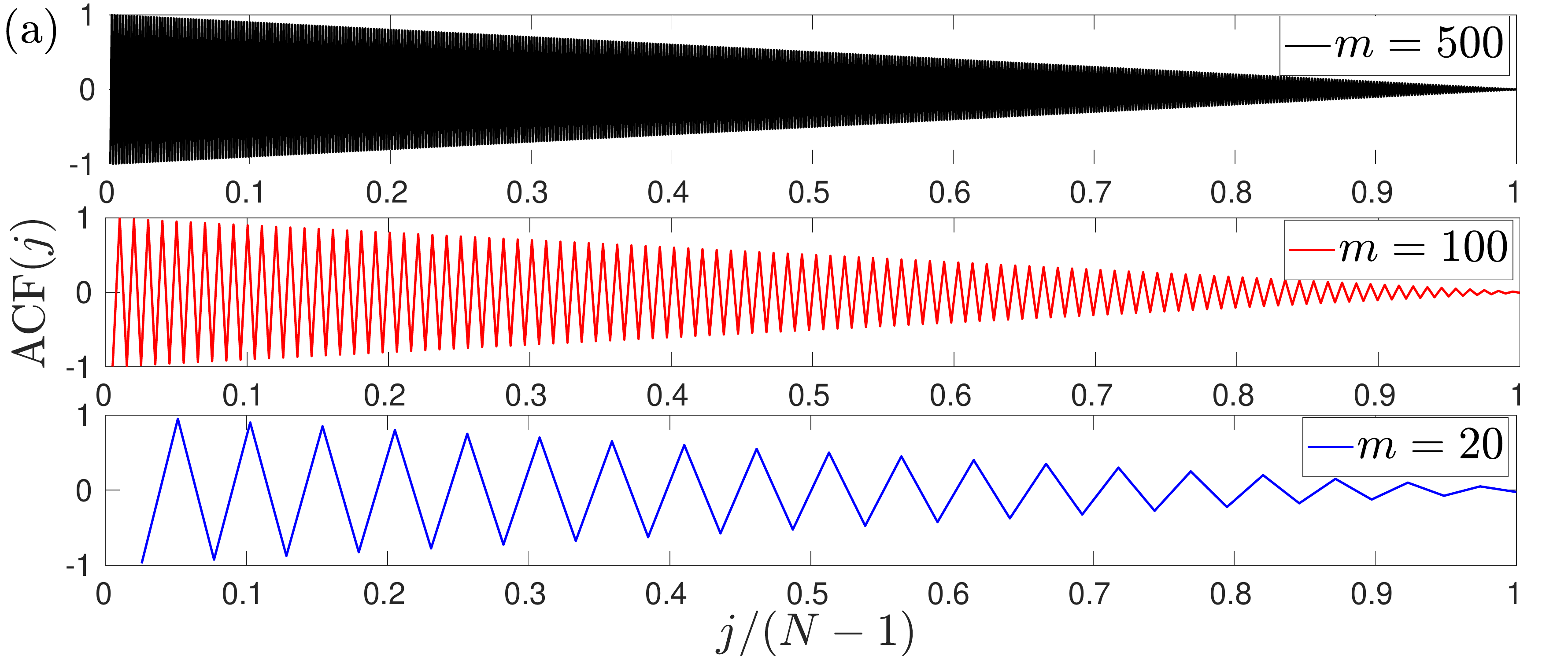}\includegraphics[width=0.95\columnwidth]{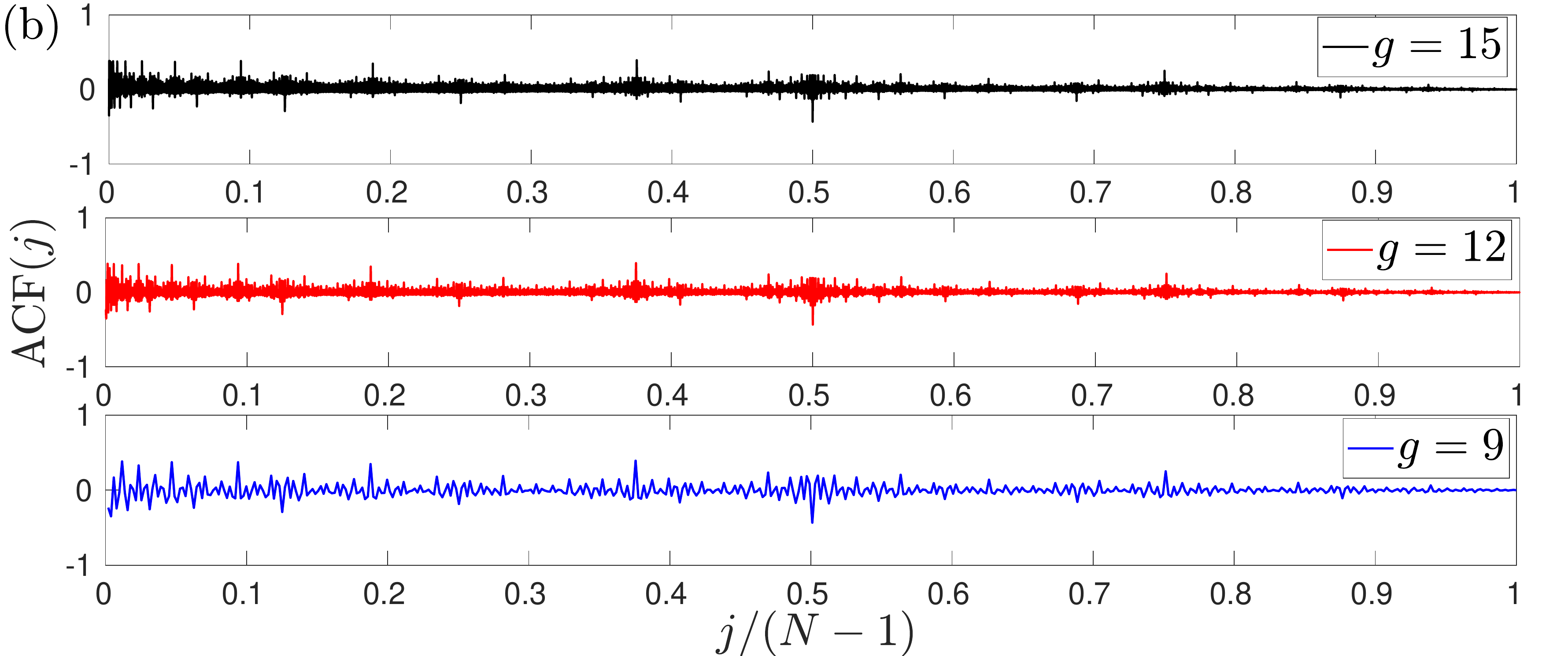}\\
	\includegraphics[width=0.95\columnwidth]{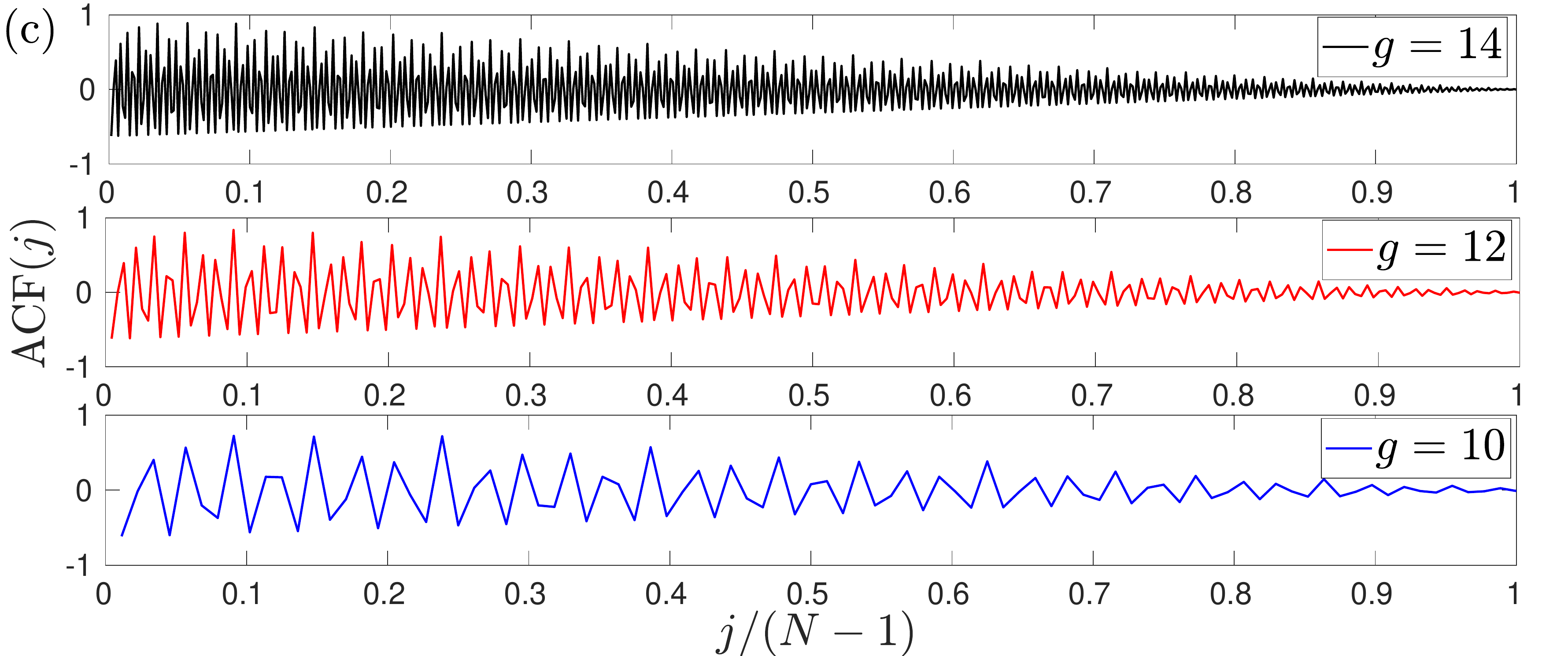}\includegraphics[width=0.95\columnwidth]{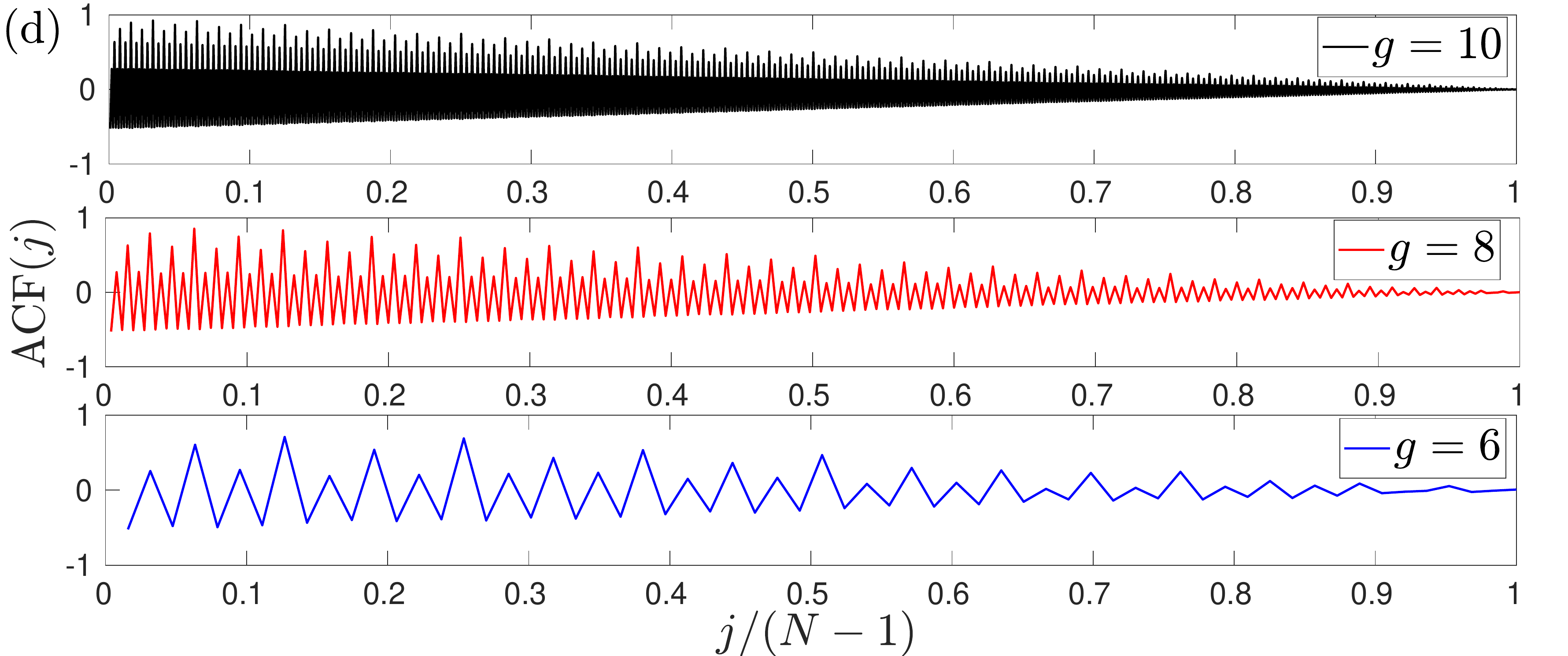}\\
	\includegraphics[width=0.95\columnwidth]{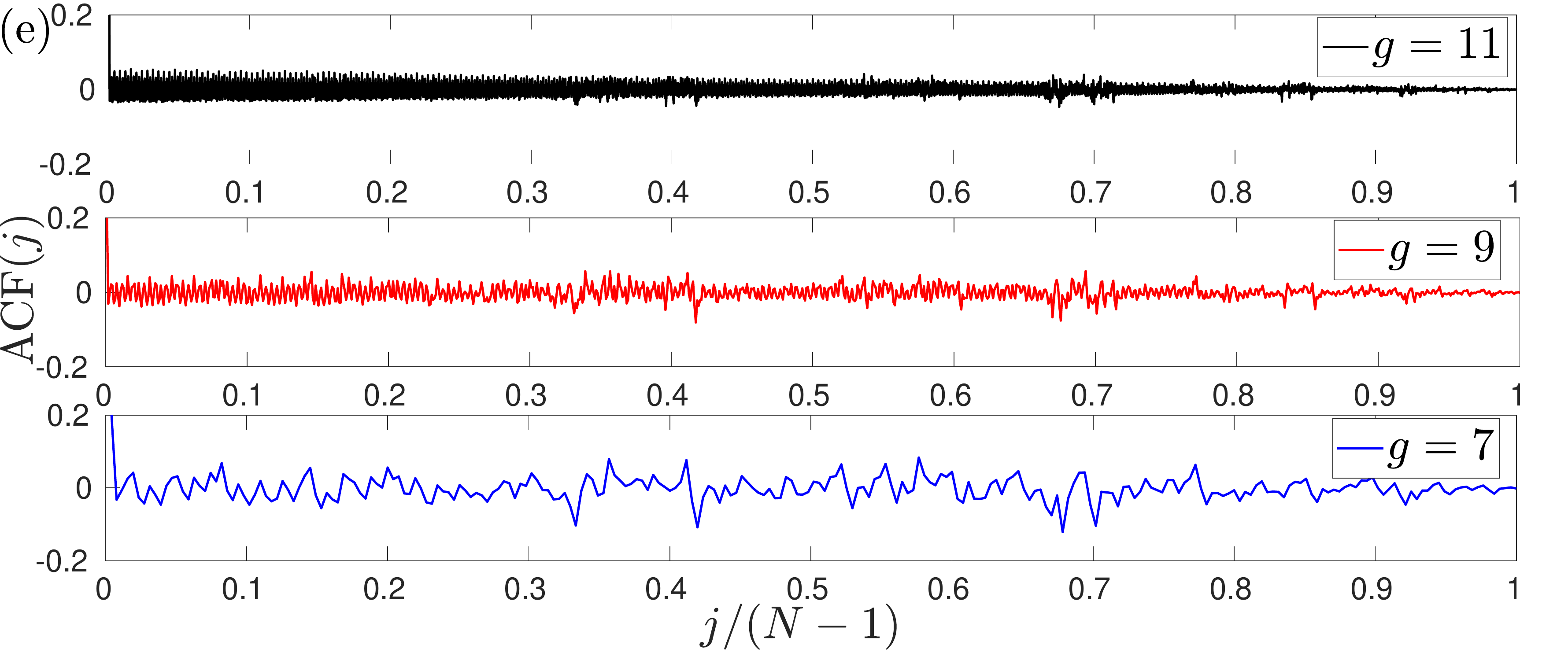}\includegraphics[width=0.95\columnwidth]{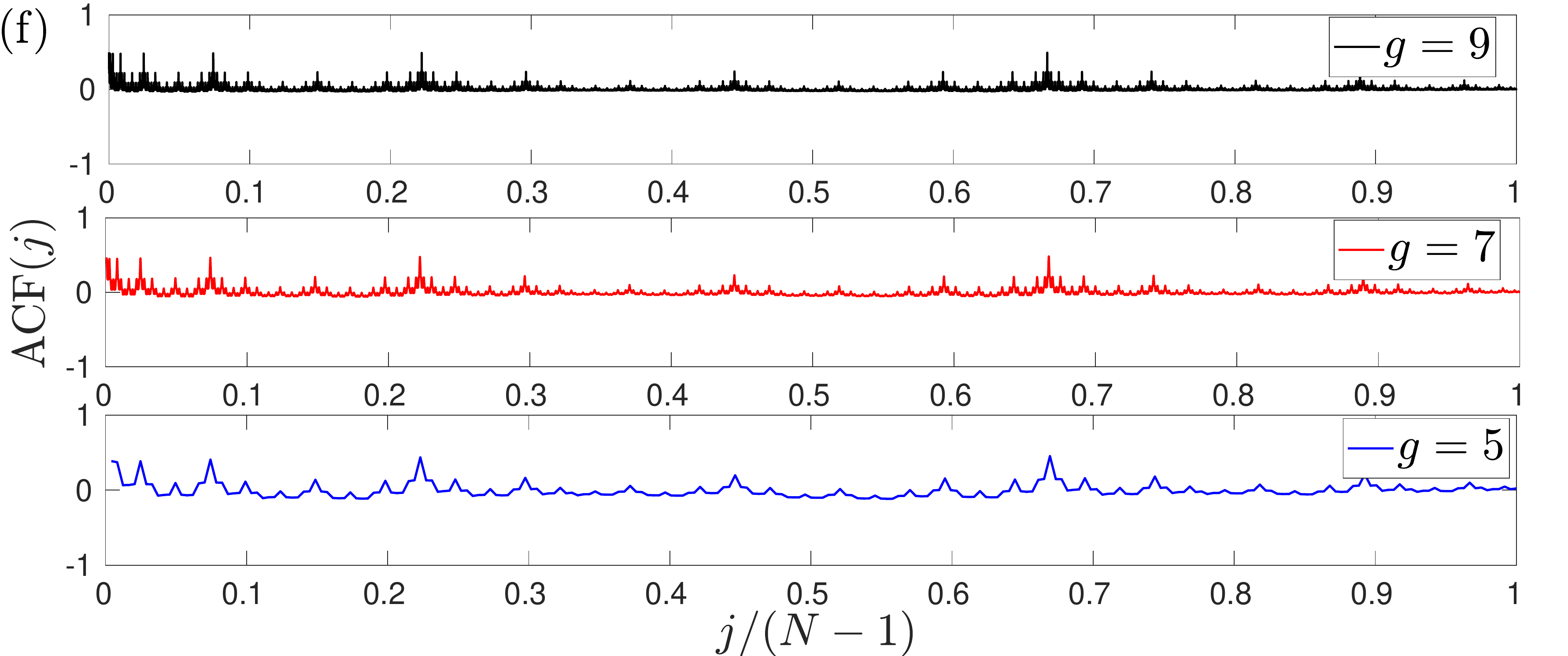}\\
	\includegraphics[width=0.95\columnwidth]{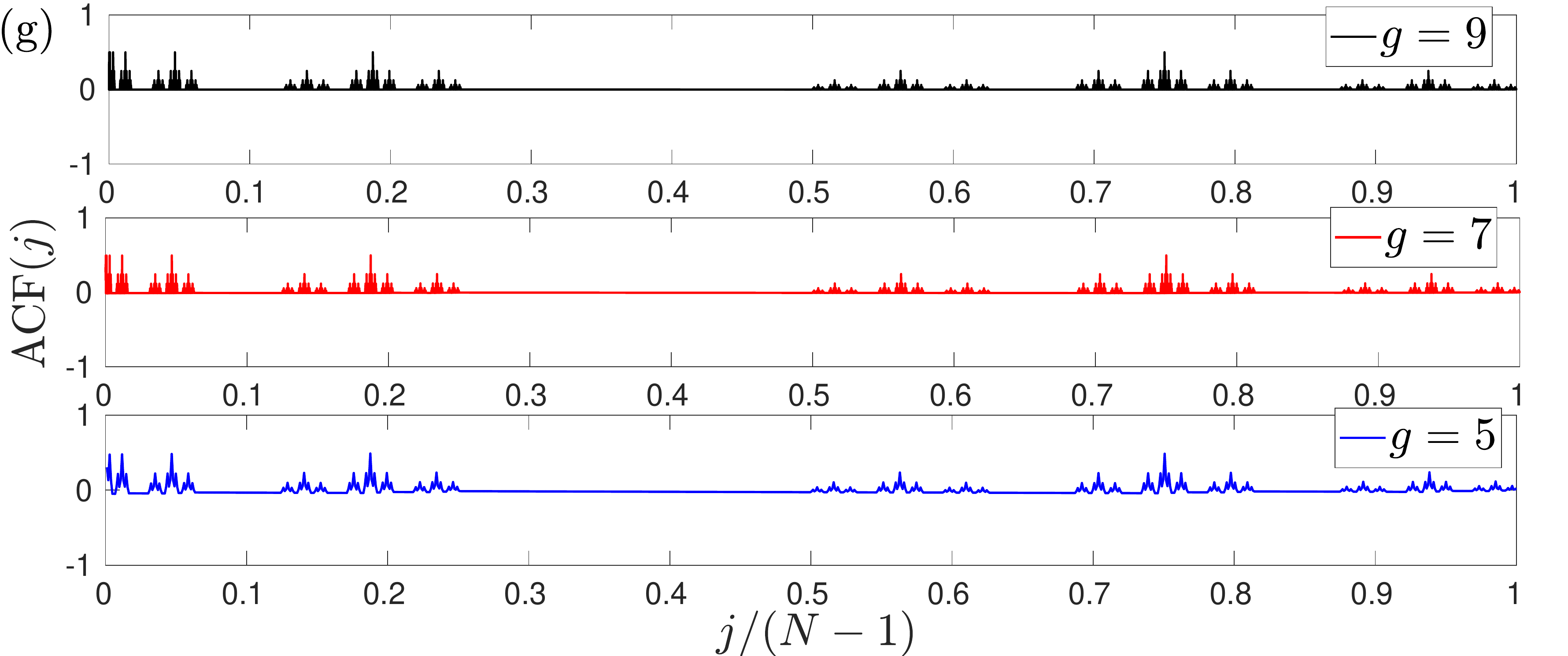}\includegraphics[width=0.95\columnwidth]{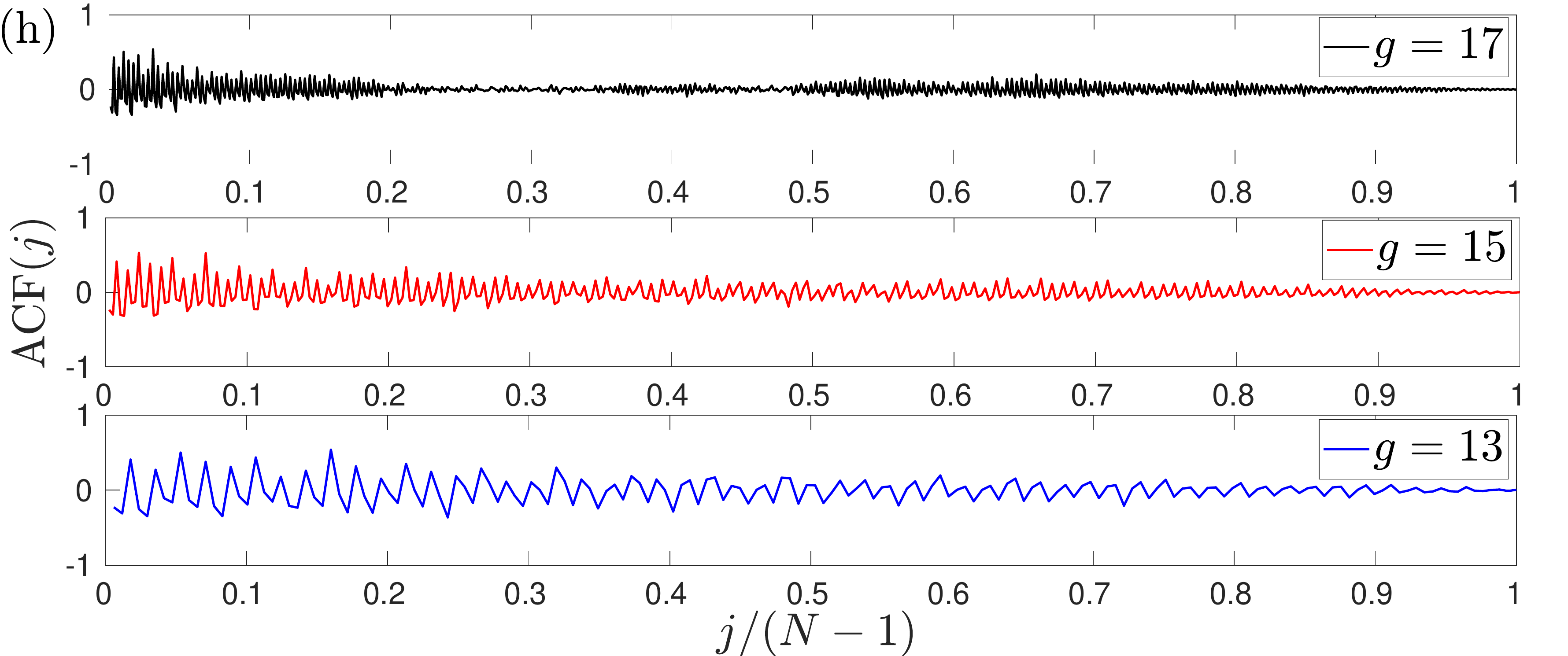}\\
	\includegraphics[width=0.95\columnwidth]{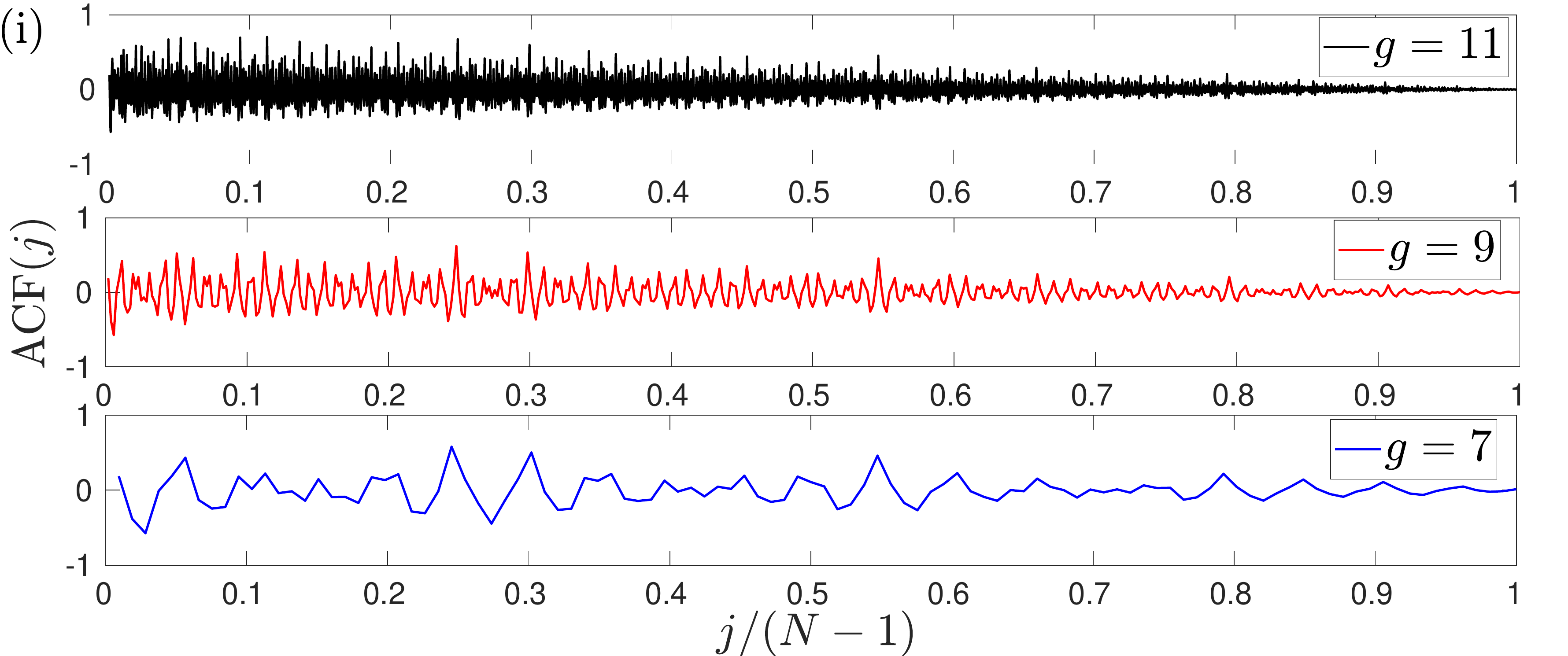}\includegraphics[width=0.95\columnwidth]{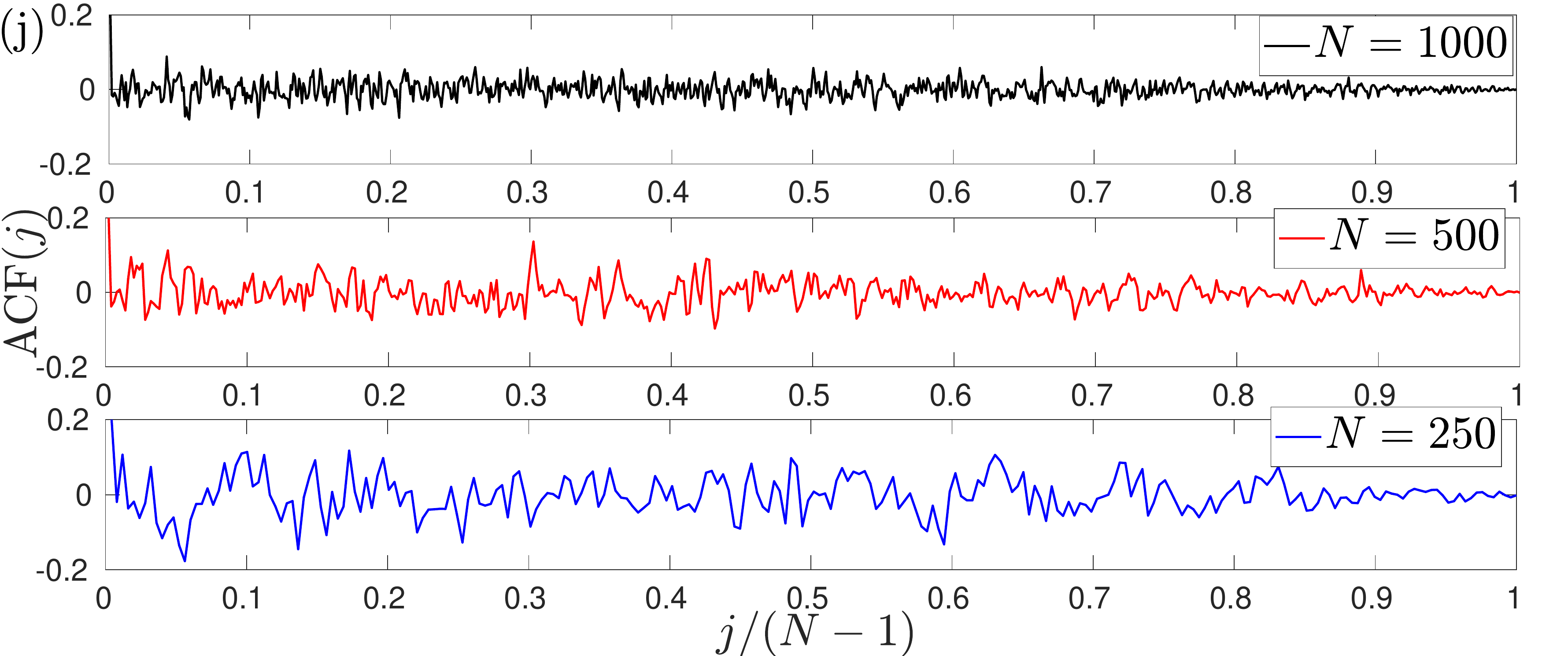}
	\caption{Scaling of the autocorrelation function of various categories of DNA segments. (a) Periodic (GA)$_m$. (b)  $\textit{TM}_g$. (c) $\textit{F}_g$. (d) $\textit{PD}_g$ (e) $\textit{RS}_g$. (f) $\textit{CS}_g$. (j) $\textit{GCS}_g(4,2)$. (h) $\textit{KOL}_g(1,2)$. (i) $\textit{KOL}_g(1,3)$. (j) Random ($50\%$ G content, $50\%$ A content).}
	\label{fig:ACF}
\end{figure*}

\begin{equation}
ACF(j)= \frac{\displaystyle\sum_{k=1}^{N-j}(y_k-\bar{y})(y_{j+k}-\bar{y})}{\displaystyle\sum_{k=1}^{N}(y_k-\bar{y})^2},
\end{equation}
where, $\bar{y}$ is the mean value of $y_{\{j\}}$.

In Fig.~\ref{fig:ACF}, we present the ACF all the categories of studied segments, for three different lengths for each. The horizontal axes are normalized over the total number of neighbors $(N-1)$, thus corresponding to the \textit{relative} neighbor distances. 
We notice that the ACF of each category has a characteristic shape. Furthermore, from the inspection of Fig.~\ref{fig:ACF}, we observe that there is a correspondence between the degree of intricacy of the segments and the strength of correlations. 
Random and $\textit{RS}$ sequences, which posses $8$ equidistributed triplets, display weak correlations. $\textit{KOL}(1,2)$ and $\textit{TM}$ sequences, which posses $6$ equidistributed triplets, display somehow stronger correlations. Then follow $\textit{KOL}(1,3)$, $\textit{CS}$, and $\textit{GCS}(4,2)$ sequences, which posses predominant triplets. The fractal sequences of the Cantor Set family possess strong correlations in the regions where G is present, interrupted by long, largely homogeneous, regions where it is not present. Deterministic aperiodic segments with the least possible triplets ($\textit{F}$ and $\textit{PD}$, with $4$ and $5$ triplets, respectively) display strong correlations, and the periodic case is the dominant one. 

Finally, we mention that by comparing the $ACF$ of each category for different $N$, we can come to conclusions about their inflation/deflation symmetry. Sequences with this symmetry have similar autocorrelations at similar relative neighbor distances. This is the case for all studied aperiodic sequences, apart from $\textit{KOL}(1,2)$ and the random ones [cf. Fig.~\ref{fig:ACF}(h) and (j), respectively]. 
As far as the $\textit{KOL}(p,q)$ family segments are concerned, we have checked no inflation or deflation symmetry exists when $\abs{p-q} = 2\nu+1$, $\nu \in \mathcal{N}$, in contrast with the cases $\abs{p-q} = 2\nu$, such as $\textit{KOL}(1,3)$, shown in  Fig.~\ref{fig:ACF}(i).


\section{Eigenspectra and density of states} \label{sec:ES-DOS-IDOS}
For fixed boundary conditions $(\psi_{N+1} = \psi_0 = 0)$, the eigenspectrum, i.e. the eigenenergies $E_j$, $j=1,2,\dots,N$ of a sequence, can be given by the roots of the polynomial $M_N^{11}(E)$~\cite{Molinari:1997,LS:2018}. For periodic segments, the eigenspectrum can be recursively obtained with the help of the Chebyshev polynomials of the second kind~\cite{LS:2018}. Here, the eigenspectra of the sequences have been calculated by  numerical diagonalization of the Hamiltonian matrix, which is real, tridiagonal and symmetric. In the periodic case, the matrix is $u$-Toeplitz, where $u$ is the size of the repetition unit. The DOS can be obtained by
\begin{equation} \label{Eq:DOS}
g(E) = \frac{N}{\pi}\dv{E}\abs{\acos(\frac{\Tr(M_N(E))}{2})}.
\end{equation}
IDOS is given by the expression
\begin{equation} \label{Eq:IDOS}
IDOS(E) = \int_{E_1}^{E_2} g(E) dE .
\end{equation}

\begin{figure*}
\centering
\includegraphics[width=0.7\textwidth]{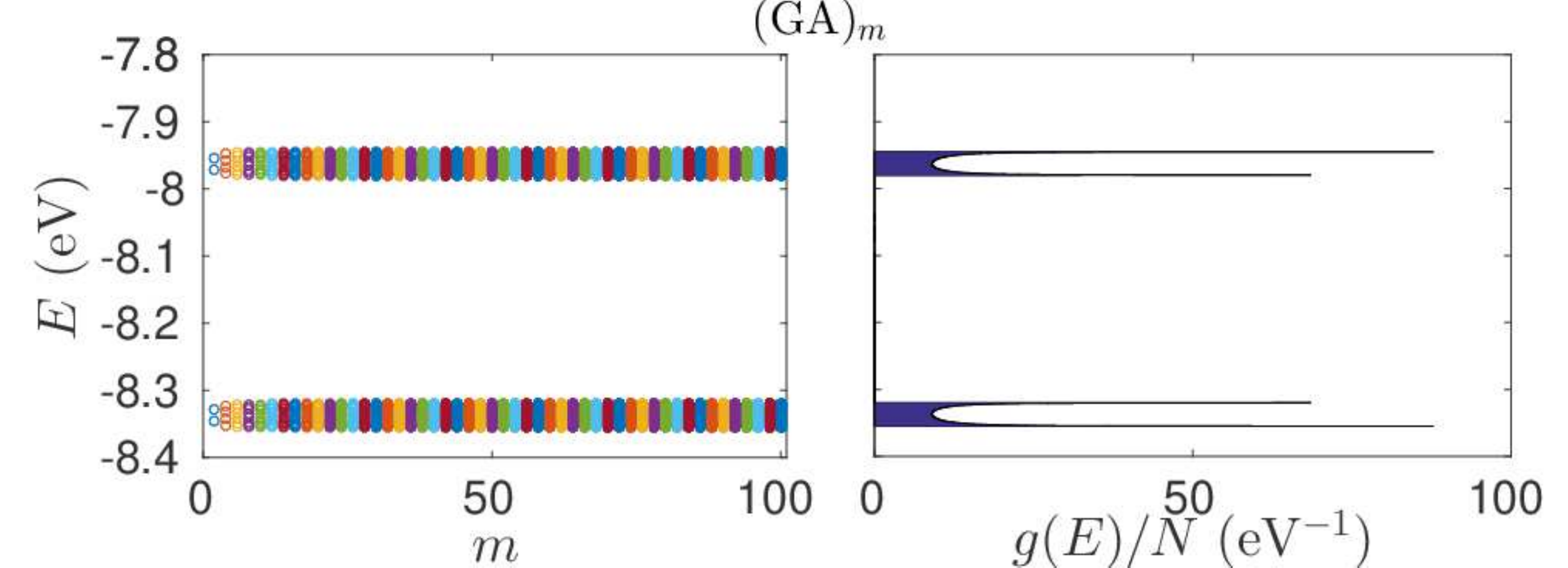}
\includegraphics[width=0.7\textwidth]{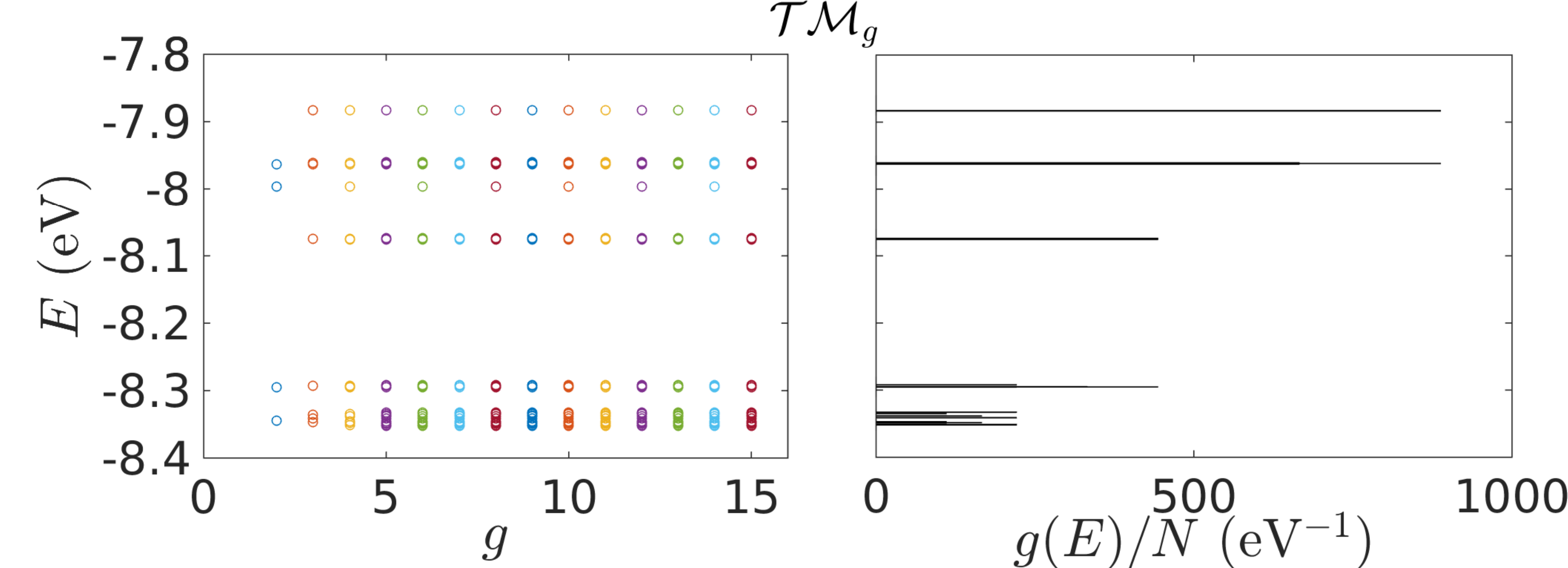}
\includegraphics[width=0.7\textwidth]{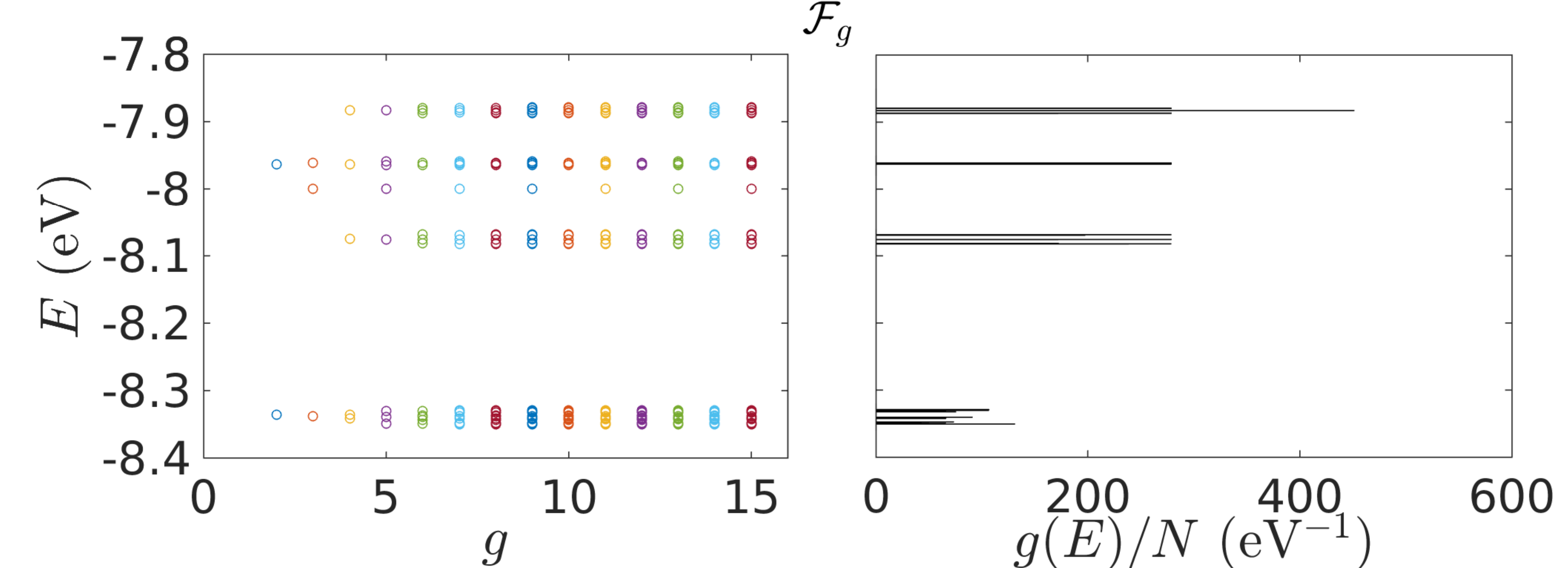}
\includegraphics[width=0.7\textwidth]{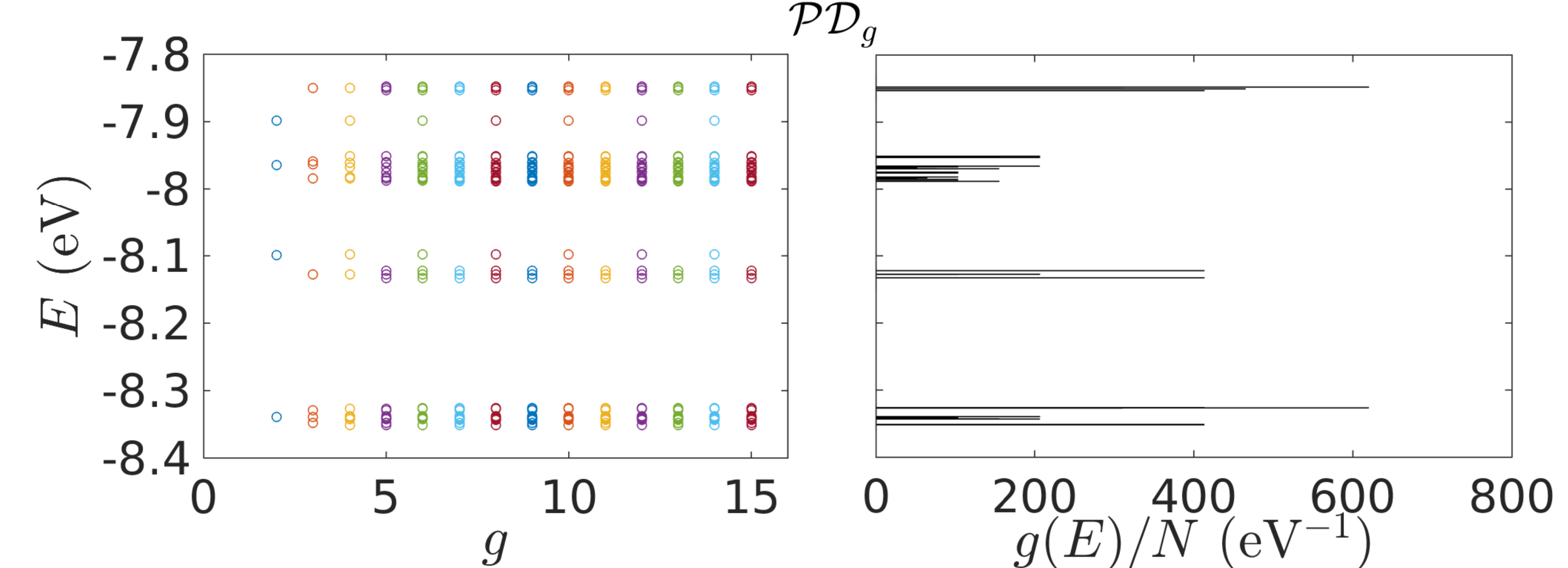}
\includegraphics[width=0.7\textwidth]{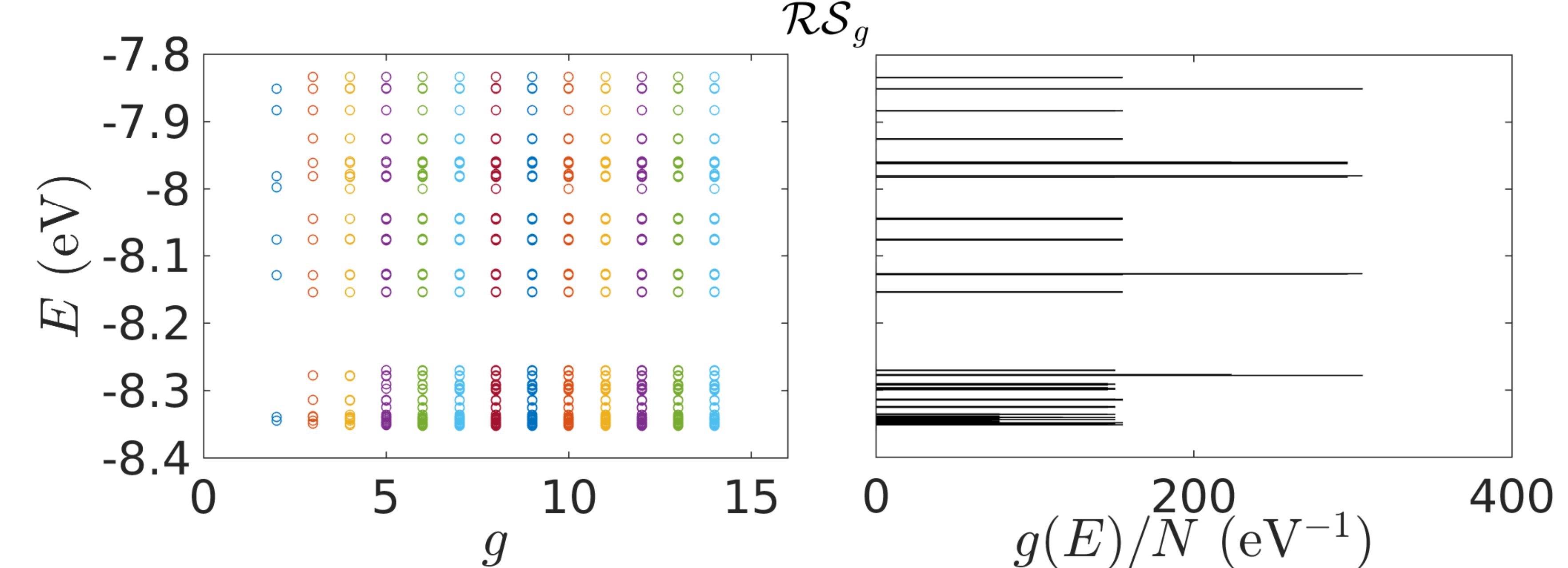}
\caption{Eigenspectra and DOS of various periodic and quasi-periodic DNA sequences: Periodic, Thue-Morse, Fibonacci, Period Doubling, Rudin-Shapiro.}
\label{fig:ESDOS-1}
\end{figure*}

\begin{figure*}
\centering
\includegraphics[width=0.7\textwidth]{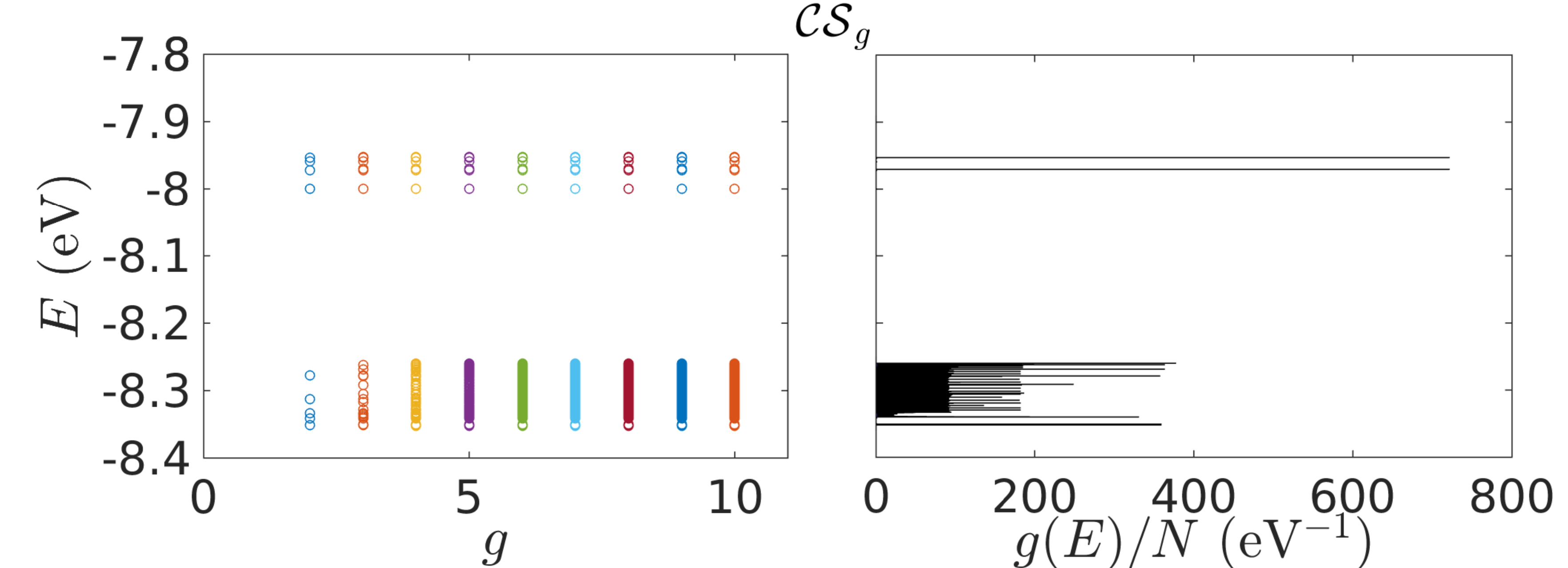}
\includegraphics[width=0.7\textwidth]{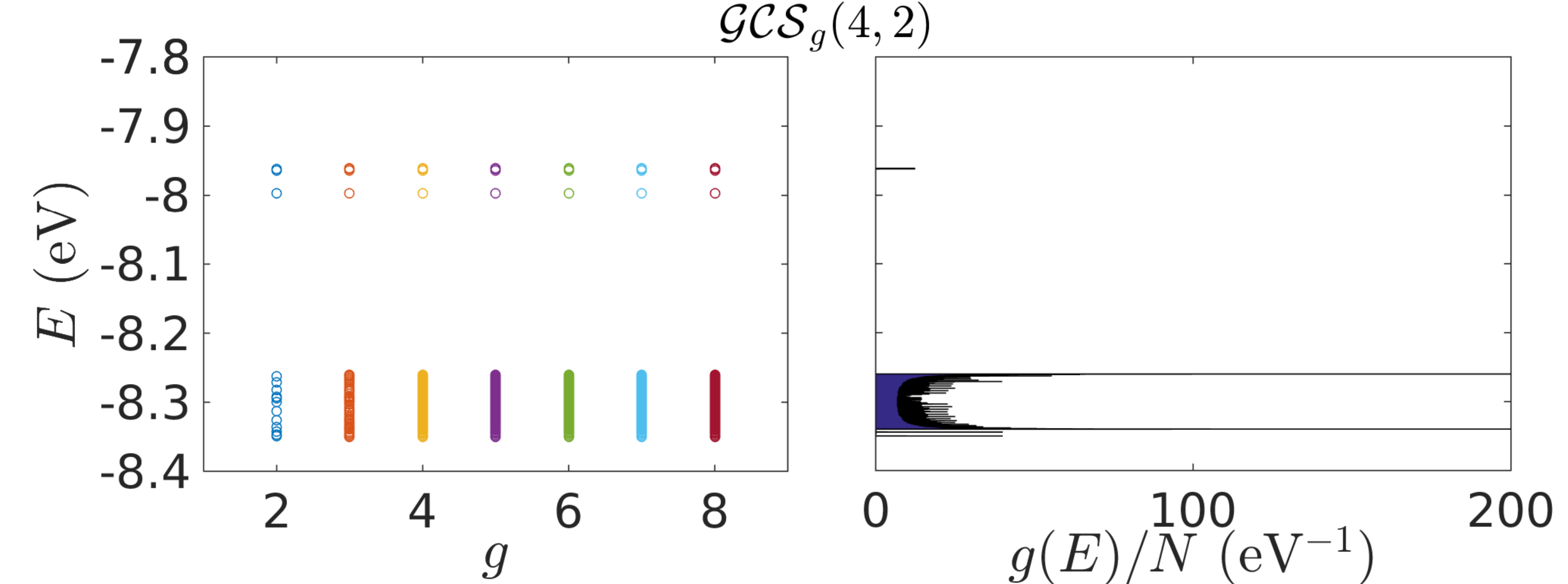}
\includegraphics[width=0.7\textwidth]{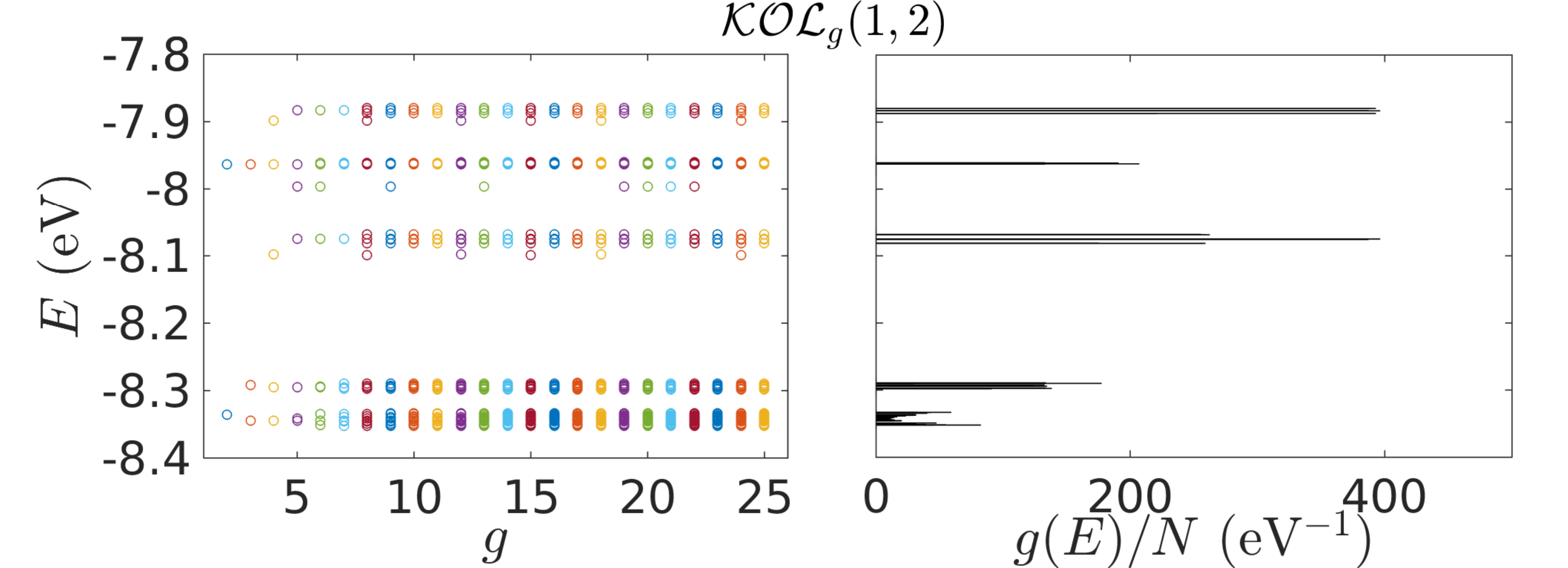}
\includegraphics[width=0.7\textwidth]{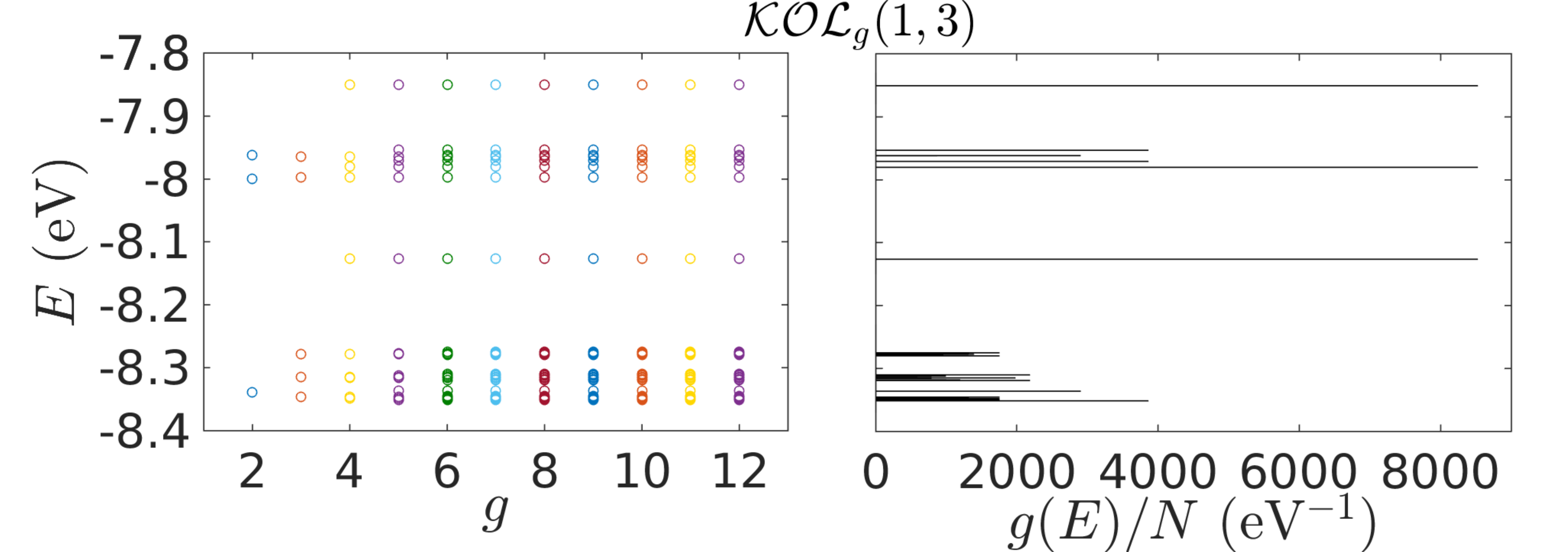}
\includegraphics[width=0.7\textwidth]{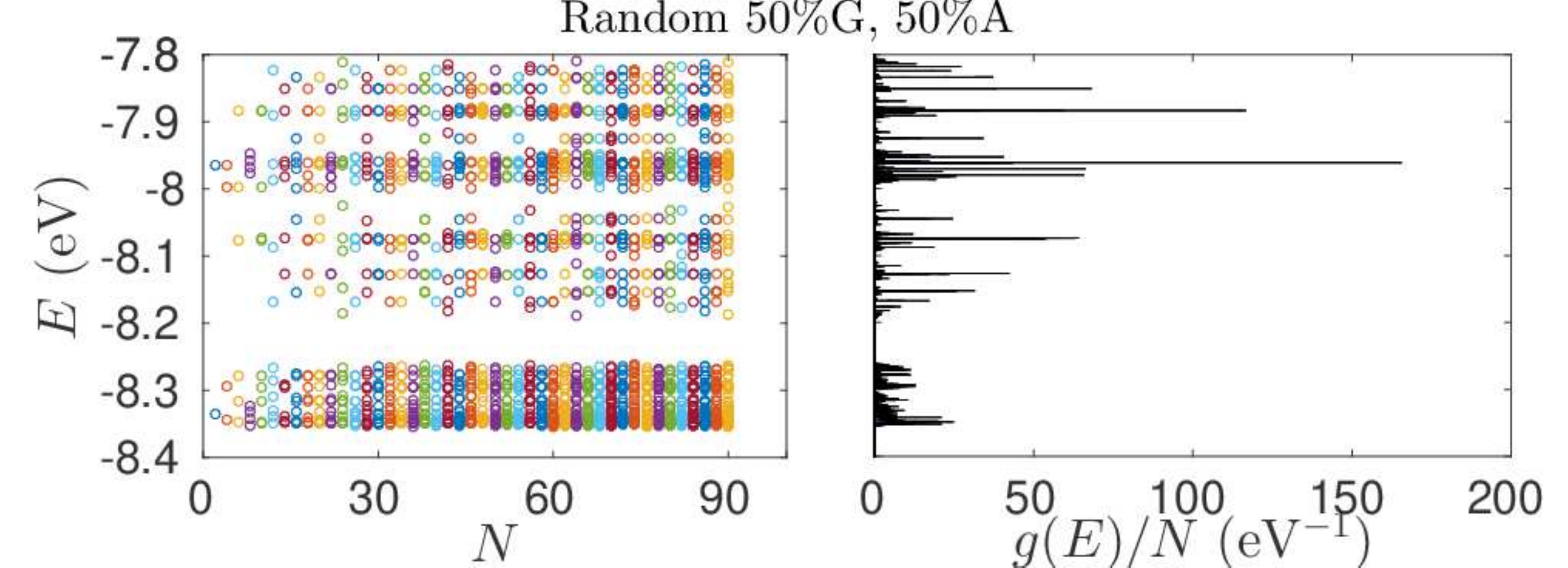}
\caption{Eigenspectra and DOS of various fractal, Kolakoski and random DNA sequences: Cantor set, Generalized $(4,2)$ Cantor set, Kolakoski $(1,2)$, Kolakoski $(1,3)$, and Random ($50\%$ G and $50\%$ A).}
\label{fig:ESDOS-2}
\end{figure*}

The eigenspectra and the corresponding DOS for all the categories of DNA segments studied in this work are presented in Figs.~\ref{fig:ESDOS-1}-\ref{fig:ESDOS-2}. We notice that for all studied deterministic aperiodic sequences, the allowed energies do not exceed the energy interval defined by the eigenspectrum of the random sequence. This also holds for periodic polymers with only G and A in the $5'$-$3'$ strand, as their repetition unit increases~\cite{LVBMS:2018}. Hence, the above mentioned interval of the random sequence represents a limit. Two subsets of the aforementioned interval gather around the on-site energies of G and A, so will be henceforth referred to as G and A energy regions.  
Comparing Fig.~\ref{fig:ESDOS-1} which shows periodic and quasi-periodic sequences with Fig.~\ref{fig:ESDOS-2} which shows fractal, Kolakoski and random sequences, we observe that the former form subbbands which are rather acute in the quasi-periodic cases, while in the latter the DOS is more fragmented and spiky.

\begin{figure*}
	\centering
	\subfloat[Periodic (GA)$_m$ segments.] 
	{\includegraphics[width=0.7\columnwidth]{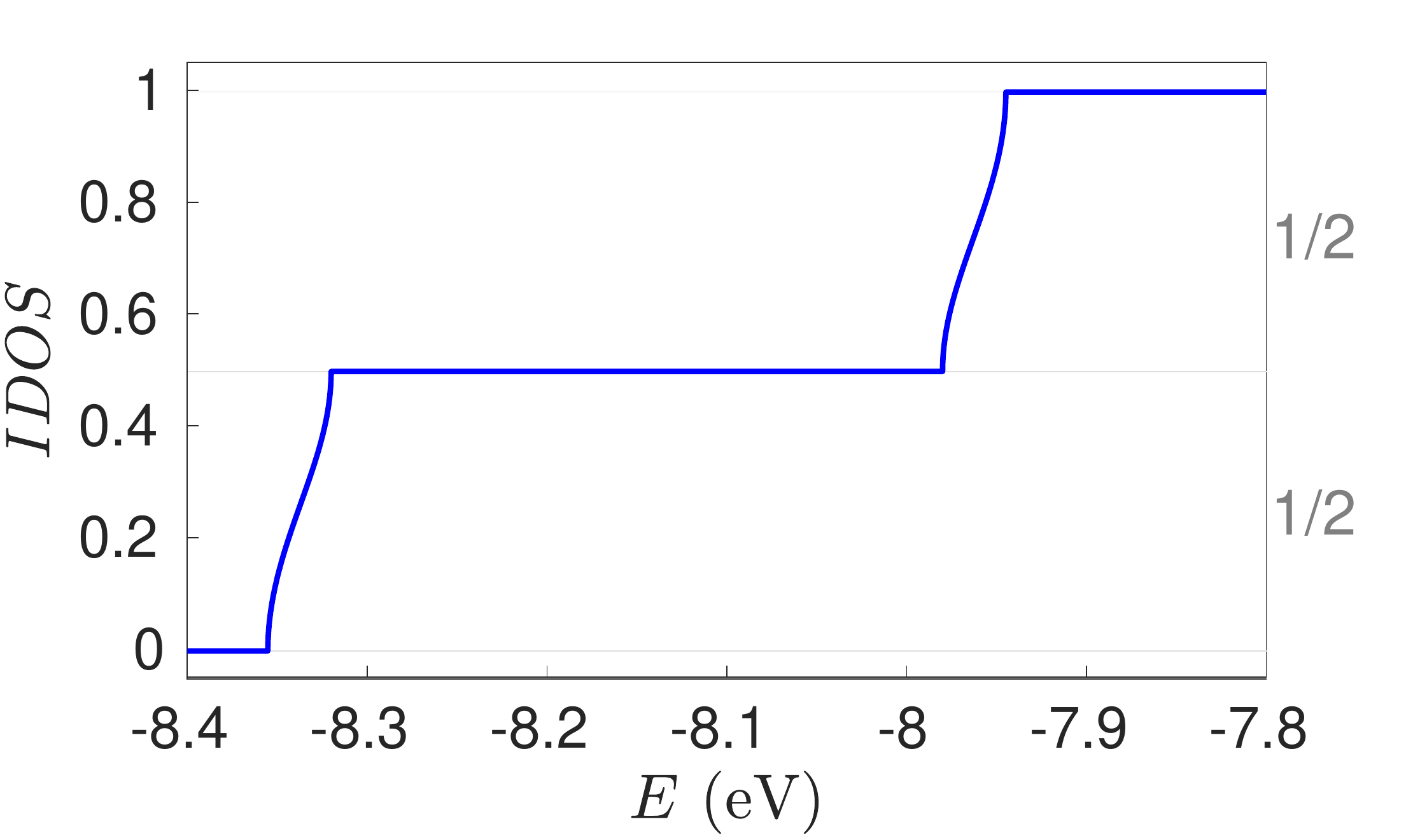}}
	\subfloat[$\textit{TM}$ segments.]
	{\includegraphics[width=0.7\columnwidth]{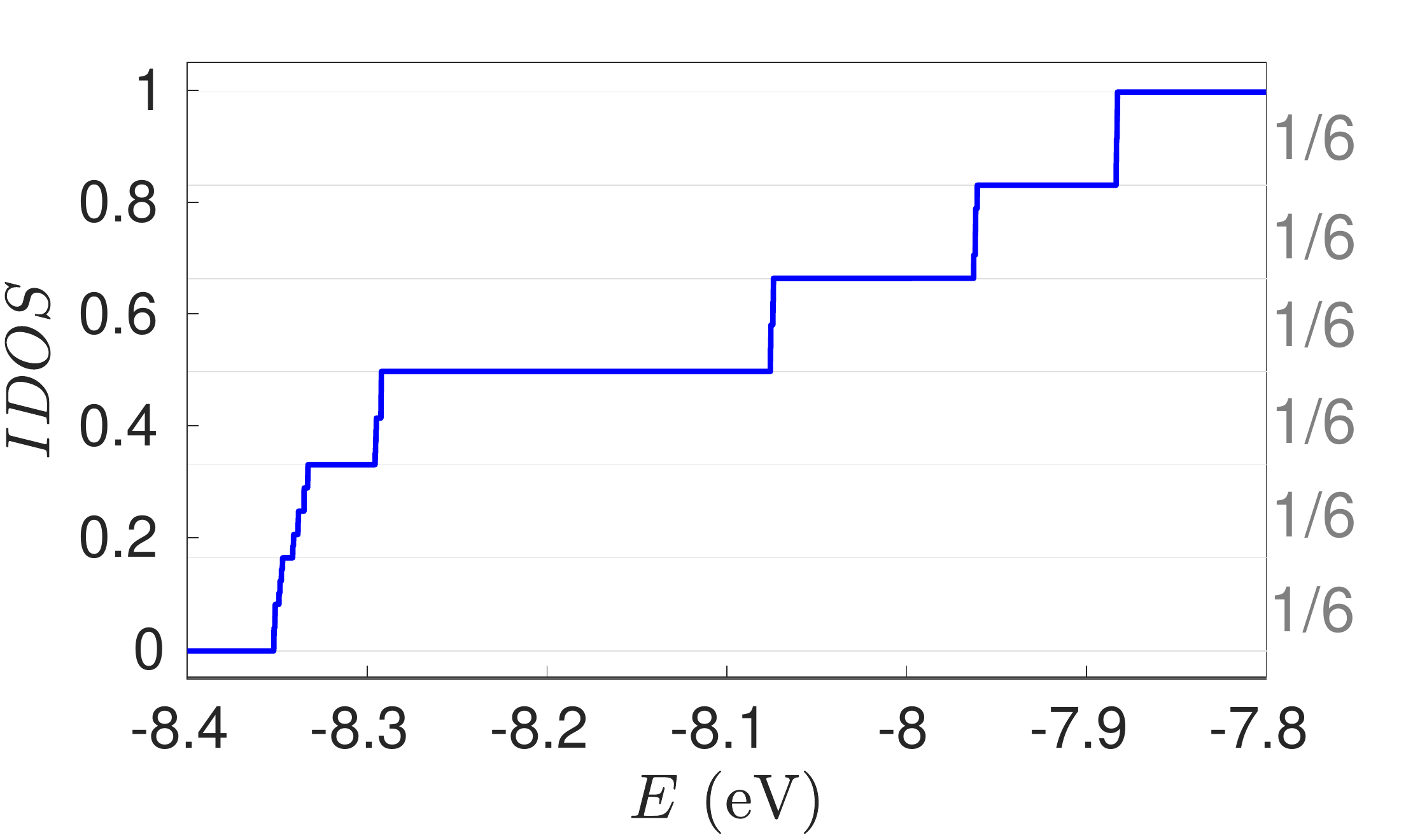}}
	
	\subfloat[$\textit{F}$ segments.]
	{\includegraphics[width=0.7\columnwidth]{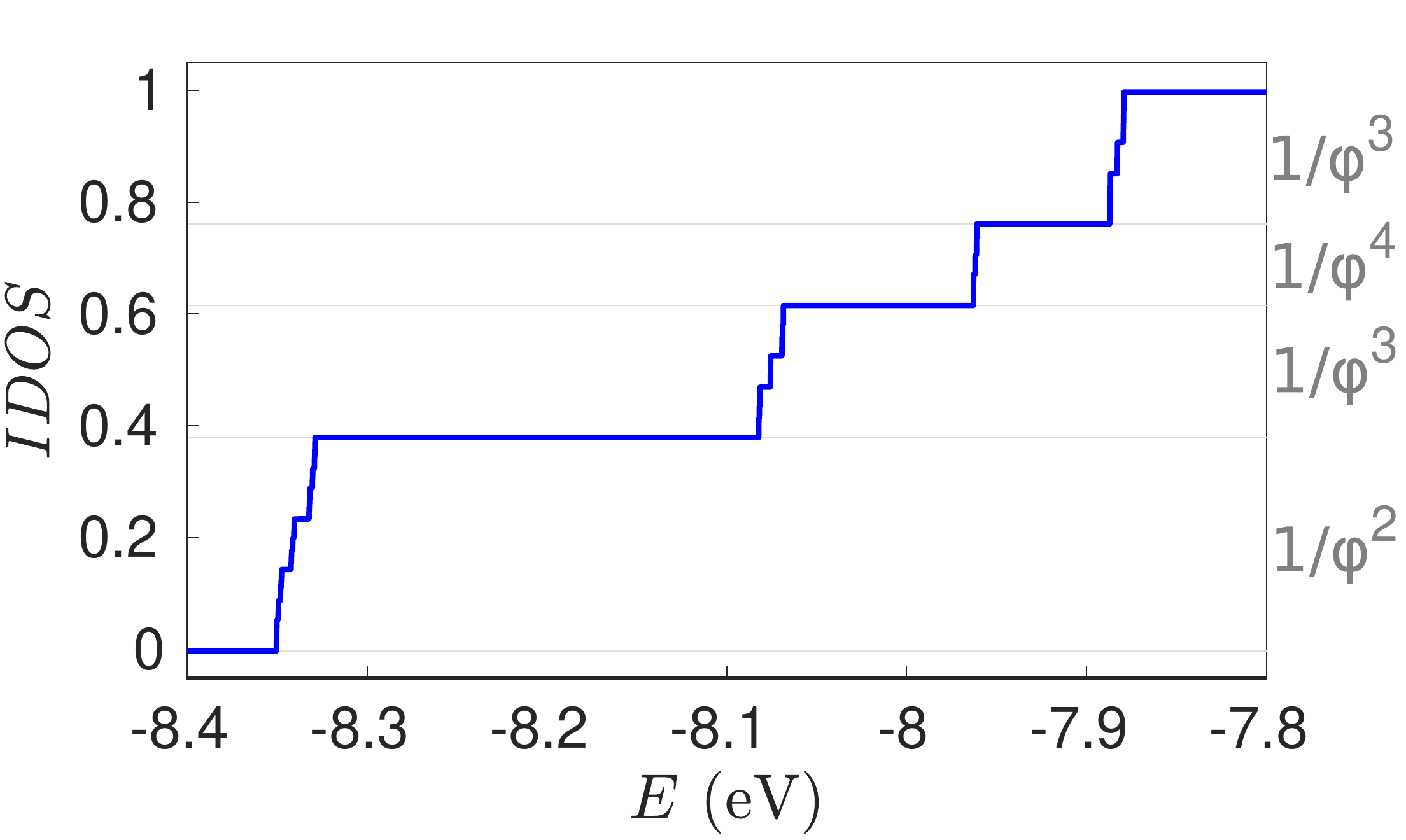}}
	\subfloat[$\textit{PD}$ segments.] 
	{\includegraphics[width=0.7\columnwidth]{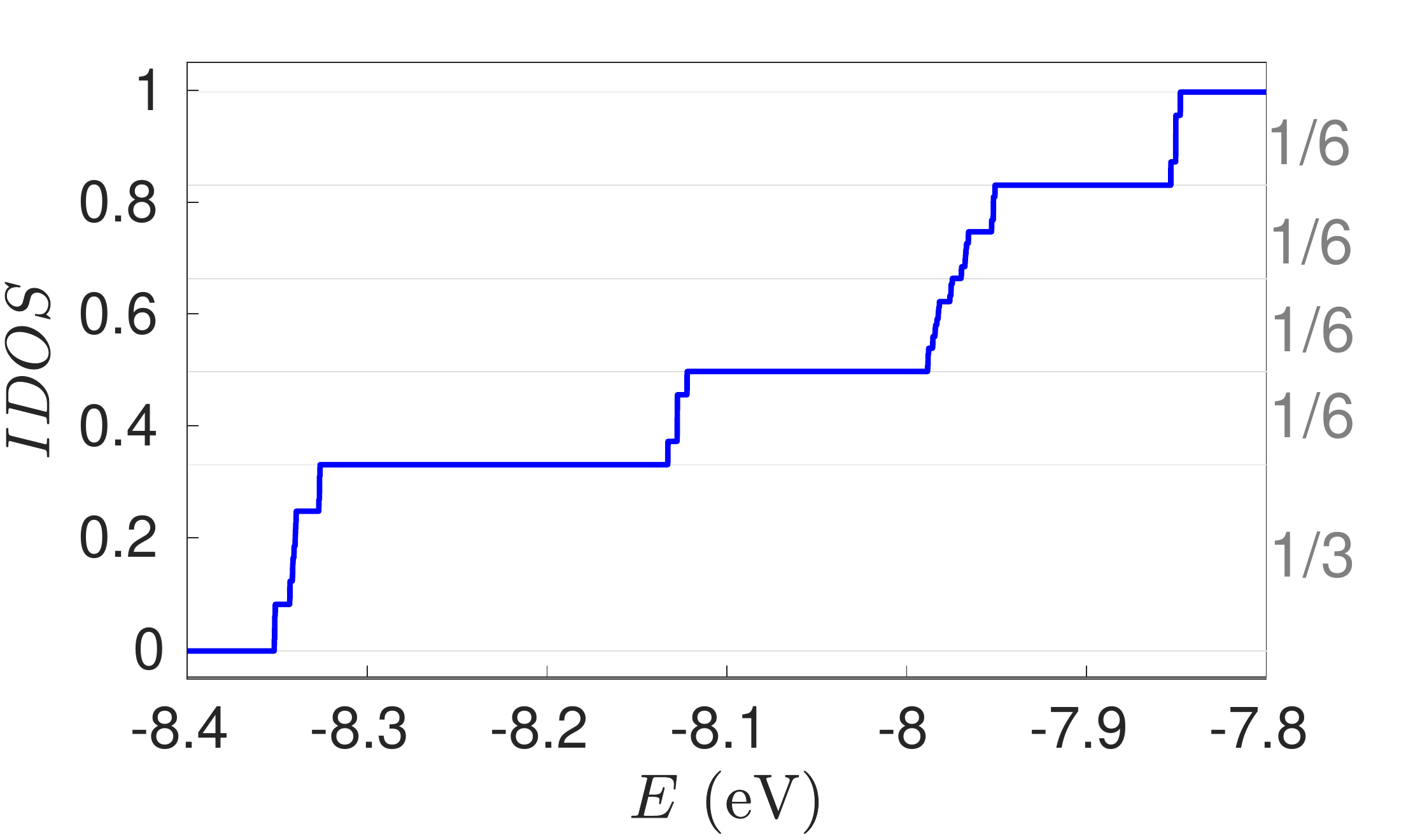}}
	
	\subfloat[$\textit{RS}$ segments.]
	{\includegraphics[width=0.7\columnwidth]{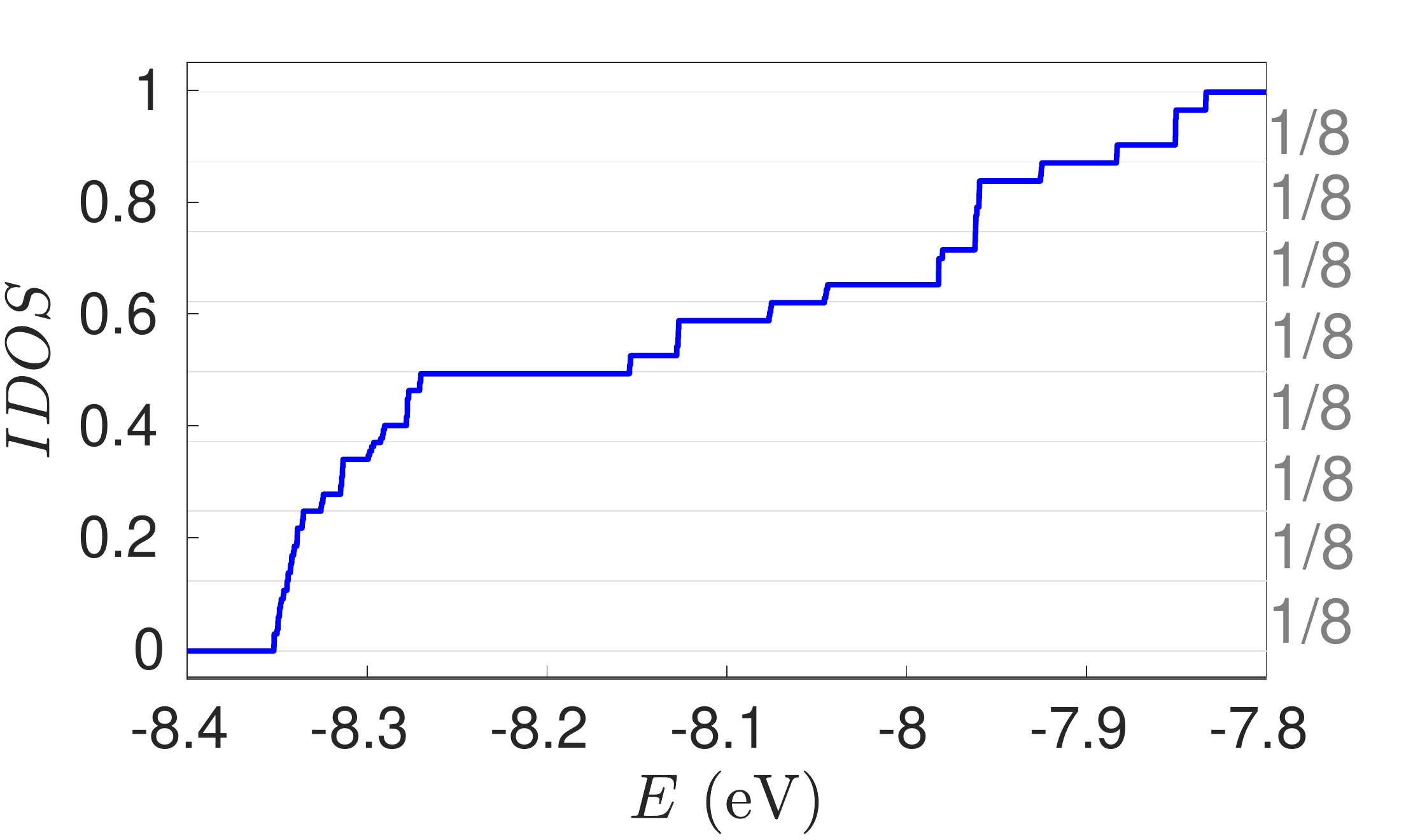}}
	\subfloat[$\textit{CS}$ segments.] {\includegraphics[width=0.7\columnwidth]{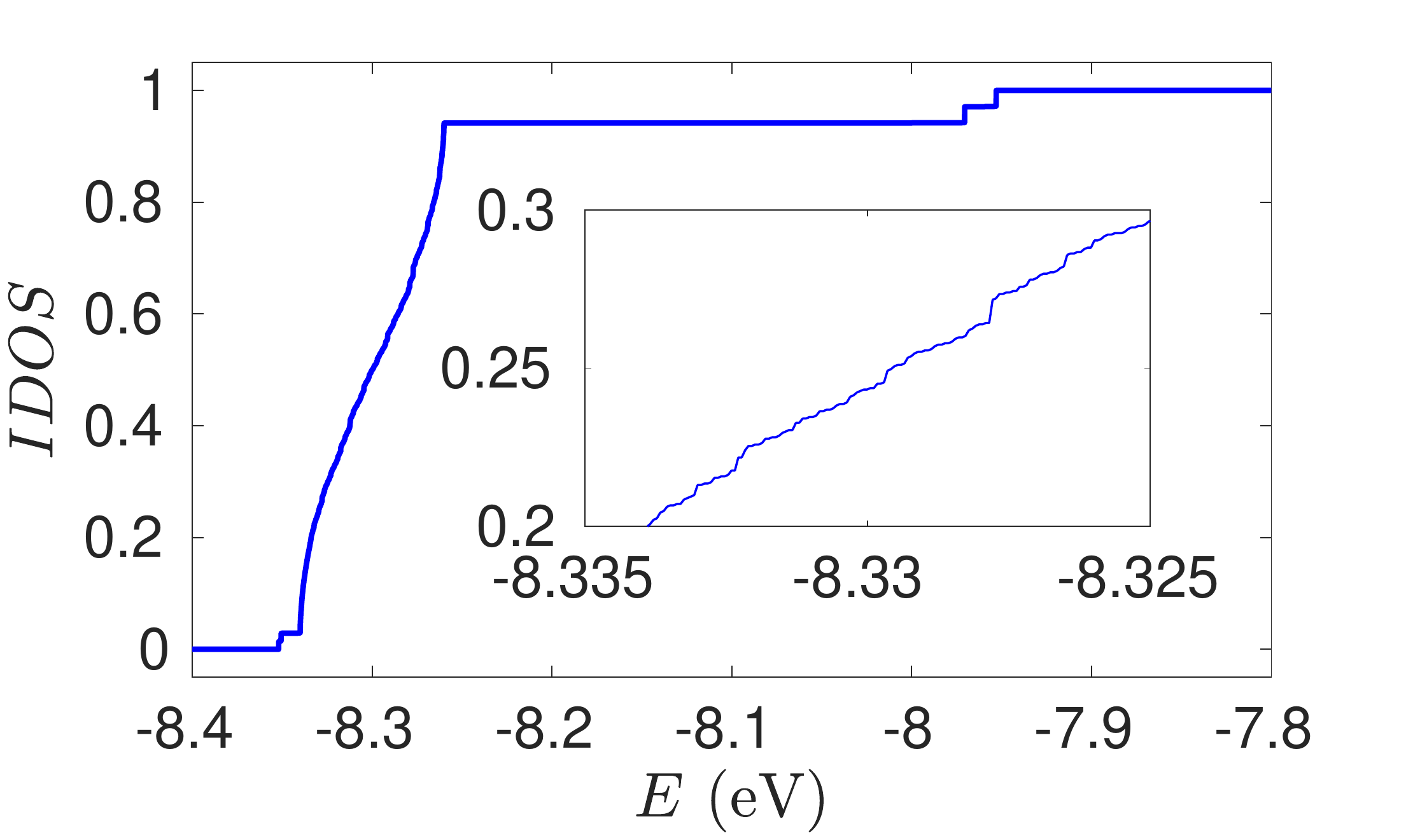}}
	
	\subfloat[$\textit{CGS}(4,2)$ segments.] {\includegraphics[width=0.7\columnwidth]{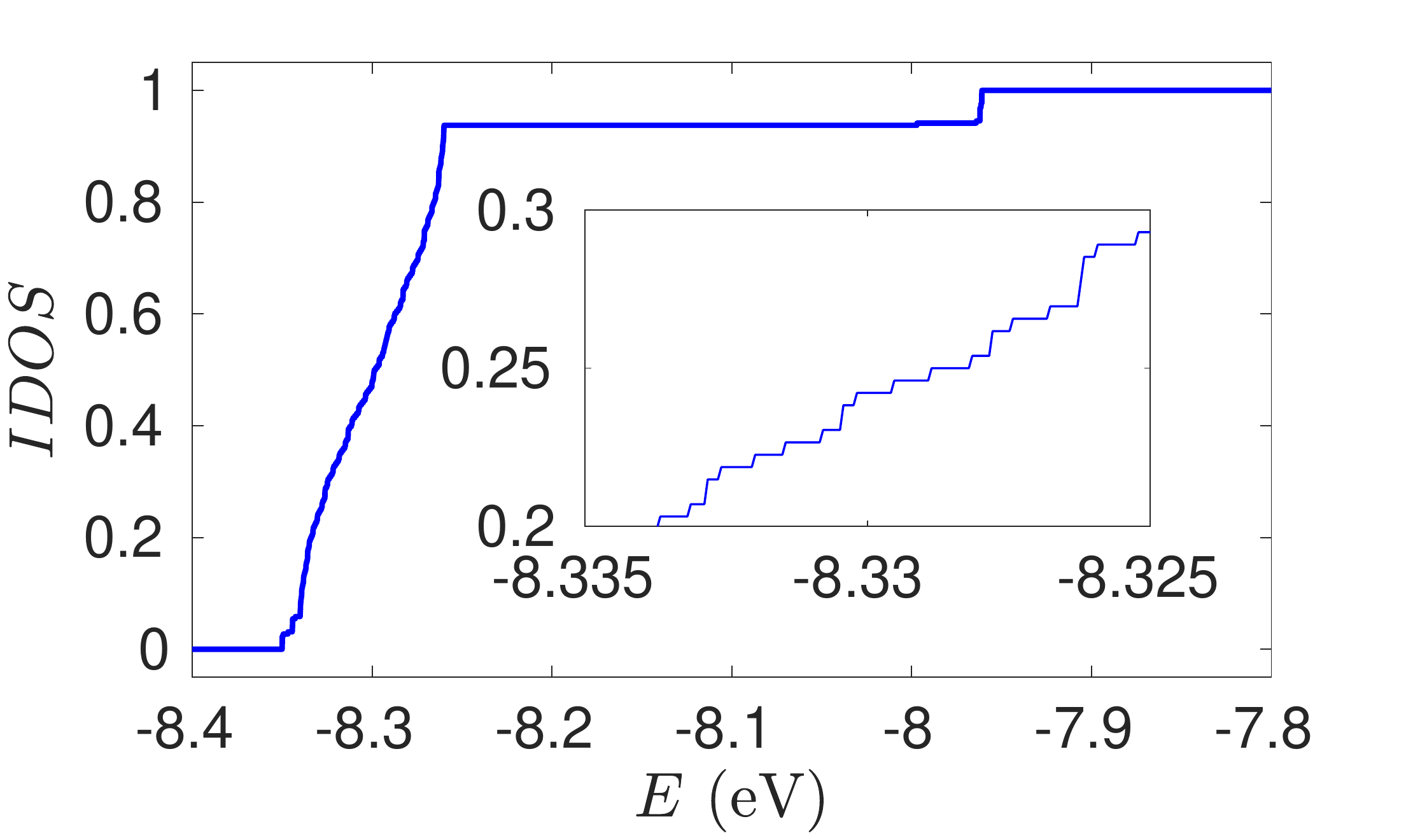}}
	\subfloat[$\textit{KOL}(1,2)$ segments.] {\includegraphics[width=0.7\columnwidth]{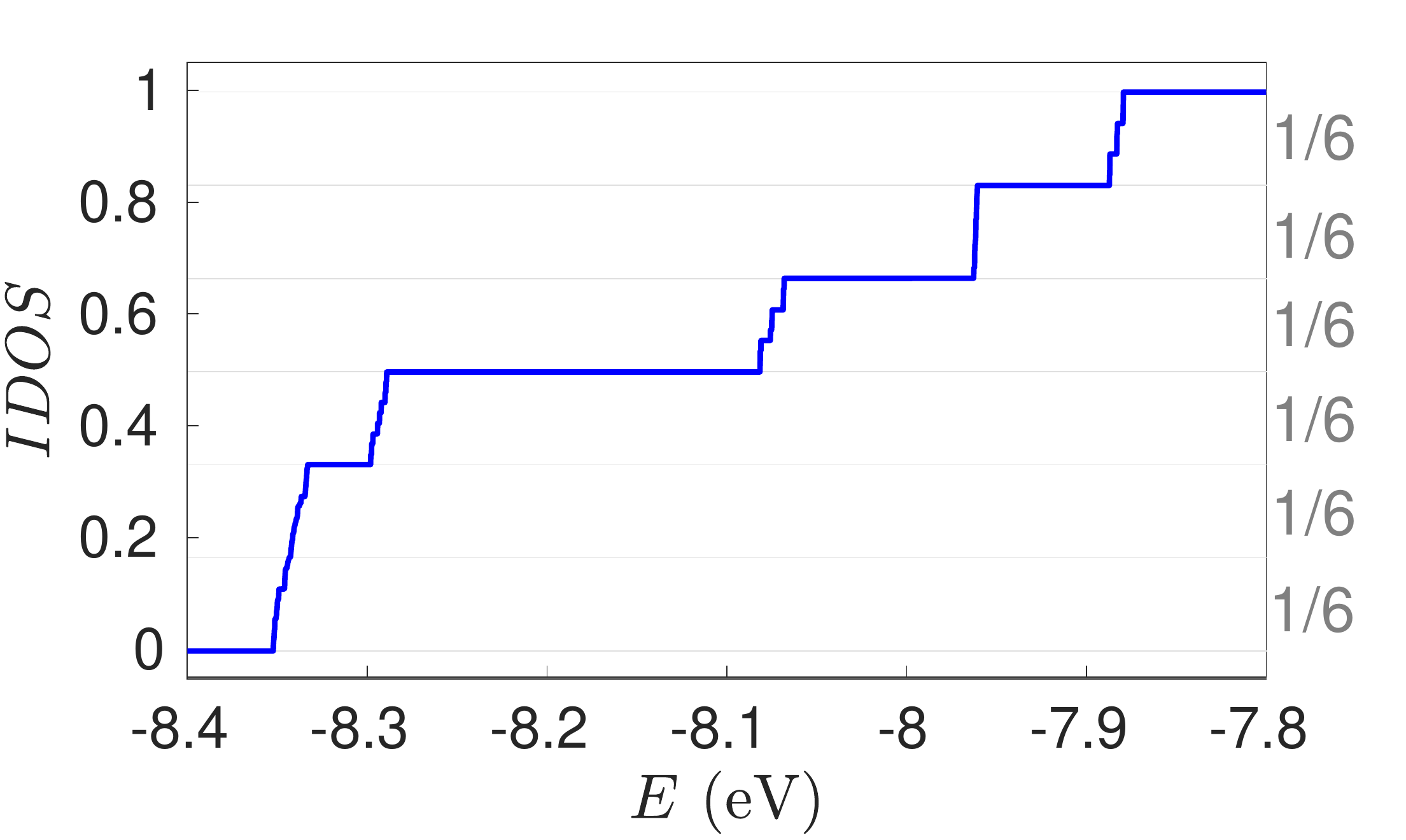}}
	
	\subfloat[$\textit{KOL}(1,3)$ segments.] {\includegraphics[width=0.7\columnwidth]{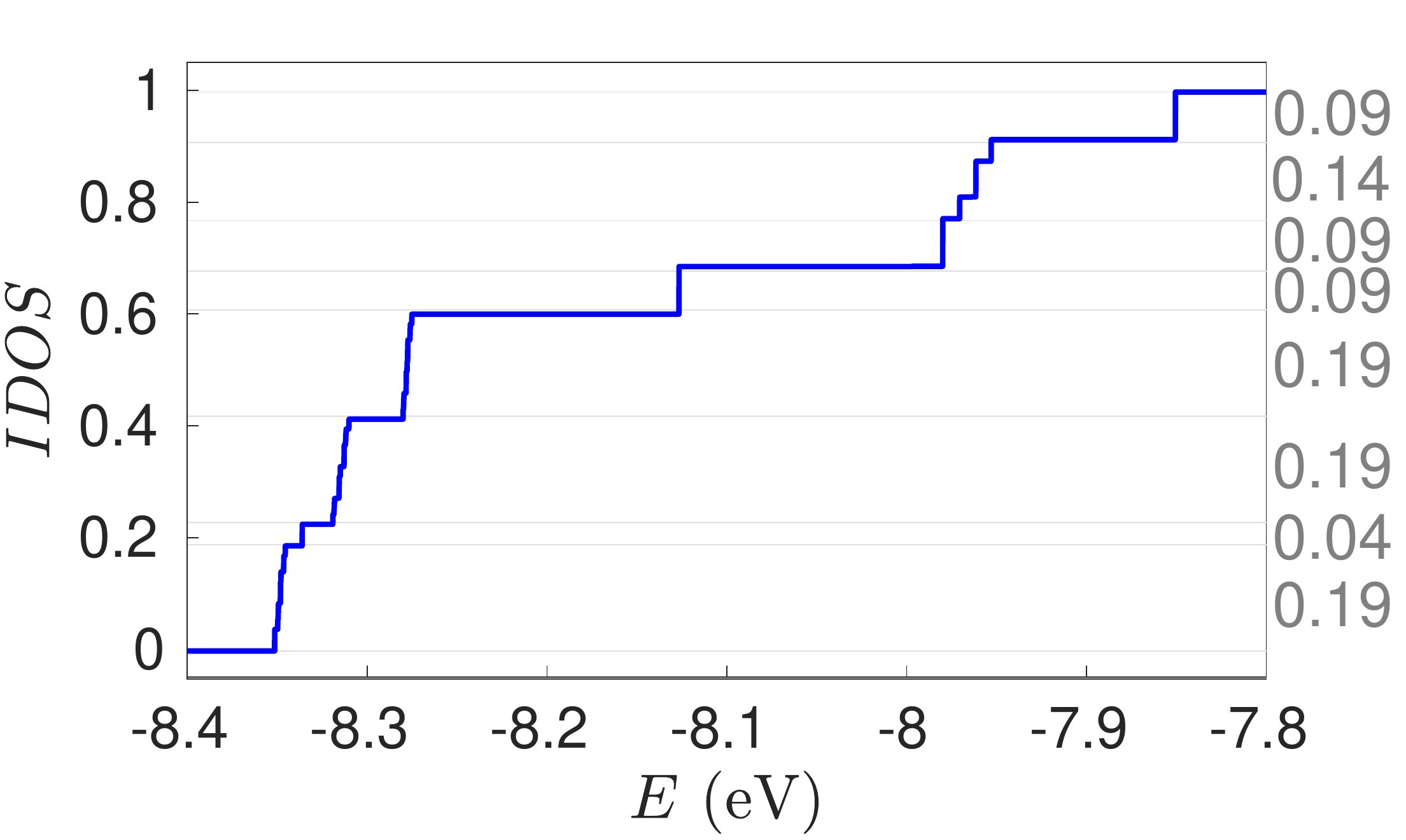}}
	\subfloat[Random segments ($50\%$ G content, $50\%$ A content).] {\includegraphics[width=0.7\columnwidth]{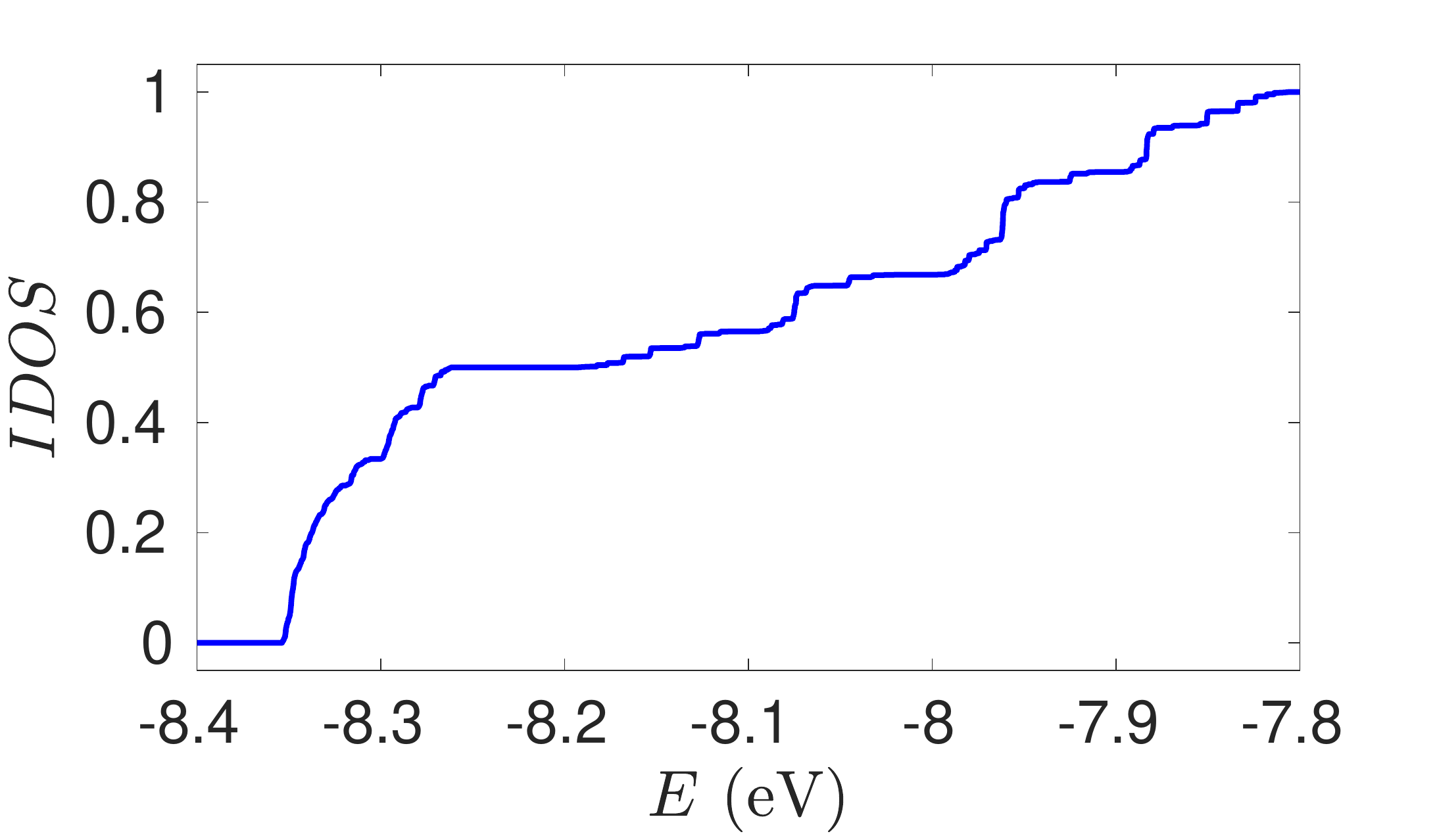}}
	\caption{Normalized IDOS of various categories of DNA segments. In (c), $\phi$ is the golden ratio.}
	\label{fig:IDOS}
\end{figure*}

The normalized IDOS for all categories of DNA segments, for large $N$, is presented in Fig.~\ref{fig:IDOS}. In each panel, the largest energy gap, which is the region between two consecutive discontinuities of the DOS, corresponds to the separation between the upper limit of the allowed energies in the A region and the lower limit of the allowed energies in the G region. The value of the normalized IDOS in this gap corresponds to the relative number of A inside the sequence. Periodic (GA)$_m$ segments possess two narrow, continuous bands, which can be recursively obtained; also, an analytical expression for the DOS exists~\cite{LS:2018}. $\textit{TM}$, $\textit{F}$, $\textit{PD}$, $\textit{RS}$, and $\textit{KOL}$ family sequences posses step-like IDOS, which indicates that the eigenenergies concentrate at specific energy regimes, separated by small gaps. Cantor set family sequences have allowed energies predominantly in the A region. Although at fist glance, the IDOS in this region may seem rather homogeneous, it can be seen from the insets of Fig.~\ref{fig:IDOS}(f)-(g), that the spectrum is very rough. The random sequence IDOS has a shape that resembles to that of the $\textit{RS}$ sequence, although it is much more disrupted. We have also observed that all periodic and deterministic aperiodic segments possess IDOS steps such that their relative value is equal to the occurrence percentages of the possible base-pair triplets (cf. Fig. \ref{fig:triplets}). These steps and the corresponding relative IDOS values are marked in the corresponding panels of Fig. \ref{fig:IDOS} (except for the fractal segments in which the non-AAA triplets have very small occurrence percentages and cannot be depicted). For example, in the $\textit{F}$ segments there are four clear IDOS steps with relative heights $\phi^{-2}, \phi^{-3}, \phi^{-4}, \phi^{-3}$, respectively, where $\phi$ is the golden ratio; this has also been reported before~\cite{Macia:1994}. Our observation connects these relative heights of the IDOS with the occurrence percentage of the possible triplets, further substantiating the relation between the sequence structure and the spectral properties of deterministic aperiodic segments.


\section{Localization} \label{sec:LE}

For the GTM of a given segment, $M_N(E)$, there exists a limiting matrix $L(E)$ such that

\begin{equation}
L(E) = \lim\limits_{N\rightarrow\infty} [M_N(E)^TM_N(E)]^{\frac{1}{2N}}.
\end{equation}
The existence of $L(E)$ is guaranteed by the Oseledec multiplicative ergodic theorem~\cite{Oseledec:1968}. The Lyapunov Exponents of the segment are connected with the $\nu$-th eigenvalue of $L(E)$, $L_\nu(E)$, through 

\begin{equation}
\gamma_\nu(E) = \lim\limits_{N\rightarrow\infty} \frac{1}{2N} \ln[L_\nu(E)].
\end{equation}
If the GTM is a $2d\times2d$ symplectic matrix, as in our case ($d=1$), the Lyapunov exponents are distinct and have the property $-\gamma_1<-\gamma_2<\dots<-\gamma_d<\gamma_d<\dots<\gamma_2<\gamma_1$, hence $\sum\limits_{\nu=1}^{2d}\gamma_\nu = 0$\cite{Crisanti:1993,Scales:1997}. Since the Lyapunov exponents control the growth/decay rate of the solutions of Eq.~\eqref{Eq:TBsystem}, they are associated with the system's inverse localization length. In the case of symplectic GTMs, the localization length is given by the inverse of the smallest positive Lyapunov exponent, $\gamma_d(E)$~\cite{Scales:1997}. 

Since we deal with finite segments, the numerical Lyapunov exponents presented below correspond to finite values of $N$, hence the limit is dropped. To avoid numerical overflows when the matrix product is constructed, we use a QR decomposition scheme: We start with the initial matrix $M_N(E)^TM_N(E) = P_1^T P_2^T \dots P_N^TP_N \dots P_2 P_1$. We perform a QR decomposition of $P_1$, i.e. $P_1 = Q_1^{(1)}R_1^{(1)}$, so that $M_N(E)^TM_N(E) =  P_1^T P_2^T \dots P_N^TP_N \dots (P_2 Q_1^{(1)})R_1^{(1)}$. By consecutively performing QR decompositions at $P_jQ_{j-1}^{(1)}$, we arrive at $M_N(E)^TM_N(E) = Q_{2N}^{(1)}\prod\limits_{j=2N}^{1} R_j^{(1)} := Q^{(1)} R^{(1)}$. Hence, the matrix $R^{(1)} Q^{(1)}$ and the initial matrix are similar, i.e., they have the same eigenvalues. By iterating this procedure, we arrive at a form $R^{(k)}Q^{(k)}$, where $Q^{(k)}$ converges to a unit matrix and $R^{(k)} = \prod\limits_{j=2N}^{1} R_j^{(k)}$, i.e.,  a product of upper triangular matrices with positive diagonal entries in descending order. Hence, the eigenvalue $L_\nu(E)$ is given by the $\frac{1}{2N}$-th power of the diagonal elements of $R^{(k)}$, $R^{(k)\nu \nu}$. The Lyapunov exponents are thus
\begin{equation}
\gamma_\nu(E) =\frac{1}{2N} \sum_{j=1}^{2N}\ln[R_j^{(k)\nu \nu}].
\end{equation}
In our case, where $d=1$, the only exponent to be determined is $\gamma_1(E)$. The index 1 will be dropped below.

\begin{figure} [!h]
\centering
\subfloat[Periodic (GA)$_{m}$ (black/dotted), $\textit{TM}$ (blue/filled), $\textit{KOL}$ $(1,2)$ (magenta/dashed) and random (red/dashed-dotted) segments. All segments have $50\%$ G content.]{\includegraphics[width=8.25cm]{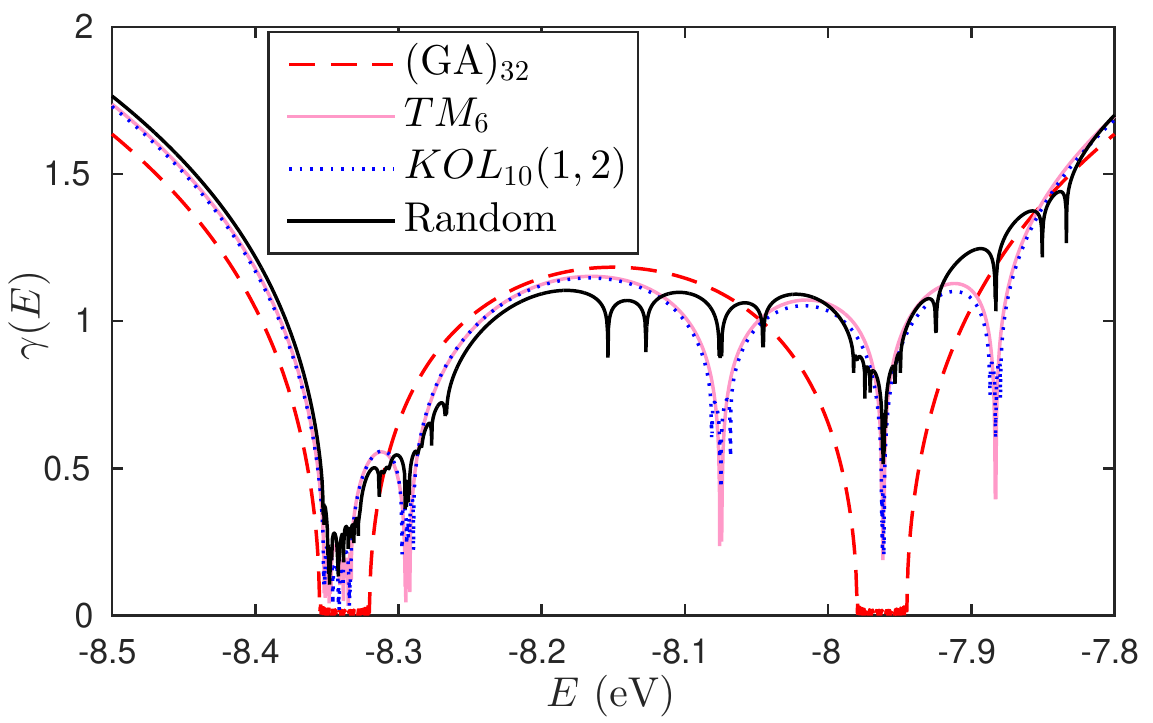}}
	
\subfloat[$\textit{F}$ ($61.82\%$, black/dotted), $\textit{RS}$ ($56.25\%$, blue/filled), $\textit{PD}$ ($67.19\%$, magenta/dashed) and random ($56.25\%$, red/dashed-dotted) segments. Percentages in parentheses denote G content.] {\includegraphics[width=8.25cm]{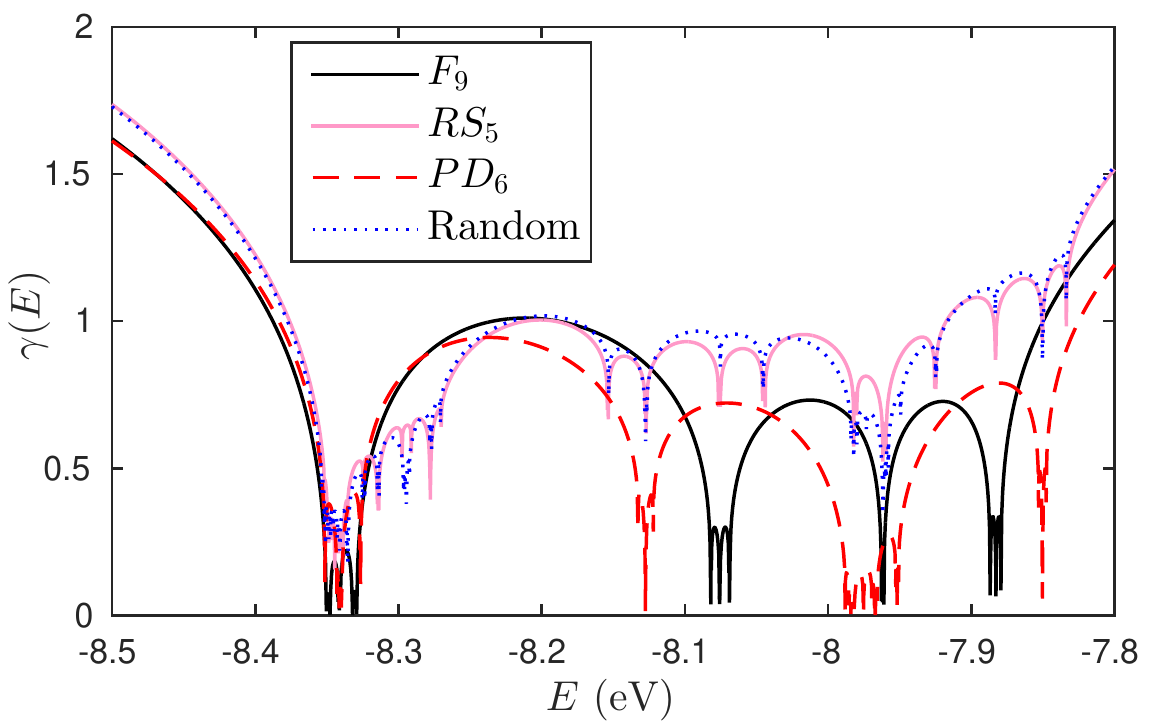}}
		
\subfloat[ $\textit{KOL}(1,3)$ ($40.00\%$, black/dotted), $\textit{CS}$ ($13.17\%$, magenta), $\textit{GCS}(4,2)$ ($6.25\%$, red/dashed-dotted), and two random sequences ($40.00\%$ blue, $10.00\%$ green/dashed). Percentages in parentheses denote G content.]{\includegraphics[width=8.25cm]{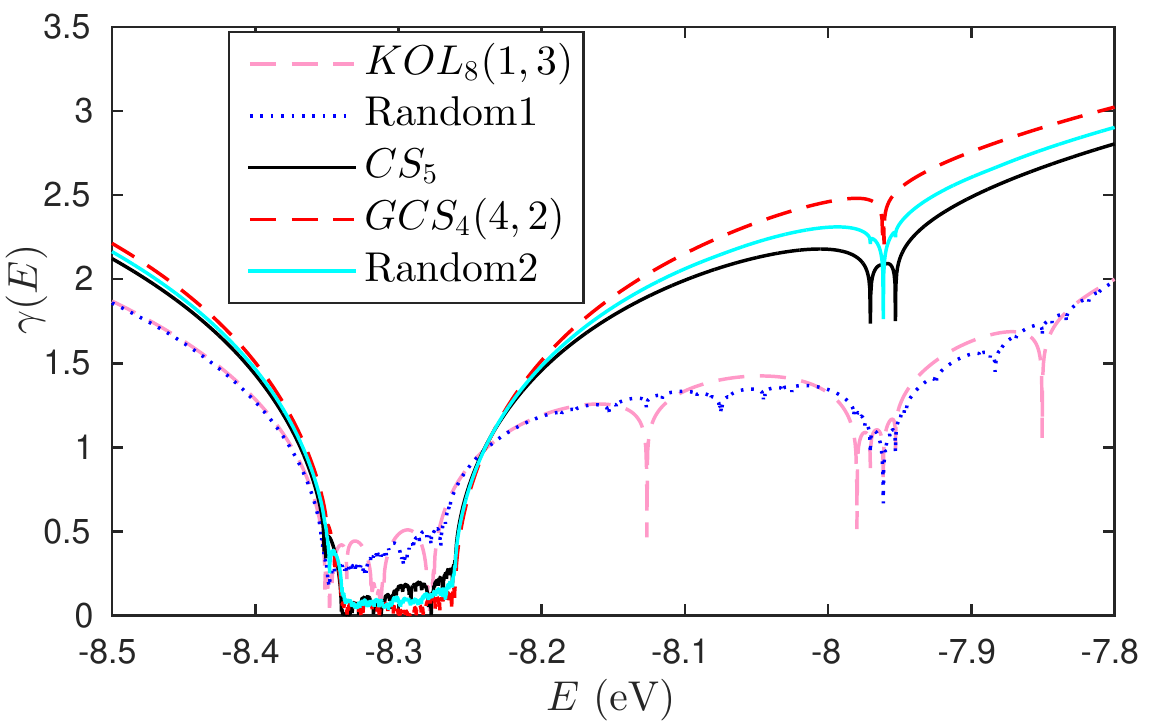}}
	
\caption{Lyapunov exponents of various categories of DNA segments.}
\label{fig:LE}
\end{figure}

The Lyapunov exponents of all categories of periodic and deterministic aperiodic DNA segments, for large $N$, are presented in Fig.~\ref{fig:LE}, together with some sequences with randomly rearranged base pairs. We have grouped together the segments according to the percentages of G and A they posses. Cases with similar G and A content are depicted in Fig.~\ref{fig:LE}(a), with dominant G content in Fig.~\ref{fig:LE}(b) and with dominant A content in Fig.~\ref{fig:LE}(c). Segments grouped together have similar sizes where possible. 

Starting with Fig.~\ref{fig:LE}(a), we notice that the Lyapunov exponents follow the trend of the autocorrelation functions; stronger correlations lead generally to less localized states. Periodic (GA)$_{m}$ segments have vanishing exponents inside their bands; this is a signature of the Bloch character of the wavefunctions. $\textit{TM}$, and $\textit{KOL}(1,2)$ sequences have non-vanishing exponents of similar magnitude. This similarity is direct consequence of the similar base-pair triplet distribution those two categories possess (cf.  Fig.~\ref{fig:triplets}). The random sequence has generally much more localized states. As a general remark, we notice that the Lyapunov exponents in the A energy region are rather smaller than the ones in the G energy region.

The conclusion that segments with stronger correlations possess less localized states is also evident from Fig.~\ref{fig:LE}(b). Furthermore, the Lyapunov exponents of $\textit{F}$ and $\textit{PD}$ segments reach very small values in both base-pair energy regions, while those of $\textit{RS}$ and random segments do not. $\textit{F}$ ($\textit{PD}$) segments posses larger energy intervals of less localized states in the A (G) region than $\textit{PD}$ ($\textit{F}$), while for $\textit{RS}$ and random segments the exponents follow resembling trends. The dominance of smaller exponents in $\textit{PD}$ segments over $\textit{F}$ segments in the G region can be explained by the enhanced presence of $t_{GG}$ (which are of large magnitude) in the former, induced by the occurrence of GGG triplets (cf. Fig.~\ref{fig:triplets}).

In segments with dominant A content, which are depicted in Fig.~\ref{fig:LE}(c), the Lyapunov exponents in the A energy region are much smaller than those in the G region. $\textit{KOL}(1,3)$ segments posses less localized states than random ones with similar G content in their common allowed energy intervals. The more dominant A becomes, the less (more) localized are the states in the A (G) region; this is the case for segments $\textit{CS}$, $\textit{GCS}(4,2)$ and random sequences with similar G content. In these cases, there are large A-rich regions within the segments, interrupted by Gs, which act like a disorder. The more homogeneous regions the segments possess, the less localized their eigenstates will be in the A energy region. Comparing these segments in Fig.~\ref{fig:LE}(c), we can see that, generally, as the percantage of G decreases, the exponents become smaller in the A region; however, there are always energies at which the fractal sequences, which possess stronger correlations, are more delocalized than the random one. The very small percentage of G leads to highly localized states in the corresponding energy interval.

\section{Transmission coefficient} \label{sec:TC}

The transmission coefficient describes the probability of an incident wave to be transmitted through a specific segment. We connect the segment to semi-infinite homogeneous metallic leads, which act as carrier baths. The leads' energy spectrum is given by the dispersion relation $E = E_M + 2t_M\cos(qa)$, where $E_M$ is the on-site energy of the leads and $t_M$ is the hopping integral between the leads' sites. The coupling between the segment and the left (right) lead is described by the effective parameters $t_{cL(R)}$. Assuming incident waves from the left, we have
\begin{equation}
\psi_{\{n\}\le 1} = e^{iqna} + re^{-iqna}, \quad  \psi_{\{n\}\ge N} = te^{iqna}.
\end{equation}
The transmission coefficient is defined as $T(E) = \abs{t}^2$. 
The GTM of the scattering region obeys the equation
\begin{equation} \label{Eq:MNs}
\begin{pmatrix}
\psi_{N+1} \\ \psi_N
\end{pmatrix} = P_R M_N P_L
\begin{pmatrix}
\psi_{1} \\ \psi_0
\end{pmatrix}.
\end{equation}
\begin{equation} \label{Eq:PLR}
P_R = 
\begin{pmatrix}
\frac{t_N}{t_{cR}} & 0 \\ 0 &  \frac{t_{cR}}{t_M}
\end{pmatrix}, \quad
P_L =
\begin{pmatrix}
\frac{t_M}{t_{cL}} & 0 \\ 0 & \frac{t_{cL}}{t_N}
\end{pmatrix}
\end{equation}
are the matrices that describe the coupling of the three subsystems. After some manipulations, we arrive at the following expression for the transmission coefficient
\begin{equation} \label{Eq:TCdef}
T(E) = \frac{1}{1 + \Lambda(E)},
\end{equation}
\begin{equation} \label{Eq:Lambdadef}
\Lambda(E) = \frac{\left[W_N(E) + X_N^+(E) \cos(q a)\right]^2}{4 \sin[2](q a)} + \frac{X_N^-(E)^2}{4}.
\end{equation}

\begin{figure} [h!]
\centering
\subfloat[Periodic (GA)$_{m}$ (black/dotted), $\textit{TM}$ (blue/filled), $\textit{KOL}$ $(1,2)$ (magenta/dashed) and random (red/dashed-dotted) segments. 
All segments have $50\%$ G content.]{\includegraphics[width=8.25cm]{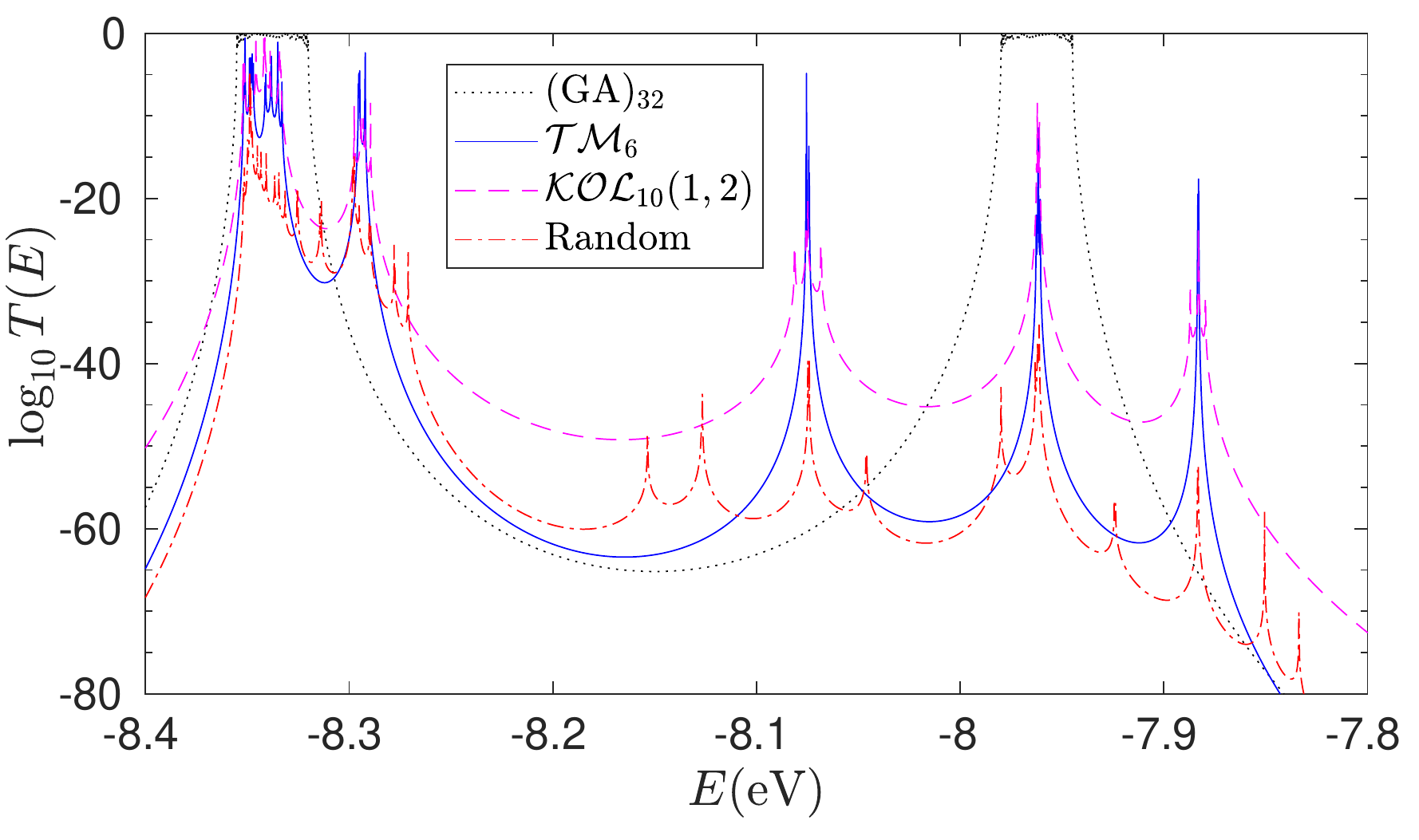}}
	
\subfloat[$\textit{F}$ ($61.82\%$, black/dotted), $\textit{RS}$ ($56.25\%$, blue/filled), $\textit{PD}$ ($67.19\%$, magenta/dashed) and random ($56.25\%$, red/dashed-dotted) segments. Percentages in parentheses denote G content.] {\includegraphics[width=8.25cm]{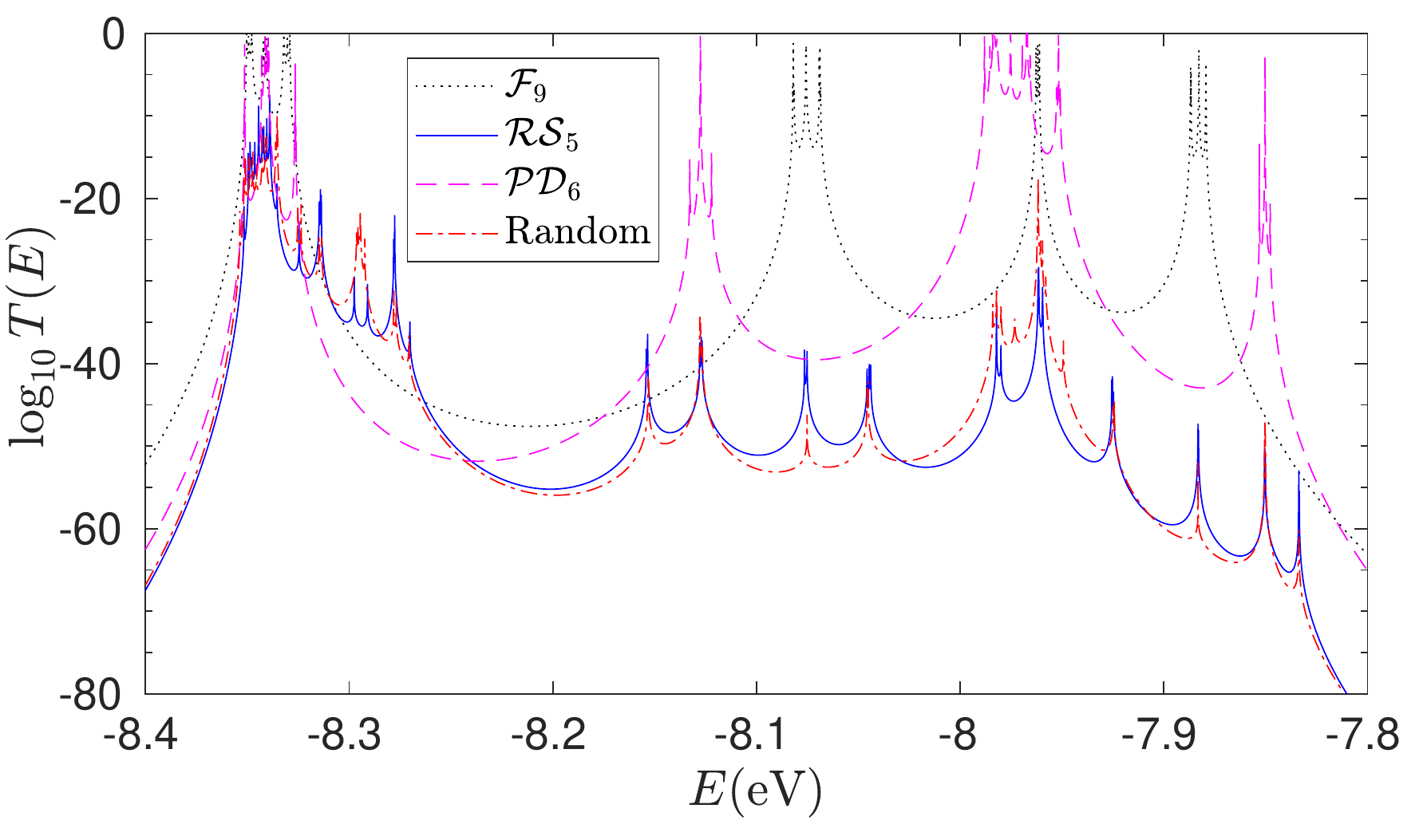}}

\subfloat[ $\textit{KOL}(1,3)$ ($40.00\%$, black/dotted), $\textit{CS}$ ($13.17\%$, magenta), $\textit{GCS}(4,2)$ ($6.25\%$, red/dashed-dotted), and two random sequences ($40.00\%$ blue, $10.00\%$ green/dashed). Percentages in parentheses denote G content.]{\includegraphics[width=8.25cm]{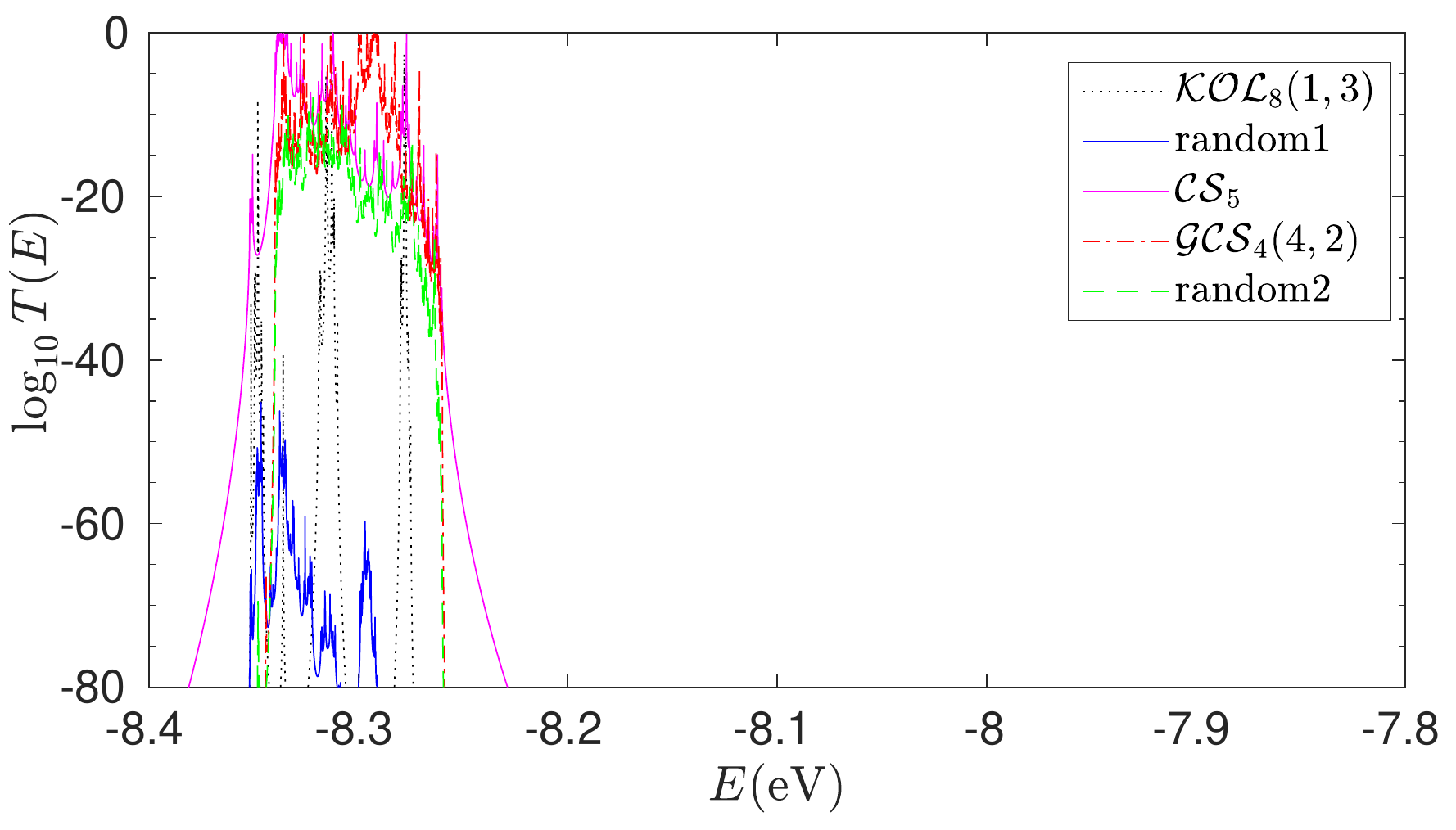}}

\caption{Transmission coefficients of various categories of DNA segments.}
\label{fig:TC}
\end{figure}

\begin{subequations}
	\begin{equation} \label{Eq:Wdef}
	W_N(E) = M_N^{11} \omega-M_N^{22}\omega^{-1},
	\end{equation}
	\begin{equation} \label{Eq:Xdef}
	X_N^\pm(E) = M_N^{12} \chi \pm M_N^{21} \chi^{-1},
	\end{equation}
	\begin{equation} \label{Eq:omegadef}
	\omega = \frac{t_Mt_{N}}{t_{cR}t_{cL}}, \quad \chi = \frac{t_{cL}}{t_{cR}}.
	\end{equation}
\end{subequations}
$\omega$, included only in $W_N(E)$, expresses the deviation of the coupling of the system to the leads from the ideal case in which they are interconnected as if they were connected to themselves; hence $\omega$ is a coupling strength factor. $\chi$, included only in $X_N^\pm(E)$, expresses the difference of the coupling between the leads and each end of the system; hence, $\chi$ is a coupling asymmetry factor. In Ref.~\cite{LS:2018} we discuss the effects of $\omega$ and $\chi$, as well as of the leads properties, to the transmission profiles of periodic segments. In the following, we choose the coupling parameters to satisfy the ideal and symmetric coupling conditions, $\abs{\omega} = \abs{\chi} = 1$. These have been shown to be the optimal coupling conditions for periodic segments~\cite{LS:2018}. 
We choose $E_M = \tfrac{(E_{A-T} + E_{G-C})}{2} = -8.15$ eV and $t_M = -0.25$ eV, so that all eigenstates of the systems under examination are contained within the leads' bandwidth.

In Fig.~\ref{fig:TC} we present the transmission coefficients. At first glance, the transmission coefficients qualitatively follow the trend of the Lyapunov exponents (cf. Fig.~\ref{fig:LE}). The less localized the eigenstates are, the more transparent the segments are to the incident waves at their energy region. Periodic (GA)$_m$ segments display the most enhanced transmission, and reach the full transmission condition at specific energies~\cite{LS:2018}; this does not hold in general for deterministic aperiodic and random segments. Furthermore, apart from periodic (GA)$_m$, $\textit{F}$, and $\textit{PD}$ segments, transmission in the G energy region is from very small to negligible. These categories, together with the Cantor Set family ones, display the most enhanced transmission. $\textit{TM}$ and $\textit{KOL}(1,2)$ sequences display some energies at which transmission is rather significant. Deterministic aperiodic segments are more transparent than random ones with similar base-pair content, with the exception of $\textit{RS}$, that generally follows the trend of its randomly redistributed counterpart. Finally, we notice that the sequences shown in Fig. \ref{fig:TC}(c) have negligible transmission in the G energy region. This is due to the small role $t_{GG}$ plays, since it rarely occurs within the segments.

\section{Current-Voltage Curves} \label{sec:IV}

We apply a constant bias voltage $V_b$ between the leads, so that their chemical potential takes the form $\mu_{\substack{L \\ R}} = E_M \pm \frac{V_b}{2}$. Then, a linear voltage drop within the DNA segment is induced and the transmission coefficient becomes bias-dependent. The energy regime between the leads' chemical potentials defines the conductance channel. 
The electrical current at zero temperature can be computed using the Landauer-B\"{u}ttiker formalism~\cite{Landauer:1957,Buttiker:1988,Datta:1995} as
\begin{equation} \label{Eq:IV}
I(V) = \frac{2e}{h} \int\limits_{E_M-\frac{V_b}{2}}^{E_M+\frac{V_b}{2}} T(E,V_b)\ dE, 
\end{equation}
since the Fermi-Dirac distributions, $f(E_M\pm\frac{V_b}{2})$, are Heaviside step-functions. The factor $2$ in Eq.~\eqref{Eq:IV} comes from the double spin-degeneracy of each electronic level.

\begin{figure} [!h]
	\centering
	\includegraphics[width=\columnwidth]{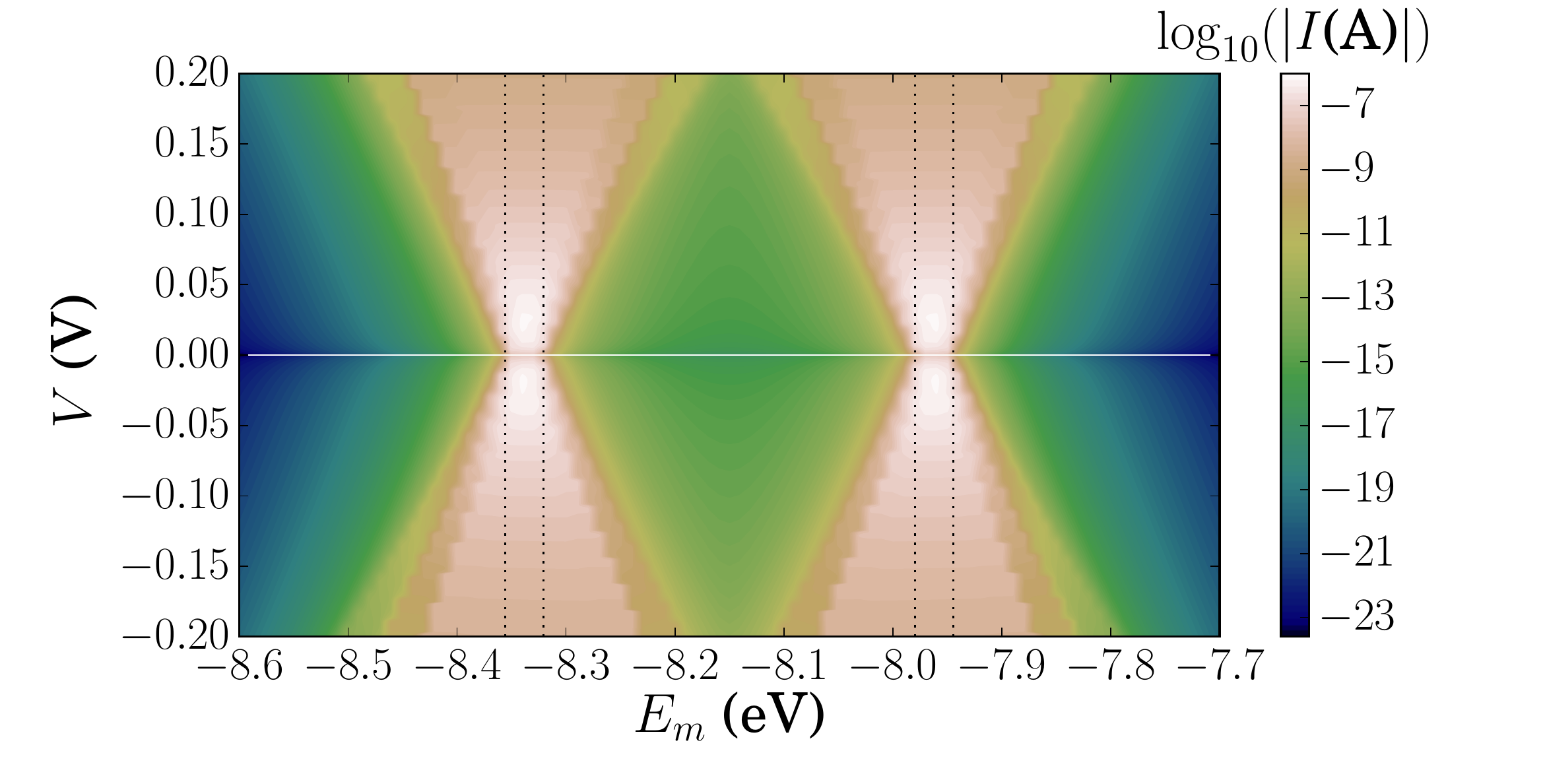}
	\caption{The role of the leads' Fermi level, $E_M$, to the I-V curve of a (GA)$_{16}$ segment. The vertical dotted lines encompass the bands of the segment.}
	\label{fig:IVleads}
\end{figure}

Again, we choose the coupling parameters to satisfy the ideal and symmetric coupling conditions, $\abs{\omega} = \abs{\chi} = 1$.
We set the leads hopping integral $t_M = -0.5$ eV to ensure that the leads' bands are wide enough to capture the whole picture. The choice of the leads Fermi level, $E_M$, plays a major role in both the shape of the I-V curves and the magnitude of the currents. This is demonstrated in Fig.~\ref{fig:IVleads}, where the I-V curve of a periodic (GA)$_{16}$ segment is determined as a function of $E_M$. 
It is evident that larger currents ($\sim 0.1$ $\mu$A) occur at small biases when $E_M$ lies within the bands of the segment. When this is not the case, voltage thresholds appear, and the (smaller in magnitude) turn-on currents emerge at biases that increase in a linear fashion with changing $E_M$. The magnitude of the currents becomes gradually smaller as $E_M$ moves further away from the segments' bands, and is negligible when $E_M$ lies well outside the bands. Finally, we should mention that the I-V curves are symmetric with respect to the difference between $E_M$ and $\tfrac{(E_{A-T} + E_{G-C})}{2}$. The above mentioned conclusions hold also qualitatively for segments consisting of identical monomers with crosswise purines, such as (GC)$_m$, where only one on-site energy ($E_{G-C}$) is involved, with the difference that the curves are symmetric with respect to the difference between $E_M$ and $E_{G-C}$.

\begin{figure} [!h]
	\centering
	\includegraphics[width=8.25cm]{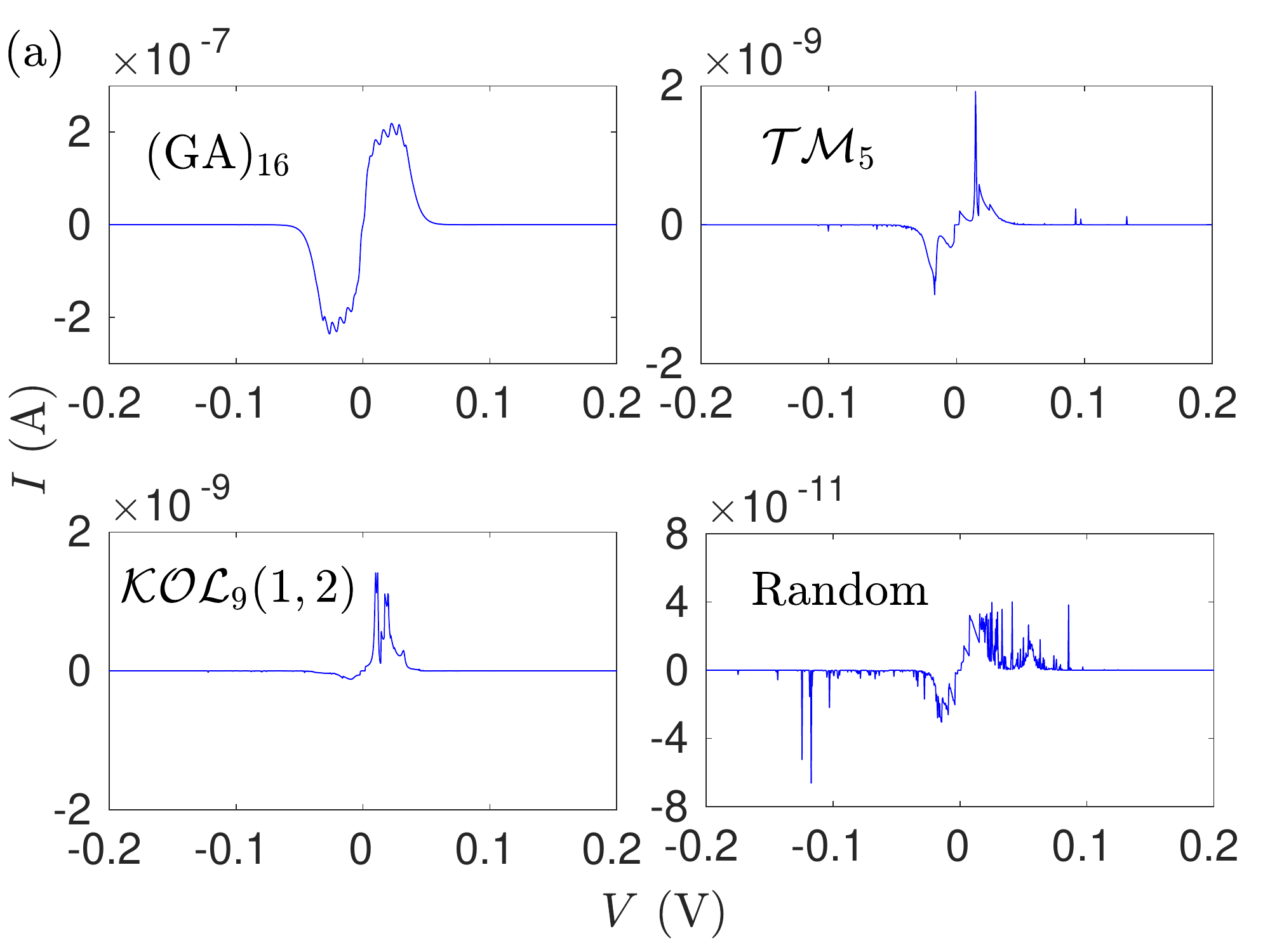}
	\includegraphics[width=8.25cm]{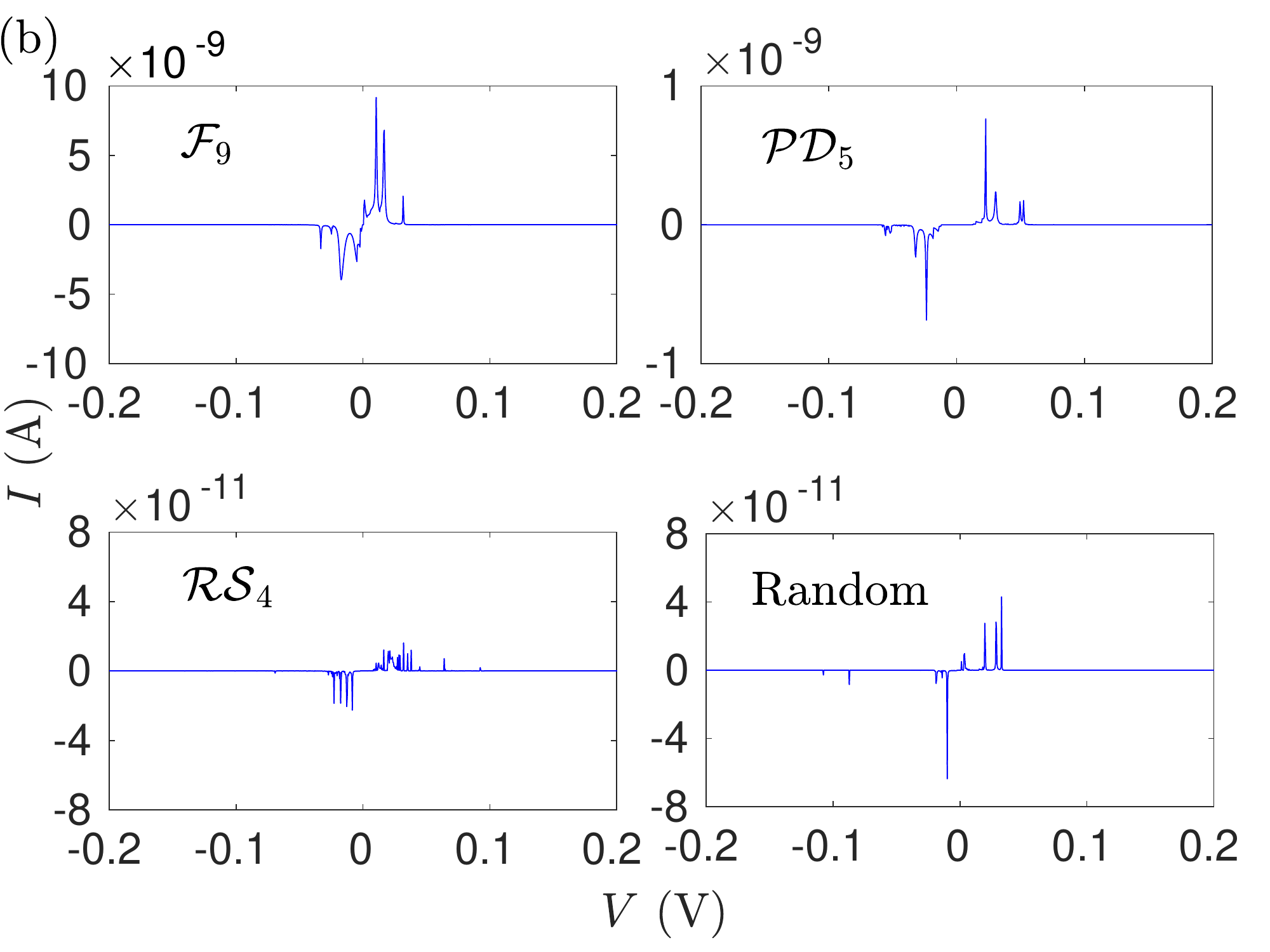}
	\includegraphics[width=8.25cm]{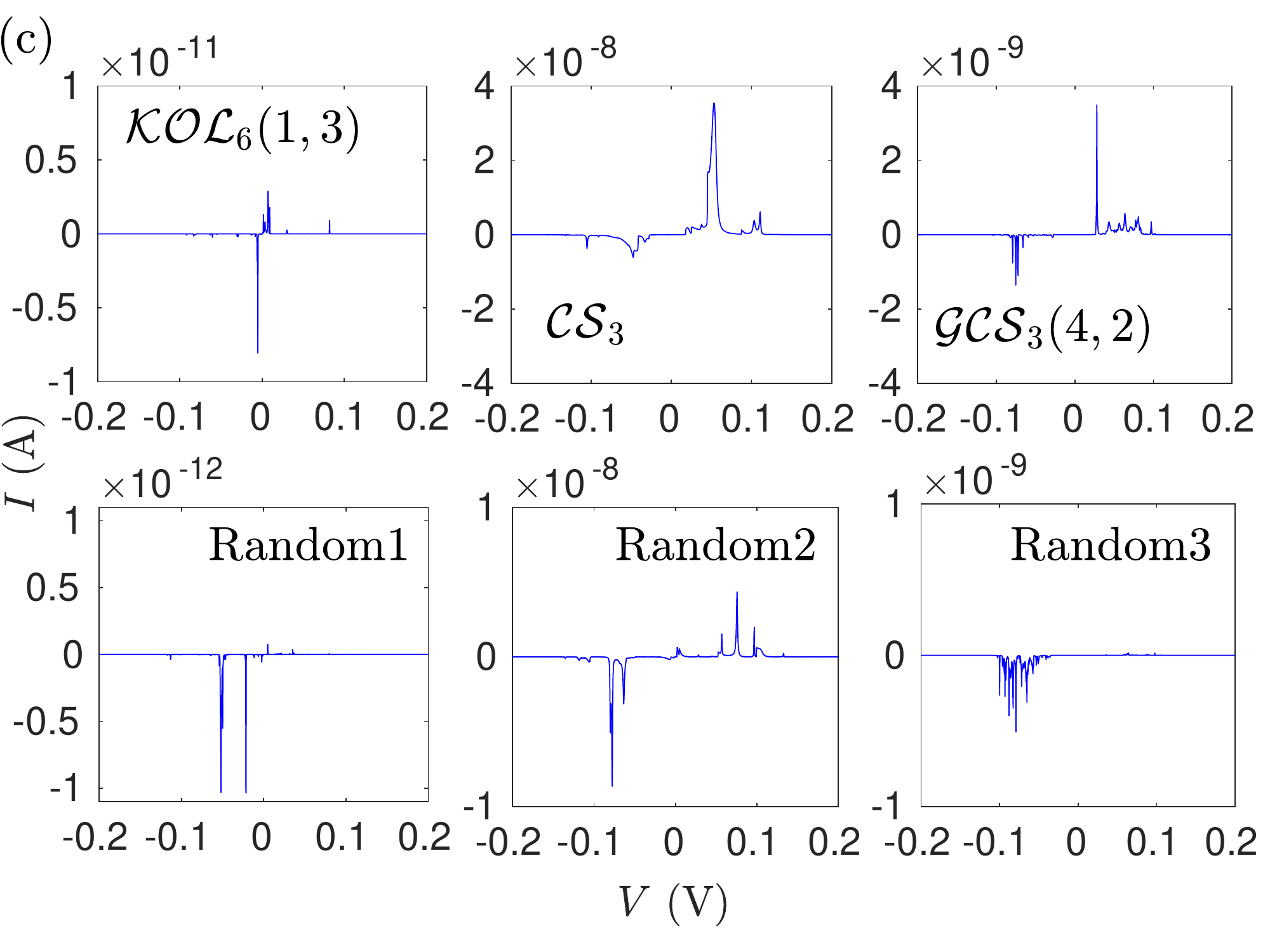}
	
	\caption{I-V curves of various categories of DNA segments for $E_M = -8.35$ eV.
		Categories as in Figs. \ref{fig:LE} and \ref{fig:TC}. (a) Periodic (GA)$_{m}$, $\textit{TM}$, $\textit{KOL}$ $(1,2)$ segments and a random segment with similar G content. (b) $\textit{F}$, $\textit{PD}$, $\textit{RS}$ segments, and a random segment with similar G content. (c)  (top) $\textit{KOL}(1,3)$, $\textit{CS}$, $\textit{GCS}(4,2)$ segments. (Bottom) Random rearrangements of $\textit{KOL}(1,3)$, $\textit{CS}$, $\textit{GCS}(4,2)$ segments, respectively.}
	\label{fig:IV835}
\end{figure}

\begin{figure} [!h]
	\centering
	\includegraphics[width=8.25cm]{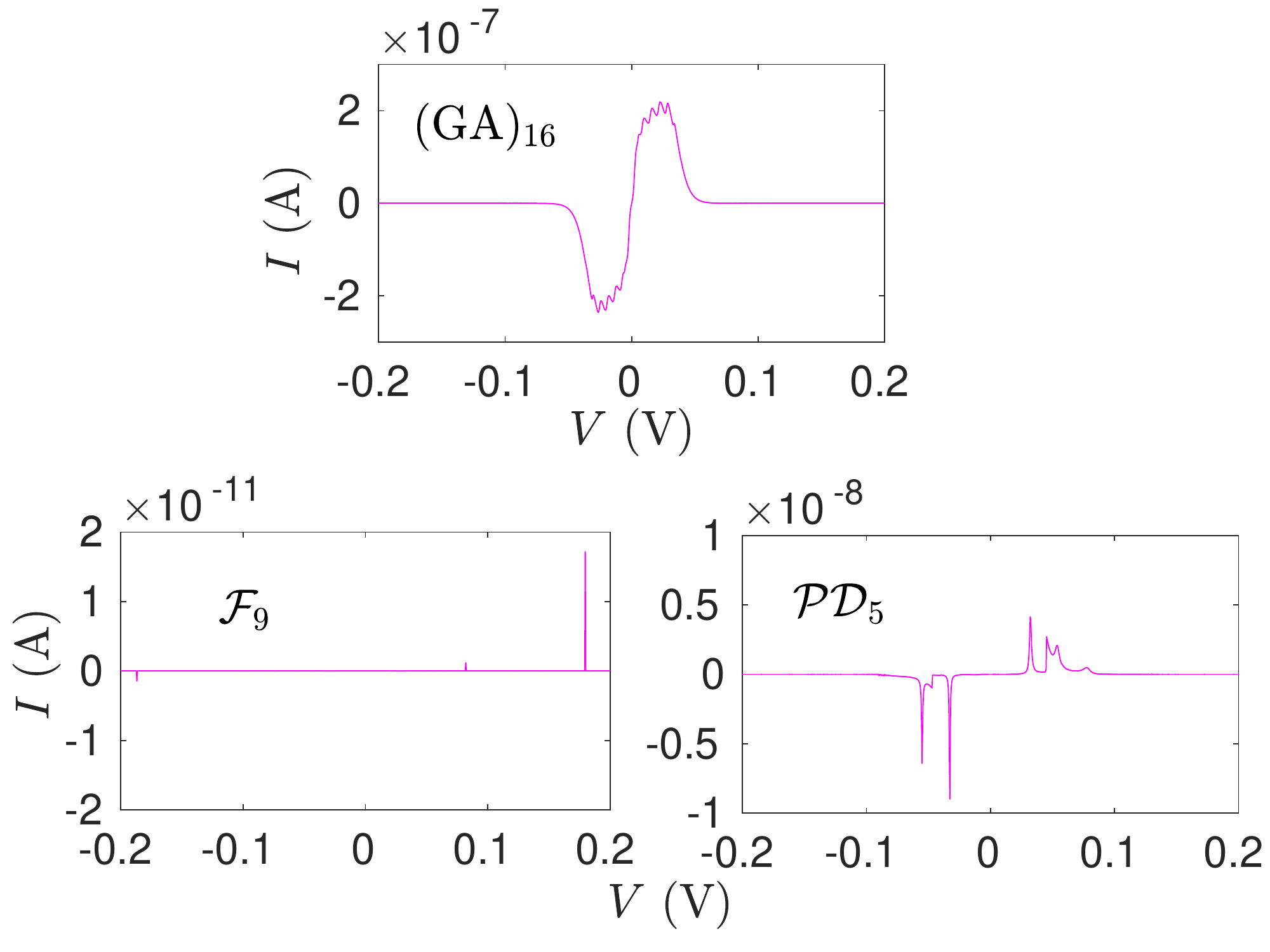}
	\caption{I-V curves of various categories of DNA segments for $E_M = -7.95$ eV.}
	\label{fig:IV795}
\end{figure}

Given the previous discussion and Fig.~\ref{fig:IVleads}, we chose to study the I-V curves of all segments for two values of $E_M$, specifically $-7.95$ eV and $-8.35$ eV (i.e. at the center of the periodic segment's bands), to capture both G and A energy regions. 
In the following, we will only present curves the currents of which reach the pA regime. Our results are depicted in Figs.~\ref{fig:IV835} and~\ref{fig:IV795}, for $E_M = -8.35$ eV and $E_M = -7.95$ eV, respectively. 

From Fig~\ref{fig:IV835}(a), it is evident that periodic segments can carry significantly larger currents ($\sim 0.1$ $\mu$A) than other categories. The deterministic aperiodic $\textit{TM}$ and $\textit{KOL}(1,2)$ segments display quite smaller currents than the periodic ones, of similar magnitude ($\sim 1$ nA), but with clearly distinct shapes. The similarity of current magnitudes between $\textit{TM}$ and $\textit{KOL}(1,2)$ segments is in accordance with the similarity in the values of the Lyapunov exponents and zero-bias transmission coefficient for these cases, cf. Figs.~\ref{fig:LE}(a) and \ref{fig:TC}(a), respectively. The random segment displays significantly smaller currents compared to the rest categories, reaching $\sim 10$ pA.

As far as segments with dominant G content are concerned, we can see in Fig. \ref{fig:IV835}(b) that $\textit{F}$ and $\textit{PD}$ segments can carry significantly larger currents than the $\textit{RS}$ and random ones. This is again in accordance with the magnitude of the Lyapunov exponents and the transmission coefficients for these cases, cf. Figs. \ref{fig:LE}(b) and \ref{fig:TC}(b). In the A energy region, there is a larger energy range in which $\textit{F}$ segments display less localized states and higher transmission than $\textit{PD}$ ones. This is fact is reflected on the magnitude of the currents ($\sim 1$ nA for $\textit{F}$, $\sim 0.1$ nA for $\textit{PD}$). $\textit{RS}$ and random segments display currents in the $\sim10$ pA regime, but their curves have different shapes.

Sequences with dominant A content are depicted in Fig. \ref{fig:IV835}(c). $\textit{KOL}(1,3)$ sequences display rather small currents, that hardly reach $10$ pA, due to the fact that the hopping integral with the largest occurrence percentage, i.e. $t_{AA}$, is of rather small value. Albeit their small magnitude, the currents of $\textit{KOL}(1,3)$ sequences are larger than of their random rearrangement, which hardly reach $1$ pA. In Cantor set family sequences, A content is much larger than G content, leading to large parts of the segment being essentially homogeneous. Hence, although $t_{AA}$ has a small value, rather large currents occur ($\sim 10$ nA for $\textit{CS}$, $\sim 1$ nA for $\textit{GCS}(4,2)$). In this class of sequences, G, which, due to its small presence acts as a disorder in an otherwise homogeneous segment, is gathered in specific regions. Therefore, the currents they display are about one order of magnitude larger than their random rearrangements ($\sim$ 1 nA and $\sim 10$ nA, respectively).

As discussed in previous sections, in the G energy region the eigenstates of most segment categories are highly localized and display very small or negligible transmission. This, for $E_M = -7.95$ eV, leads to currents that lie well below the pA regime. The only cases that do not follow this trend are the periodic, $\textit{F}$, and $\textit{PD}$ segments, the I-V curves of which are depicted in Fig.~\ref{fig:IV795}. The periodic segments curve in this case is identical to the one for $E_M = -8.35$ eV, due to the symmetry of the I-V curves with respect to the difference between $E_M$ and $\tfrac{(E_{A-T} + E_{G-C})}{2}$, cf. Fig.~\ref{fig:IVleads}. The rest two cases display energy intervals in the G region for which less localized states and enhanced transmission occur, as shown in previous sections. Close to $E_M$, the interval for $\textit{F}$ segments is much smaller than the one for $\textit{PD}$ segments, leading to a great difference in the current magnitudes between the two cases: a single spike of $\sim 100$ pA for $\textit{F}$ segments, currents in the $\sim 10$ nA regime for $\textit{PD}$ segments. This is due to the presence of GGG triplets in $\textit{PD}$ segments, which leads to enhanced presence of $t_{GG}$ (the magnitude of which is large), compared to $\textit{F}$ segments, cf. Fig~\ref{fig:percentage}(c)-(d).

\section{Effect of parameters} \label{sec:parametrization}

It is common in the literature that all hopping parameters between different moieties are considered equal, for simplicity. Let us provide some example results occurring for identical hopping parameters, with reference to the Lyapunov exponents: In this case, $\textit{F}$ segments posses more delocalized states in the G region (results not presented here), in contrast with the discussion of Fig.~\ref{fig:LE}(b). Additionally, for all studied sequences, if we take equal hopping parameters, the act of substituting G with A and vice versa leads to a mere reflection of $\gamma(E)$ relative to the mean value of the on-site energies,  
$\tfrac{(E_{A-T} + E_{G-C})}{2}$ (results not presented here). 
This is not the case when different hopping parameters are considered. Their relative presence and magnitude can lead to significant differences in the electronic properties. Another example is the $\textit{TM}$ sequence. If we equalize all hopping parameters, the Lyapunov exponent is also symmetric relative to 
$\tfrac{(E_{A-T} + E_{G-C})}{2}$ (results not presented here), 
a scenario that does not hold for different hopping parameters, cf. Fig.~\ref{fig:LE}(a). 
Of course, the inclusion of different hopping parameters plays significant role not only in the Lyapunov exponents, but also in all properties that are determined by the electronic structure, such as the transmission coefficient and the I-V curves. To conclude, besides the fact that, in terms of chemical complexity, taking identical hopping parameters is unrealistic, our treatment reveals that considering different hopping parameters leads to a better understanding of the interplay between sequence intricacy and transport properties, both quantitatively and qualitatively.

\begin{figure} [!h]
	\centering
	\includegraphics[width=8.25cm]{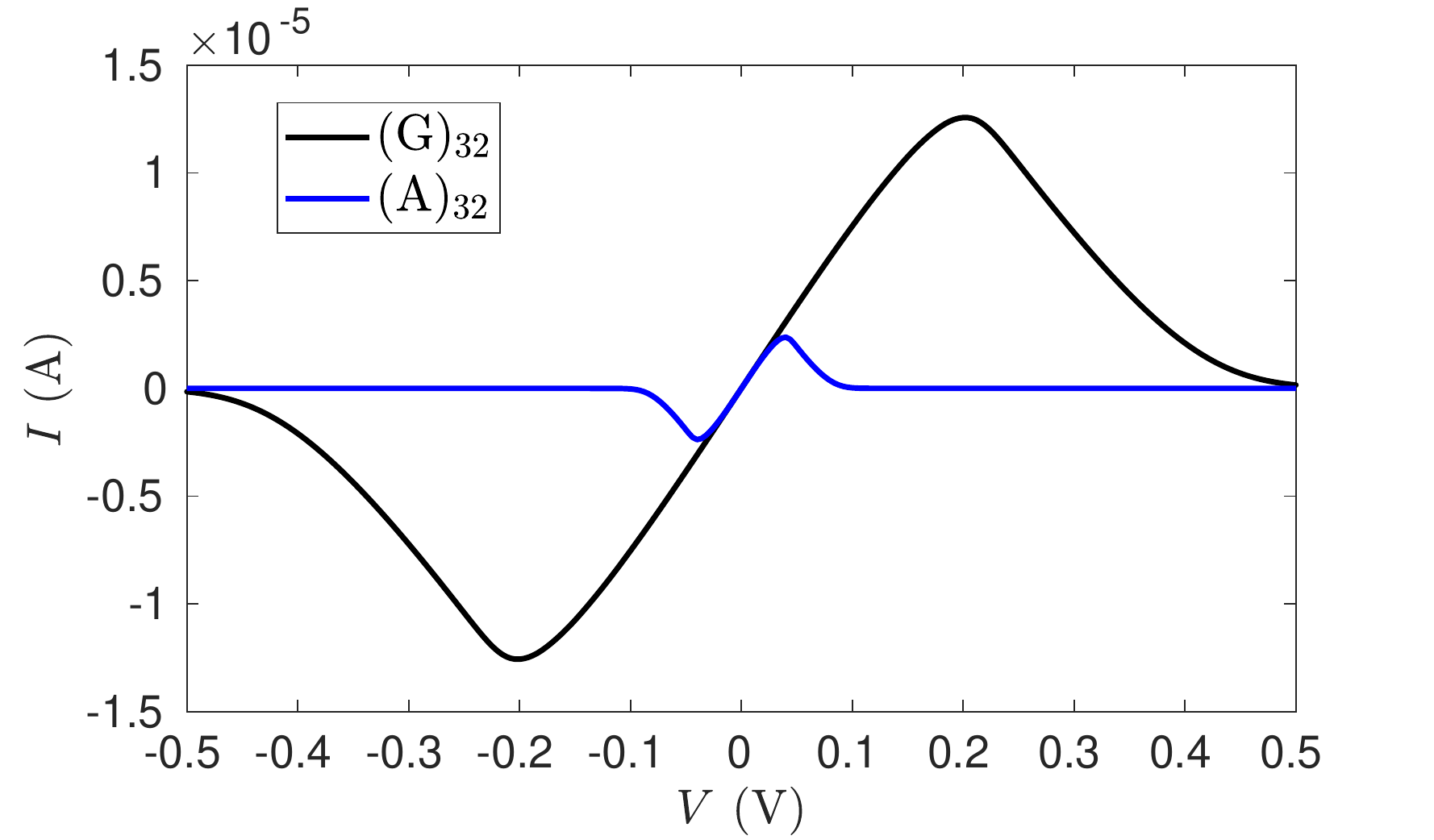}
	\caption{I-V curves of (G)$_{32}$ and (A)$_{32}$ segments.}
	\label{fig:polyGpolyA}
\end{figure}

Furthermore, as far as transport properties are concerned, different results  occur for different parameter values. For example, we have been able to reproduce the results reported for the transmission coefficients in Refs.~\cite{Roche:2003, Macia:2005, Guo:2007}, and for the I-V curves in Ref.~\cite{Macia:2005}, using the corresponding parametrizations, which are different from the one used here (all with equal hopping integrals). Different shapes as well as current-voltage regimes can be obtained, if the parameters are modified. For example, in Ref.~\cite{Oliveira:2014} where microRNA chains are studied, taking different hopping integrals between nucleotides but of significantly larger magnitude than the ones used here, the authors report currents in the nA regime for voltages up to $16$ V. These curves have been reproduced as well. The difference in the current-voltage regimes can also be seen be comparing the I-V curves of the homogeneous (G)$_m$ and (A)$_m$ segments (Fig.\ref{fig:polyGpolyA}), which, due to their sequential simplicity, represent the most efficient cases for charge transport. The curves have been calculated for $E_M = E_{G-C}$ ($E_{A-T}$) for the former (latter) case, i.e., in the center of the bands, with $t_M = -0.5$ eV, and ideal and symmetric coupling conditions. Since the leads are aligned with the band centers, the only defining factor of the current-voltage regime is the value of the hopping parameter $t_{GG}$ ($t_{AA}$). Since $t_{GG} > t_{AA}$, (G)$_m$ segments display greater currents than (A)$_m$ segments ($\sim 10 $ $\mu$A vs. $\sim 1 $ $\mu$A) and lie in a larger bias regime. Generally, increasing the value of the hopping parameter results in increase of both the current magnitude and the voltage regime, until the states of the segment reach the bandwidth of the leads. For both I-V curves, the conductance at zero bias is equal to the quantum of conductance, i.e., $\pdv{I}{V}\big|_{V=0} = G_0 = \tfrac{2e^2}{h} \approx$ 7.748$\times 10^{-5}$ S.

As discussed in Sec. \ref{sec:IV} (cf. Fig. \ref{fig:IVleads}), the occurrence of voltage gaps in the I-V curves depends on the relative position of the Fermi level of the leads and the eigenenergies of the segments. For example, a typical semiconducting I-V curve occurs for (G)$_{30}$ segments, if we set $E_M - E_{G-C} = 0.3$ eV (i.e. for $E_M$ lying outside the band of the segment), with a voltage gap of $\approx 0.7$ V and currents $\sim 1$ nA. This is in accordance with the experimental I-V curves reported for the same system in Ref.~\cite{Porath:2000}, where the authors also attribute the voltage gap to the offset between the Fermi level of the electrode and the energy levels of the (G)$_{30}$ segment.

\section{Conclusion} \label{sec:Conclusion}

We comparatively studied periodic and deterministic aperiodic sequences including quasi-periodic (Thue-Morse, Fibonacci, Period Doubling, Rudin-Shapiro), fractals (Cantor, generalized Cantor), Kolakoski and random binary sequences within the framework of the Tight-Binding wire model. 
We used B-DNA and as a prototype system and the binary alphabet $\{$G,A$\}$.
All segments had their purines on the same strand. 
We gained a better understanding of the interplay between the intricacy of the segments and their spectral, localization and charge transport properties. 
We took differences in hopping parameters between successive monomers into account. 
This led to a more realistic evaluation of the role the sequence intricacy plays in the aforementioned properties.

We determined the number and occurrence percentage of all possible base-pair triplets that can be found within these segments, as well as their autocorrelation functions. Our results showed that there is a relation between the number of possible triplets, the existence of dominant triplets and the strength of correlations. 

We calculated the eigenenergies, the density of states, and the integrated density of states. The allowed eigenenergies of all studied deterministic aperiodic segments lie within the interval defined by the eigenspectrum of random sequences. In all deterministic aperiodic segments, there exist energy steps in the relative normalized IDOS, equal to the occurrence percentages of the possible monomer triplets. This observation establishes a clear relation between the sequence intricacy and the spectral properties.

Furthermore, we calculated the Lyapunov exponents and showed that the sequence intricacy, the relative presence of each monomer, and the values of the TB parameters play major role in the degree of eigenstates localization. Generally, sequences with strong correlations posses less localized states. 

Next, we connected the segments to semi-infinite homogeneous leads and studied the zero-bias transmission coefficients, reaching similar conclusions regarding their transparency to incident carriers. 

We also studied the current-voltage characteristics of the segments, using the 
Landauer-B\"{u}ttiker formalism. We showed that the shape of the curves and the magnitude of the currents strongly depends on the leads' on-site energy (Fermi level). The current-voltage characteristics were calculated for two values of the latter, corresponding to positions that catch the energy regions of interest. For the parametrization used, we found that periodic binary segments can carry currents in the $\mu$A regime. Several deterministic aperiodic segments (specifically, Fibonacci, Period-doubling, Cantor and generalized Cantor) can also display rather large currents, namely in the nA regime, depending on the Fermi level of the leads. Random sequences hold the smallest currents, in accordance with the weak correlations they posses.  

Finally, the I-V curves of the homogeneous (G)$_m$ and (A)$_m$ segments, due to their sequential simplicity, represent the most efficient cases for charge transport with  conductance at zero bias equal to the quantum of conductance. Typical semiconducting I-V curves occur for these segments when there is a mismatch between their eigenstates and Fermi level of the leads, in accordance with experimental results.
\\
\section*{Acknowledgements}

K. Lambropoulos wishes to acknowledge support by the Hellenic Foundation for Research and Innovation (HFRI) and the General Secretariat for Research and Technology (GSRT), under the HFRI PhD Fellowship grant (GA no 260). The authors would also like to thank Professor E. Maci\'a for his comments on the relation between the IDOS steps and the occurrence percentages of the base-pair triplets.


\bibliography{LS.bib}

\end{document}